\documentclass[12pt,a4paper]{article}
\usepackage[T2A]{fontenc}
\usepackage[cp1251]{inputenc}
\usepackage{cite}
\usepackage{mathtext}
\usepackage{amssymb,amsthm,amsmath}
\usepackage{array}
\usepackage[dvips]{epsfig}
\usepackage[russian, english]{babel}
\usepackage[title,titletoc]{appendix}

\begin{document}
\author{\bf Yu.A. Markov$\!\,$\thanks{e-mail:markov@icc.ru}\,,
M.A. Markova$\!\,$\thanks{e-mail:markova@icc.ru},
and A.I. Bondarenko$\!\,$\thanks{e-mail:370omega@mail.ru}}
\title{Fourth order wave equation\\
in Bhabha-Madhavarao spin\hspace{0.02cm}-$\frac{3}{2}$ theory}
%
%
\date{\it\normalsize Matrosov Institute for System Dynamics and Control Theory SB RAS\\
P.O. Box 1233, 664033 Irkutsk, Russia}
\thispagestyle{empty}
\maketitle{}
\def\theequation{\arabic{section}.\arabic{equation}}
\[
{\bf Abstract}
\]

{\noindent
Within the framework of the Bhabha-Madhavarao formalism, a consistent approach to the derivation of a system of the fourth order wave equations for the description of a spin-$\frac{3}{2}$ particle is suggested. For this purpose an additional algebraic object, the so-called $q$\hspace{0.02cm}-\hspace{0.02cm}commutator ($q$ is a primitive fourth root of unity) and a new set of matrices $\eta_{\mu}$, instead of the original matrices $\beta_{\mu}$ of the Bhabha-Madhavarao algebra, are introduced. It is shown that in terms of the $\eta_{\mu}$ matrices we have succeeded in reducing a procedure of the construction of fourth root of the fourth order wave operator to a few simple algebraic transformations and to some operation of the passage to the limit $z \rightarrow q$, where $z$ is some (complex) deformation parameter entering into the definition of the $\eta$\hspace{0.02cm}-\hspace{0.02cm}matrices. In addition, a set of the matrices ${\cal P}_{1/2}$ and ${\cal P}_{3/2}^{(\pm)}(q)$ possessing the properties of projectors is introduced. These operators project the matrices $\eta_{\mu}$ onto the spins 1/2- and 3/2-\hspace{0.01cm}sectors in the theory under consideration. A corresponding generalization of the obtained results to the case of the interaction with an external electromagnetic field introduced through the minimal coupling scheme is carried out. The application to the problem of construction of the path integral representation in parasuperspace for the propagator of a massive spin-$\frac{3}{2}$ particle in a background gauge field within the Bhabha-Madhavarao approach is discussed.
}



\newpage

\section{Introduction}
\setcounter{equation}{0}

In our previous paper \cite{markov_2015} we have considered a question of the construction of cubic root of the third order wave operator for a massive spin-1 particle within the framework of the Duffin-Kemmer-Petiau (DKP) theory. For this purpose we have introduced a new set of the matrices $\eta_{\mu}$, instead of the original matrices $\beta_{\mu}$ of the DKP-algebra. We have shown that in terms of these matrices a procedure of the construction of cubic root of the third order wave operator is reduced to a few simple algebraic transformations and to a certain operation of the passage to the limit $z \rightarrow q$, where $z$ is some (complex) deformation parameter and $q$ is a primitive cubic root of unity. A corresponding generalization of the obtained result to the case of the interaction with an external electromagnetic field introduced through the minimal coupling scheme was also suggested.\\
\indent
In this paper we would like to expand the ideas of paper \cite{markov_2015} to the case of a massive particle with the spin 3/2. There are a long history of the spin-3/2 theory and a substantial body of publications starting with the pioneer papers by Dirac \cite{dirac_1936} and Fierz and Pauli \cite{fierz_1939, fierz_pauli_1939}. Below we will consider in the main only those works that immediately concern to the subject of our research.\\
\indent
It is common knowledge that the three-index spinors $a_{\alpha \beta}^{\dot{\gamma}}$ and $b_{\gamma}^{\dot{\alpha} \dot{\beta}}$ symmetric in their dotted and undotted indices supplemented by the auxiliary spin-1/2 spinors $c_{\alpha}$ and $d^{\dot{\alpha}}$ are the basis of the Dirac-Fierz-Pauli approach in the description of the massive spin-3/2 particle. Further, the field equations for these spinors were rewritten by K.K. Gupta \cite{gupta_1952} and S.N. Gupta \cite{gupta_1954} in the form analogous to that of the Dirac equation for a spin-1/2 particle:
\begin{equation}
(\beta_{\mu}\partial_{\mu} + m I)\hspace{0.02cm}\Psi(x) = 0,
\label{eq:1q}
\end{equation}
where $\Psi(x)$ is the 16-component wave function; $m$ is the mass of particle; $I$ is the unity matrix and the matrices $\beta_{\mu}$ satisfy the relation
\begin{equation}
\sum_{({\cal P})} (\beta_{\mu}\beta_{\nu} - \delta_{\mu\nu})\beta_{\lambda}\beta_{\sigma} = 0.
\label{eq:1w}
\end{equation}
Here, the symbol $\sum_{(\cal P)}$ denotes permutation over all the free indices $\mu, \nu, \lambda$ and $\sigma$. Throughout all the paper we put $\hbar\!=\!c\!=\!1$ and use Euclidean metric $\delta_{\mu \nu} = (+, +, +, +)$. The Greek letters $\mu, \nu,\ldots$ run from 1 to 4. It follows from Eqs.\,(\ref{eq:1q}) and (\ref{eq:1w}) that the function $\Psi$ satisfies the second order wave equation
\begin{equation}
(\hspace{0.02cm}\Box - m^2)\hspace{0.02cm}\Psi(x) = 0,
\label{eq:1e}
\end{equation}
where $\Box \equiv \partial_{\mu}\partial_{\mu}$ is  the d'Alembert operator and a summation over a repeated index is understood. In spite of relatively low order of the matrices $\beta_{\mu}$ and the fact that the algebra (\ref{eq:1w}) leads to the standard single-mass Klein-Gordon-Fock equation (\ref{eq:1e}) with only one spin state, one of essential drawback of this algebra is that the latter is not finite and probably here there exists an infinite number of inequivalent irreducible representations. Here, it is necessary to impose an additional stronger algebraic relation for the matrices $\beta_{\mu}$ compatible with original one (\ref{eq:1w}) to make this algebra finite\footnote{\,The situation here is completely similar to the Duffin-Kemmer-Petiau case. The DKP algebra
\begin{equation}
\beta_{\mu}\beta_{\nu}\beta_{\lambda} + \beta_{\lambda}\beta_{\nu}\beta_{\mu} =
\delta_{\mu\nu}\beta_{\lambda} + \delta_{\lambda\nu}\beta_{\mu}
\label{eq:1r}
\end{equation}
is unique finite-dimensional subalgebra of the abstract infinite algebra generated by the relation similar to (\ref{eq:1w})
\[
\sum_{({\cal P})} (\beta_{\mu}\beta_{\nu} - \delta_{\mu\nu})\beta_{\lambda} = 0.
\]
This infinite algebra was considered in papers \cite{harish-chandra_1947, bhabha_1949, tzou_1957, carpi_1969, solovyov_1995, gaidukevich_2012}, and it was analyzed in greater detail in \cite{hurley_1974}.}. Another disadvantage of the algebra (\ref{eq:1w}) is that the
infinitesimal generators of rotation $I_{\mu\nu}$ can not be represented by means of the commutator $[\hspace{0.02cm}\beta_{\mu},\beta_{\nu}]$ and ipso facto the algebra does not lead to the relation
\begin{equation}
[\hspace{0.03cm}[\hspace{0.03cm}\beta_{\mu},\beta_{\nu}],\beta_{\lambda}\hspace{0.02cm}]
= \beta_{\mu}\hspace{0.02cm}\delta_{\nu\lambda} - \beta_{\nu}\hspace{0.02cm}\delta_{\mu\lambda}
\label{eq:1t}
\end{equation}
required for the relativistic covariance of the corresponding wave equation (\ref{eq:1q}).\\
\indent
Nontrivial example of a finite algebra for the spin-3/2 matrices $\beta_{\mu}$ was given by Harish-Chandra \cite{harish-chandra_1948}. Within the framework of the Harish-Chandra algebra,
the matrices $\beta_{\mu}$ have the following structure:
\begin{equation}
\beta_{\mu} = {\it \Gamma}_{\!\hspace{0.02cm}\mu}\hspace{0.01cm} R + i\hspace{0.01cm}B_{\mu},
\label{eq:1tt}
\end{equation}
where ${\it \Gamma}_{\!\hspace{0.02cm}\mu}$ and $B_{\mu}$ commute with each other and satisfy the Dirac and the Duffin-Kemmer-Petiau commutation rules, respectively. The matrix $R$ is expressed in terms of the $B_{\mu}$ and satisfies, in turn, the following relations:
\[
R^{2} = 1,
\qquad
R\hspace{0.015cm}B_{\mu} + B_{\mu}R =0,
\qquad
R\hspace{0.015cm}{\it \Gamma}_{\!\hspace{0.025cm}\mu} =
{\it \Gamma}_{\!\hspace{0.025cm}\mu}\hspace{0.01cm} R.
\]
The algebra suggested in fact represents the Kronecker product of the Dirac and DKP algebras.
Sometimes the algebras of such type are called the ``parametric'' ones. The $\beta_{\mu}$ matrices are expressed covariantly in terms of other matrices whose commutation rules are known, or can be easily found. By elimination of these ``parameters'' the explicit commutation rules for the $\beta_{\mu}$ matrices can be obtained. The Harish-Chandra approach can be considered as a  certain analog of the Rarita-Schwinger one \cite{rarita_1941}. In the former case the ``vector'' part of the spin degree of freedom is described within the framework of the DKP formalism rather than the usual vector formalism as it takes place in the paper by Rarita and Schwinger. The disadvantage of the proposed algebra is its high order, namely $16\times126=2016$. For this reason and also for a number of the other reasons of a more fundamental character, the Harish-Chandra algebra is not suitable in practical respect for the description of a particle with the spin 3/2. In the paper \cite{petras_1955} by Petr\`{a}\v{s} a somewhat different version of the parametric algebra was suggested, where the ``vector'' part of the spin degree of freedom is described by the use of ``tensor'' matrices $B_{\mu\nu}$ that are subject to the commutation rules
\[
B_{\mu\nu}B_{\lambda\sigma} = \delta_{\nu\lambda}B_{\mu\sigma},
\qquad
{\it \Gamma}_{\!\hspace{0.025cm}\mu}\hspace{0.01cm}B_{\nu\lambda} =
B_{\nu\lambda}{\it \Gamma}_{\!\hspace{0.025cm}\mu}.
\]
Instead of (\ref{eq:1tt}), we now have
\[
\beta_{\mu} = {\it \Gamma}_{\!\hspace{0.02cm}\mu} +
\frac{1}{\sqrt{3}}\,{\it \Gamma}_{\!\hspace{0.02cm}\nu}\bigl(B_{\nu\mu} - B_{\mu\nu}\bigr).
\]
Within the framework of the Petr\`{a}\v{s} approach, Fradkin \cite{fradkin_1957} has considered the interaction of a spin-3/2 particle with an electromagnetic field. However, the final relativistic invariant equation turns out to be rather cumbersome, essentially non-linear relative to the electromagnetic field and involves the interaction terms with dipole and quadrupole kinematic moments of the
spin-3/2 particle.\\
\indent
Finally, note one further approach in the description of the spin-3/2 massive particle, which in the subsequent discussion we will use as a basis for our consideration. In the papers \cite{bhabha_1944, bhabha_1_1945, bhabha_2_1945, bhabha_3_1945, bhabha_1949, bhabha_1951, bhabha_1952} Bhabha set up a new theory for relativistic particles of any spin. Bhabha studied in detail the algebraic aspects of a first order wave equation in the form (\ref{eq:1q}) with the only assumption that the transformation properties of the wave function, and hence the spin of the particle, are determined entirely by the infinitesimal transformations $I_{\mu \nu}$ given by the following expression:
\begin{equation}
I_{\mu\nu} = [\hspace{0.02cm}\beta_{\mu},\beta_{\nu}].
\label{eq:1y}
\end{equation}
Equivalently, the $\beta$-matrices must satisfy (\ref{eq:1t}) for all spins.\\
\indent
Although equation (\ref{eq:1q}) has a compact expression without subsidiary conditions, it has drawback: its solutions correspond not to unique spins and masses, but to several spins and masses. Thus, for example, in the case of spin 3/2, the wave function $\Psi(x)$ must satisfy the {\it multimass} Klein-Gordon-Fock equation, instead of (\ref{eq:1e})
\begin{equation}
\Biggl[\,\Box - \biggl(\frac{m}{3\!\hspace{0.03cm}/\hspace{0.04cm}\!2}\biggr)^{\!\!2}\hspace{0.03cm}\Biggr]
\Biggl[\,\Box - \biggl(\frac{m}{1\!\hspace{0.03cm}/\hspace{0.04cm}\!2}\biggr)^{\!\!2}\hspace{0.03cm}\Biggr]
\hspace{0.01cm}\Psi(x) = 0
\label{eq:1u}
\end{equation}
and, in addition to the spin $s=3/2$ of interest, the wave function also contains the lower spin component $(s-1)=1/2$. It is necessary to select the spin 3/2 and spin 1/2 sectors of the theory from a general representation of the $\beta$-matrices through a set of projection operators. Such a set of operators will be introduced in section 5 of this work.\\
\indent
In addition, it was found that it has been extremely difficult task to find the explicit expressions for algebras, to which the matrices $\beta_{\mu}$ in Bhabha's theory have to satisfy. The paper by Madhavarao \cite{madhavarao_1942} materially simplifies the construction of the algebras. Madhavarao was the first who has defined an explicit form of these algebras for the special cases of the 3/2 and 2 spins. Many years later, some particular commutation relations of the $\beta$-matrices were derived by Baisya \cite{baisya_1995} for the case of spin 5/2.\\
\indent
For the spin 3/2, the algebra Bhabha-Madhavarao is of the following form:
\[
2\hspace{0.015cm}\bigl(\beta_{\mu}\beta_{\nu}\beta_{\lambda}\beta_{\sigma} +
\beta_{\mu}\beta_{\sigma}\beta_{\lambda}\beta_{\nu} +
\beta_{\nu}\beta_{\lambda}\beta_{\sigma}\beta_{\mu} +
\beta_{\sigma}\beta_{\lambda}\beta_{\nu}\beta_{\mu}\bigr)
\]
\begin{equation}
\begin{split}
=\hspace{0.02cm} 3\hspace{0.015cm}&\bigl(\beta_{\mu}\beta_{\nu} + \beta_{\nu}\beta_{\mu}\bigr)\delta_{\lambda\sigma} +
3\hspace{0.02cm}\bigl(\beta_{\mu}\beta_{\sigma} + \beta_{\sigma}\beta_{\mu}\bigr)\delta_{\nu\lambda} \\[1ex]
+\, &\bigl(\beta_{\lambda}\beta_{\sigma} + \beta_{\sigma}\beta_{\lambda}\bigr)\delta_{\mu\nu} +
\bigl(\beta_{\lambda}\beta_{\nu} + \beta_{\nu}\beta_{\lambda}\bigr)\delta_{\mu\sigma} \\[1ex]
+\, &\bigl(\beta_{\nu}\beta_{\sigma} + \beta_{\sigma}\beta_{\nu}\bigr)\delta_{\mu\lambda} +
\bigl(\beta_{\mu}\beta_{\lambda} + \beta_{\lambda}\beta_{\mu}\bigr)\delta_{\nu\sigma}
\end{split}
\label{eq:1i}
\end{equation}
\[
-\,\frac{3}{2}\,\bigl(\delta_{\mu\nu}\delta_{\lambda\sigma} + \delta_{\mu\lambda}\delta_{\nu\sigma} + \delta_{\nu\lambda}\delta_{\mu\sigma}\bigr)I.
\]
The algebra has considerably more complicated structure in comparison with the Duffin-Kemmer-Petiau algebra, Eq.\,(\ref{eq:1r}), and is its immediate extension to the case of spin 3/2. In spite of the awkwardness of the expression (\ref{eq:1i}), the order of this algebra is sufficiently small relative to the Harish-Chandra algebra ($16\times42=672$ versus 2016). Besides, the structure of this algebra perfectly coincides with the structure of the so-called para-Fermi algebra of order $p=3$ obtained by Kamefuchi and Takahashi \cite{kamefuchi_1962}, and Scharfstein \cite{scharfstein_1963, scharfstein_1966} (see also Ryan and Sudarshan \cite{ryan_1963}). This circumstance may be very helpful in the construction of the path integral representation for the spin-3/2 particle propagator interacting with a background gauge field.\\
\indent
The algebra (\ref{eq:1i}) has been analyzed in detail in a few papers \cite{madhavarao_1946, madhavarao_1947, venkatachaliengar_1954, srinivasarao_1965}. In particular, one of the most
important conclusion which can be done based on the paper by Madhavarao {\it et al.} \cite{madhavarao_1946} lies in the fact that this algebra is the direct product of the corresponding Clifford-Dirac algebra $D_{\gamma}$ and the algebra called $A_{\xi}$-algebra generated by the matrices $\xi_{\mu}$. In other words the matrices $\beta_{\mu}$ can be presented as $\gamma_{\mu} \otimes \xi_{\mu}$ or $\xi_{\mu} \otimes \gamma_{\mu}$. Further, for definiteness we set
\begin{equation}
\hspace{3.5cm}
\beta_{\mu} = \gamma_{\mu} \otimes \xi_{\mu} \qquad \mbox{(no summation!)}.
\label{eq:1o}
\end{equation}
Here, $\gamma_{\mu}$ is usual (Euclide) $4 \times 4$ Dirac matrices obeying the algebra
\begin{equation}
\{\gamma_{\mu},\gamma_{\nu}\} = 2\hspace{0.02cm}\delta_{\mu\nu},
\label{eq:1p}
\end{equation}
where $\{\hspace{0.02cm},\}$ designates anticommutator. The algebraic relations for the $\xi_{\mu}$ matrices are given in Appendix A. By using the basic rule for the multiplication of matrices
\[
(A\otimes B)(C\otimes D) = (AC\otimes BD),
\]
the decomposition (\ref{eq:1o}) and relations for the $\xi_{\mu}$ matrices, Eqs.\,(\ref{ap:A1})\,--\,(\ref{ap:A4}), by straightforward substitution (\ref{eq:1o}) into (\ref{eq:1i}) one can verify that (\ref{eq:1i}) is reduced to identity. It is more simple, however, to consider various particular cases of the algebra (\ref{eq:1i}), which are written out in \cite{madhavarao_1942}. The direct product (\ref{eq:1o}) considerably simplifies the problem of determining the irreducible representations of the Bhabha-Madhavarao algebra (\ref{eq:1i}) (see Appendix A).\\
\indent
Further, H\"onl and Boerner in the fundamental paper \cite{honl_1950} came to the representation (\ref{eq:1o}) from a different point of view. The authors also analyzed the equation in the form (\ref{eq:1q}). They have put an approach suggested by Louis de Broglie \cite{broglie_1943} in the construction of the theory for particles with an arbitrary spin (so-called the {\it method of fusion}) in the basis of this analysis (see also Kramers {\it et al.} \cite{kramers_1941} and Shelepin \cite{shelepin_1958, shelepin_1960}). In the de Broglie theory the matrices $\beta_{\mu}$ are defined solely in terms of the Dirac matrices $\gamma_{\mu}$ (and the unity $4 \times 4$ matrix $I$). Thus, for instance,  for the case of a particle of the maximum spin 3/2, the matrices $\beta_{\mu}$ have the following structure:
\begin{equation}
\beta_{\mu} = \gamma_{\mu} \otimes I \otimes I + I\otimes\gamma_{\mu}\otimes I +
I\otimes I\otimes\gamma_{\mu},
\label{eq:1pp}
\end{equation}
and automatically satisfy the relation (\ref{eq:1t}). H\"onl and Boerner on the basis of the reduction method suggested by them, have shown that the $\beta$-matrices (\ref{eq:1pp}) admit the decomposition in the form (\ref{eq:1o}), where the matrices $\xi_{\mu}$ can be presented by $5 \times 5$ (once), $4 \times 4$ (twice) and $1 \times 1$ (thrice) irreducible matrices  that is in agreement with the conclusions of the paper by Madhavarao {\it et al.} \cite{madhavarao_1946}.\\
\indent
It should be also mentioned the little-known but rather interesting paper by \'{U}lehla \cite{ulehla_1954, ulehla_1958} in which the author independently concludes a possibility of representation of the matrices $\beta_{\mu}$ in the form of the direct product (\ref{eq:1o}). \'{U}lehla did not analyzed any concrete algebra of the (\ref{eq:1i}) type, but he has directly dealt with a system of matrix equations for the infinitesimal generators $I_{\mu \nu}$ (i.e. without appeal to the representation (\ref{eq:1y})):
\[
[\hspace{0.03cm}I_{\mu\nu},\beta_{\lambda}\hspace{0.02cm}]
= \beta_{\mu}\hspace{0.02cm}\delta_{\nu\lambda} - \beta_{\nu}\hspace{0.02cm}\delta_{\mu\lambda},
\]
\[
[\hspace{0.03cm}I_{\mu\nu},I_{\lambda\sigma}\hspace{0.02cm}] =
- \delta_{\mu\lambda} I_{\nu\sigma}
+ \delta_{\nu\lambda}I_{\mu\sigma}
+ \delta_{\mu\sigma}I_{\nu\lambda}
- \delta_{\nu\sigma}I_{\mu\lambda}
\]
plus two equations with the matrix $Z$ of the space inversion. This system has been added by the only requirement that the magnitude of the spin be limited from above. For the spin 3/2, in particular, this means that the spin operator $S_{\mu\nu}$ must satisfy the equation
\[
\left[S_{\mu\nu}^{2} - \biggl(\frac{3}{2}\biggr)^{\!\!2}\hspace{0.03cm}\right]\!
\left[S_{\mu\nu}^{2} - \biggl(\frac{1}{2}\biggr)^{\!\!2}\hspace{0.03cm}\right]
= 0
\]
with $S_{\mu\nu}\equiv i\hspace{0.02cm}I_{\mu\nu}$. However, the degrees of the irreducible represantations of the corresponding $A_{\xi}$\hspace{0.01cm}-\hspace{0.01cm}algebra obtained in \cite{ulehla_1954, ulehla_1958} did not coincide with those obtained in the paper \cite{madhavarao_1946}.\\
\indent
We now proceed to discuss the Bhabha-Madhavarao theory for the case of the presence of an external gauge field in the system. In the paper by Nowakowski \cite{nowakowski_1998} devoted to the problem of electromagnetic coupling in the Duffin-Kemmer-Petiau theory one quite unusual circumstance relating to a second order DKP equation has been pointed out. This circumstance is connected with the fact that the second order Kemmer equation \cite{kemmer_1939} lacks a back-transformation in the presence of a background gauge field which would allow us to obtain solutions of the first order DKP equation from solutions of the second order equation.\\
\indent
A completely similar circumstance takes place within the framework of the multimass Bhabha theory: the fourth order wave equation (\ref{eq:1u}) in the presence of a background electromagnetic field lacks a back-transformation which would allow one to obtain solutions of the first order equation (\ref{eq:1q}) from the solutions of the fourth order equation. The reason for the latter is that the multimass\footnote{\,The term {\it multimass} in this case implies that a product of the divisor (\ref{eq:1d}) and the Bhaba operator (\ref{eq:1f}), by virtue of the algebra of $\beta$-matrices (\ref{eq:1i}), leads to the multimass Klein-Gordon-Fock operator:
\begin{equation}
d(\partial)\hspace{0.02cm} L(\partial) =  L(\partial)\hspace{0.02cm}d(\partial) =
\bigl(\hspace{0.02cm}\Box - \frac{4}{9}\, m^2\bigr)
\bigl(\hspace{0.02cm}\Box - 4\hspace{0.03cm} m^2\bigr)
\hspace{0.02cm}I.
\label{eq:1a}
\end{equation}
Within the framework of the single-mass formalism by Takahashi, Umezawa and Visconti \cite{umezawa_1956, takahashi_book} the divisor $d(\partial)$ has a somewhat more simple structure
\[
d(\partial) = \frac{1}{m^{2}}\,\bigl[\hspace{0.02cm}m^{3} - m^{2}(\beta\cdot\partial) + m\hspace{0.02cm}(\beta\cdot\partial)^{2}
- m\hspace{0.03cm}\Box\hspace{0.02cm} I - (\beta\cdot\partial)^{3} + (\beta\cdot\partial)\hspace{0.02cm}\Box\hspace{0.02cm}\bigr].
\]
The divisor satisfies the relation
\begin{equation}
d(\partial)\hspace{0.02cm}L(\partial) =  L(\partial)\hspace{0.02cm}d(\partial) =
(\hspace{0.02cm}\Box - m^2) I.
\label{eq:1s}
\end{equation}
In so doing, the matrices $\beta_{\mu}$ obey the algebra (\ref{eq:1w}), instead of (\ref{eq:1i}).} Klein-Gordon-Fock divisor in the spin-3/2 case \cite{baisya_1970, nagpal_1973}
\begin{equation}
d(\partial) = - \frac{16}{9}\,\bigl[\hspace{0.02cm}m^{3} - m^{2}(\beta\cdot\partial) + m\hspace{0.02cm}(\beta\cdot\partial)^{2}
- \frac{5}{2}\,m\hspace{0.03cm}\Box\hspace{0.02cm} I - (\beta\cdot\partial)^{3} + \frac{5}{2}\,(\beta\cdot\partial)\hspace{0.02cm}\Box\hspace{0.02cm}\bigr]
\label{eq:1d}
\end{equation}
ceases to be commuted with the original Bhabha operator
\begin{equation}
L(\partial) = \beta\cdot\partial + m\hspace{0.02cm}I,
\label{eq:1f}
\end{equation}
when we introduce the interaction with an external electromagnetic field within the framework of the minimal coupling scheme: $\partial_{\mu}\rightarrow D_{\mu} = \partial_{\mu} + i\hspace{0.02cm}e A_{\mu}(x)$, i.e.
\[
[\hspace{0.04cm}d(D), L(D)\hspace{0.03cm}] \neq 0.
\]
Here, $\beta\cdot\partial \equiv \beta_{\mu}\partial_{\mu}$. To achieve the commutativity between the divisor $d(D)$ and operator $L(D)$ in the presence of an external gauge field, we have to give up the requirement that a product of these two operators is an operator of the multimass Klein-Gordon-Fock type
\[
d(D) L(D) \neq \bigl(D^{2} - \frac{4}{9}\,m^{2}\bigr)\bigl(D^2 - 4\hspace{0.02cm}m^2\bigr)I +
{\cal G}\hspace{0.02cm}[A_{\mu}],
\]
where ${\cal G}\hspace{0.02cm}[A_{\mu}]$ is a functional of the potential $A_{\mu}$, which vanishes in the interaction free case.\\
\indent
In constructing a divisor for the spin-3/2 Bhabha operator $L(D)$ that would maintain the commutative property in the presence of the external electromagnetic field, we will closely follow ideology suggested in our paper \cite{markov_2015} for the spin-1 case. It may be supposed that the construction of the desired divisor will be related to the problem of the construction of fourth root of a certain fourth-order wave operator. Besides, as in the DKP case one can expect that instead of the original matrices $\beta_{\mu}$ here its ``deformed'' variant may be required, where one of the primitive fourth roots of unity serves as a deformation parameter.\\
\indent
The lack of commutativity of the reciprocal operator $d(D)$ and the Bhabha operator $L(D)$ in the presence of an external gauge field has another negative consequence. It does not give a possibility within the framework of the Bhabha-Madhavarao approach to construct the path integral representation for the Green's function of a massive spin-3/2 particle in the background gauge field in a spirit of the approaches developed for a spin 1/2 particle (see, for example, Fradkin and Gitman \cite{fradkin_1991}). We will briefly discuss this question in section 9.\\
\indent
It should be also noted that the multimass divisor $d(D)$  with a minimal electromagnetic coupling for the spin-3/2 case was first introduced by Nagpal \cite{nagpal_1974} and Krajcik and Nieto \cite{krajcik_1976}. The divisor has been intensively used in analysis of causality violation in higher spin theories in the presence of the electromagnetic field. In particular, Nagpal in paper \cite{nagpal_1974} has used an alternative algebra of the $\beta$-matrices, instead of (\ref{eq:1w}),
\begin{equation}
\begin{split}
&\sum_{({\cal P})}\, \bigl(\beta_{\mu}\beta_{\nu} - \frac{1}{4}\,\delta_{\mu\nu}\bigr)\bigl(\beta_{\lambda}\beta_{\sigma} - \frac{9}{4}\,\delta_{\lambda\sigma}\bigr) \\
=
&\sum_{({\cal P})} \bigl(\beta_{\mu}\beta_{\nu}\beta_{\lambda}\beta_{\sigma} -
\frac{5}{2}\,\beta_{\mu}\beta_{\nu}\delta_{\lambda\sigma}
+
\frac{9}{16}\,\delta_{\mu\nu}\delta_{\lambda\sigma}\bigr) = 0.
\end{split}
\label{eq:1g}
\end{equation}
It can be shown that the matrices $\beta_{\mu}$ obeying the Bhabha-Madhavarao algebra (\ref{eq:1i}) satisfy (\ref{eq:1g}). The converse is obviously false. The algebra (\ref{eq:1g})  in view of complete symmetry in the vector indices is possible more convenient in some applications (as well as the algebra (\ref{eq:1w})). However, the algebra generated by algebraic quantities satisfying only (\ref{eq:1g}) is not finite and one would expect there to be an infinite number of inequivalent irreducible sets of matrices satisfying (\ref{eq:1g}), all except three of which (see appendix A) will not satisfy (\ref{eq:1i}). Besides, in contrast to the Bhabha-Madhavarao algebra (\ref{eq:1i}), the algebra (\ref{eq:1g}) does not lead to the relation (\ref{eq:1t}) required for relativistic covariance of the corresponding wave equation.\\
\indent
If one takes as a general guiding principle the considerations in our paper \cite{markov_2015} for the spin-1 case, then the next step to the spin-3/2 case will be the following extension: as a basis we take the fourth roots of unity $(q,\, q^{2\!},\, q^{3\!},\, 1)$, where
\begin{equation}
q = i, \qquad q^{2} = -1, \qquad q^3= -i,
\label{eq:1h}
\end{equation}
and as the matrices $\beta_{\mu}$ we take the $\beta$\hspace{0.02cm}-\hspace{0.02cm}matrices satisfying the Bhabha-Madhavarao algebra (\ref{eq:1i}). The starting point of all further considerations
will be the following expression for the fourth-order massive wave operator:
\begin{equation}
\bigl[\hspace{0.02cm}\bigl(\beta\cdot\partial\bigr) + q\hspace{0.015cm}m\hspace{0.015cm}I\bigr]
\hspace{0.02cm}
\bigl[\hspace{0.02cm}\bigl(\beta\cdot\partial\bigr) + q^{2}\hspace{0.015cm}m\hspace{0.015cm}I\bigr]
\hspace{0.02cm}
\bigl[\hspace{0.02cm}\bigl(\beta\cdot\partial\bigr) + q^{3}\hspace{0.015cm}m\hspace{0.015cm}I\bigr]
\hspace{0.02cm}
\bigl[\hspace{0.02cm}\bigl(\beta\cdot\partial\bigr) + m\hspace{0.02cm}I\bigr]
\label{eq:1j}
\end{equation}
\[
=
\bigl(\beta\cdot\partial\bigr)^{4} + m^{2}(q + q^{3}) \bigl(\beta\cdot\partial\bigr)^{2} +
q^{2}m^{4}I.
\]
Here, we have used one of the basic properties of roots of unity, namely,
\begin{equation}
1 + q + q^{2} + q^{3} = 0.
\label{eq:1k}
\end{equation}
In view of the algebra (\ref{eq:1i}) the first term on the right-hand side of (\ref{eq:1j}) can be presented as follows:
\begin{equation}
\bigl(\beta\cdot\partial\bigr)^{4} = \frac{5}{2}\,\bigl(\beta\cdot\partial\bigr)^{2\hspace{0.02cm}}\Box -
\frac{9}{16}\,{\Box}^{\hspace{0.02cm}2}.
\label{eq:1l}
\end{equation}
\indent
However, it is to be special noted that a set of the fourth roots of unity possesses a qualitative distinction from the corresponding set of the cubic roots of unity, which we have used in \cite{markov_2015}. The matter is that for the set $(q,\,q^{2\!},\,q^{3\!},\,1)$ we have two more weak properties than the general property (\ref{eq:1k})
\[
1 + q^{2} = 0, \qquad q + q^{3} = 0.
\]
As we will see from a subsequent consideration, the existence of two ``subalgebras'' $(1,\,q^2\hspace{0.02cm})$ and $(q,\,q^3\hspace{0.02cm})$ is closely connected with the presence in the $\beta$-matrices algebra of two spin sectors, one of which is associated with the spin 3/2, and another\footnote{\,A set of roots $(1, q)$ represents that of the square roots of unity. This set gives us a possibility to write the single-mass Klein-Gordon-Fock operator (\ref{eq:1s}) in the form of a product of two first order differential operators within the Dirac theory (see Eq.\,(\ref{eq:1i}) in \cite{markov_2015}).} is with the spin 1/2. It can already be seen on the example of the expression (\ref{eq:1j}). If, instead of  the $q$ in (\ref{eq:1j}), we set the primitive root $i$ (or $-i$), then the right-hand side with the use of the identity (\ref{eq:1l}) takes the form
\begin{equation}
\bigl(\beta\cdot\partial\bigr)^{4} - m^{4}I \equiv
\frac{5}{2}\,\bigl(\beta\cdot\partial\bigr)^{2\hspace{0.02cm}}\Box -
\frac{9}{16}\,{\Box}^{\hspace{0.02cm}2}I - m^{4}I.
\label{eq:1z}
\end{equation}
It is precisely this expression that we accept as the definition of the fourth order wave operator for the spin-3/2 particle.\\
\indent
From the other hand, if in (\ref{eq:1j}) we formally set $q = -1$ (the relation (\ref{eq:1k}) holds in this case also), then we would have on the right-hand side of (\ref{eq:1j})
\begin{equation}
\bigl(\beta\cdot\partial\bigr)^{4} - 2m^{2}\bigl(\beta\cdot\partial\bigr)^{2} + m^{4}I \equiv
\Bigl(\bigl[\hspace{0.02cm}(\beta\cdot\partial) - m\hspace{0.02cm}I\hspace{0.02cm}\bigr]\!
\bigl[\hspace{0.02cm}(\beta\cdot\partial) + m\hspace{0.02cm}I\hspace{0.02cm}\bigr]\Bigr)^{2}.
\label{eq:1x}
\end{equation}
The structure of this expression represents the square of the second order Dirac equation. It can serve as a hint of inevitable involvement of  the spin-1/2 component to the general theory of a particle with the spin 3/2.\\
\indent
Further we can state a question of defining a matrix $A$ such that
\begin{equation}
\bigl[\hspace{0.03cm}A(\hspace{0.02cm}\beta\cdot\partial + m\hspace{0.02cm}I)\bigr]^{4}
=
\frac{1}{m^{2}}\,\biggl\{
\frac{5}{2}\,\bigl(\beta\cdot\partial\bigr)^{2\hspace{0.04cm}}\Box -
\frac{9}{16}\;{\Box}^{\hspace{0.02cm}2}I \biggr\} - m^2I.
\label{eq:1c}
\end{equation}
The relation solves the problem of calculating the fourth root of the fourth order wave operator. In this paper we have attempted to answer this question by using the properties of the Bhabha-Madhavarao theory added by new structures generated by algebra of the fourth roots of unity. We have also performed a generalization of the resulting equations to the case of the presence in the system of an external electromagnetic field.\\
\indent
In closing, we would like to note a very interesting connection between the problem stated here, which in symbolic form is given by equation (\ref{eq:1c}) (and equation (1.14) in \cite{markov_2015} for the spin-1 case) and the mathematical problem concerning linearization of a partial differential equation
\begin{equation}
\sum\limits_{|J| \hspace{0.02cm}=\hspace{0.02cm} m} a_{J}\hspace{0.02cm} \frac{\partial^{|J|}}{\partial X^{\!\hspace{0.02cm}J}}\,\hspace{0.02cm} \psi(x) = c^{m}\hspace{0.02cm}\psi(x),
\label{eq:1v}
\end{equation}
where $J = (j_{1},\ldots,j_{n})$ is a multi-index, $\vert J \vert = j_{1} + \ldots + j_{n}$ and
$X^{J}\!\equiv\!x_{1}^{j_{1}} x_{2}^{j_{2}} \,\ldots\,x_{n}^{j_{n}}$. The coefficients $a_{J}$ and $c$ are scalars. The linearization here is meant as a possibility to present (\ref{eq:1v}) by a first order system
\[
\sum\limits_{i\hspace{0.02cm}=\hspace{0.02cm}1}^{n}\alpha_{i}\hspace{0.02cm}
\frac{\partial}{\partial x_{i}}\,\psi(x) = c\hspace{0.04cm}\psi(x)\hspace{0.02cm}I
\]
with $\alpha_{1}, \alpha_{2},\ldots,\alpha_{n}$ matrices.  This problem was stated by Japanese mathematicians Morinaga and N\={o}no \cite{morinago_1952, nono_1971} a long time ago. The authors observed that the problem is equivalent to solving the linearization problem for forms: to find such matrices $\alpha_i,\,i=1,\ldots,n$ that the equality
\[
\sum\limits_{|J| \hspace{0.02cm}=\hspace{0.02cm} m} a_{J}\hspace{0.02cm} X^{\!\hspace{0.02cm}J}I = \biggl(\sum\limits_{i\hspace{0.02cm}=\hspace{0.02cm}1}^{n}\alpha_{i}\hspace{0.04cm}
x_{i}\biggr)^{\!m}
\]
is fulfilled. Solving this problem leads in turn to necessity of introducing the so-called {\it generalized Clifford algebras} to which the matrices $\alpha$ must satisfy (see, for example \cite{jagannathan_2010}). In more simple version of this generalized algebra, the following requirements on the matrices $\alpha_{i}$
\begin{equation}
\alpha_{i}\hspace{0.03cm}\alpha_{j} = q\hspace{0.03cm}\alpha_{\!\hspace{0.02cm}j}\hspace{0.04cm}\alpha_{i}\hspace{0.02cm},
\quad (i>j),
\qquad \bigl(\alpha_{i}\bigr)^{\!\hspace{0.03cm}m} = I,
\label{eq:1b}
\end{equation}
are imposed. Here, $q$ is a primitive $m$-th root of the unity. This is closely related to our consideration. However, there are two important distinctions: we admit some number coefficients $a_{J}$ in equation (\ref{eq:1v}) themselves can be the fixed matrices and, instead of the conditions (\ref{eq:1b}), we require the fulfillment of more weak equalities of the (\ref{eq:5q}) type. Further development of the ideas of Morinaga and N\={o}no can be found in the papers by Childs \cite{childs_1978} and Pappacena \cite{pappacena_2000}.\\
\indent
The paper is organized as follows. In section 2 the construction of fourth root of the fourth order wave operator (\ref{eq:1z}) is considered. A number of expressions derived here are of decisive importance for the subsequent research. Section 3 is devoted to the derivation of an explicit form of the matrix $\Omega$ which is a kind of a spin-3/2 analog of the $\gamma_{5}$  matrix in the Dirac theory. In constructing this matrix we make use of the properties of the $A_{\xi}$\hspace{0.01cm}-\hspace{0.01cm}algebra. The commutative rules of the $\Omega$ matrix with the $\beta_{\mu}$ matrices are written out. In section 4 an explicit form of the required matrix $A$ in (\ref{eq:1c}) is written out in full. It is shown that the ``na\"{\i}ve'' approach in calculating the fourth root as it is presented by the relation (\ref{eq:1c}) ultimately results in contradiction.\\
\indent
In section 5 a new set of matrices $\eta_{\mu}$, instead of the original ones $\beta_{\mu}$, is introduced. It is shown that these matrices possess rather nontrivial commutative relations with the matrix~$A$, which enable us to reduce the problem of the construction of the desired fourth root to a number of simple algebraic operations. Besides, in this section a set of matrices ${\cal P}_{1/2}$ and
${\cal P}_{3/2}^{(\pm)}(q)$ possessing the properties of projectors is introduced. These operators project the matrices $\beta_{\mu}$ on sectors corresponding to the spins 1/2 and 3/2, correspondingly. Section 6 is concerned with the discussion of various commutation properties of the $\eta$\hspace{0.02cm}-\hspace{0.02cm}matrices. At the end of this section the structure of the projectors ${\cal P}_{3/2}^{(\pm)}(q)$ is carefully analyzed. In section 7 the construction of the fourth root of the fourth-order wave operator for a spin-3/2 particle in terms of the $\eta_{\mu}$\hspace{0.02cm}-\hspace{0.02cm}matrices is considered in detail. For this, a differential operator of the first-order in derivatives, which is singular with respect to the deformation parameter $z$, is introduced.\\
\indent
In section 8 an extension of the findings of the previous sections to the case of the presence of an external electromagnetic field in the system is performed. In section 9 a question of a possible application of the obtained results to the problem of the construction within the framework of the Bhabha-Madhavarao formalism of the path integral representation for the propagator of a spin-3/2 particle in a background gauge field is considered. In concluding section 10 a severe complication arising in the construction of the formalism under examination is briefly discussed.\\
\indent
In Appendix A all of the basic relations of the $A_{\xi}$ algebra are written out and an explicit form of the matrices $\xi_{\mu}$ for the case of the irreducible presentation of degree 4 are given. In Appendix B the solutions of an algebraic system for unknown coefficients of the expansion of the matrix $\Omega$  in the central elements of the $A_{\xi}$ algebra are given. In Appendix C the details of calculating an explicit form of the matrix $A$ to the third power are presented. It is shown that for a proper choice of parameters this matrix will represent in fact hermitian conjugation of the original matrix $A$. In Appendix D the proof of the identity (\ref{eq:8e}) for a product of four covariant derivatives is presented. In Appendix E an explicit form of the interaction terms with an external electromagnetic field containing the spin matrix $S_{\mu\nu}$ is given.

\section{\bf Fourth root of the fourth-order wave operator}
\setcounter{equation}{0}

In this section, we consider a question of the construction of fourth root of the fourth-order wave operator in the form as it was defined by the expression (\ref{eq:1z}). In this case our problem becomes one of constructing such a matrix $A$ for which the relation (\ref{eq:1c}) is identically satisfied.\\
\indent
By equating the coefficients of partial derivatives, we obtain a system of algebraic equations for the unknown matrix $A$:
\begin{align}
&A^4 = -\,\frac{1}{m^2}\,I,  \label{eq:2q}\\
&A\hspace{0.02cm}\beta_{\mu}A^{3} + A^{2}\beta_{\mu}A^{2} + A^{3}\beta_{\mu}A =
\frac{1}{m^2}\,\beta_{\mu}, \label{eq:2w}\\
&A\hspace{0.02cm}\beta_{\mu}A\hspace{0.02cm}\beta_{\nu}A^{2}
+ A\hspace{0.02cm}\beta_{\mu}A^{2}\hspace{0.02cm}\beta_{\nu}A
+ A^{2}\hspace{0.02cm}\beta_{\mu}A\hspace{0.02cm}\beta_{\nu}A
+ (\mu \rightleftarrows \nu)
=
-\,\frac{1}{m^2}\,\bigl\{\beta_{\mu},\beta_{\nu}\bigr\} \label{eq:2e}
\end{align}
%
%
and two further equations of the third and fourth degrees of nonlinearity in the $\beta$-matrices. A general remark need to be made regarding the system (\ref{eq:2q})\,--\,(\ref{eq:2e}). A similar point was made for the spin-1 case in \cite{markov_2015}. Equations (\ref{eq:2q}) and (\ref{eq:2w}) are universal in a matter. The former determines the mass term on the right-hand side of the equality (\ref{eq:1c}), and the latter makes it possible to get rid of the term of the first order in derivatives in (\ref{eq:1c}). The universality of these matrix equations consists in the fact that they must be satisfied in any case irrespective of that we take as the right part: or the operator (\ref{eq:1z}), or the operator  (\ref{eq:1a}), or (\ref{eq:1s}). We will show below that Eqs.\,(\ref{eq:2q})\,--\,(\ref{eq:2w}) uniquely determine the matrix $A$ (accurate within the choice of one of four roots of an algebraic equation for the parameter $\alpha$, see Eq.\,(\ref{eq:4t}) below). An explicit form of the matrix $A$ and also the equalities (\ref{eq:2q})\,--\,(\ref{eq:2w}) to which this matrix satisfies are of the great importance for further consideration. The third equation (\ref{eq:2e}) and two remaining equations are not already universal and completely depend on the specific choice of the right-hand side in the equalities of the (\ref{eq:1c}) type. These equations must be identically satisfied. If this does not hold, we come to contradiction.\\
\indent
Let us introduce a matrix $\Omega$ satisfying the following characteristic equation:
\begin{equation}
\Omega^{\hspace{0.02cm}4} = \frac{5}{2}\, \Omega^{\hspace{0.02cm}2} -  \frac{9}{16}\,I,
\label{eq:2r}
\end{equation}
and as a result
\begin{equation}
\Omega^{\hspace{0.02cm}5} = \frac{5}{2}\, \Omega^{\hspace{0.02cm}3} -  \frac{9}{16}\,\Omega,
\qquad
\Omega^{\hspace{0.02cm}6} = \frac{91}{16}\, \Omega^{\hspace{0.02cm}2} -  \frac{45}{32}\,I.
\label{eq:2t}
\end{equation}
An explicit form of the matrix $\Omega$ will be defined in the next section. Now only the fact of the existence of such a matrix satisfying (\ref{eq:2r}) is of our importance. We seek the matrix $A$ in the form of the most general expansion in powers of $\Omega$:
\begin{equation}
A = \alpha\hspace{0.02cm}I  + \beta\hspace{0.02cm}\Omega
+ \gamma\hspace{0.03cm}\Omega^{\hspace{0.02cm}2} + \delta\hspace{0.03cm}\Omega^{\hspace{0.02cm}3},
\label{eq:2y}
\end{equation}
where $\alpha,\,\beta,\,\gamma$, and $\delta$ are unknown, generally speaking, complex, scalar constants.\\
\indent
Let us consider the first matrix equation (\ref{eq:2q}). It is convenient to divide the construction of its solution into two steps. At the first step, instead of the matrix $A$, we consider the matrix $A^2$ which can be also written as an expansion in powers of $\Omega$:
\begin{equation}
A^{2} = a\hspace{0.02cm}I  + b\hspace{0.04cm}\Omega
+ c\hspace{0.05cm}\Omega^{\hspace{0.02cm}2} + d\hspace{0.04cm}\Omega^{\hspace{0.02cm}3}.
\label{eq:2u}
\end{equation}
Here, the coefficients of the expansion $(a,\,b,\,c,\,d)$ are associated with the initial ones $(\alpha,\,\beta,\,\gamma,\,\delta)$ by the fixed nonlinear algebraic relations which can be easily defined by making use of (\ref{eq:2r})\,--\,(\ref{eq:2t}). These relations will be written just below, and now we restrict our attention to calculating an explicit form of the coefficients in the expansion (\ref{eq:2u}). A system of the algebraic equations for these coefficients
\begin{align}
&a^{2} - \frac{9}{16}\,c^{\hspace{0.02cm}2} - \frac{45}{32}\,d^{\hspace{0.03cm}2} - \frac{9}{8}\,b\hspace{0.02cm}d = -\,\frac{1}{m^2},  \label{eq:2i}\\[1ex]
&2\hspace{0.03cm}a\hspace{0.015cm}b - \frac{9}{8}\,c\hspace{0.03cm}d = 0, \label{eq:2o}\\[1ex]
&b^{2} + \frac{5}{2}\,c^{\hspace{0.02cm}2} + \frac{91}{16}\,d^{\hspace{0.03cm}2} + 2\hspace{0.02cm}a\hspace{0.015cm}c + 5\hspace{0.02cm}b\hspace{0.02cm}d = 0, \label{eq:2p}\\[1ex]
&2\hspace{0.015cm}a\hspace{0.015cm}d + 2\hspace{0.015cm}b\hspace{0.015cm}c + 5\hspace{0.015cm}c\hspace{0.02cm}d = 0 \label{eq:2a}
\end{align}
follows from the matrix equation (\ref{eq:2q}). An immediate consequence of equation (\ref{eq:2o}) is
\begin{equation}
d = \frac{16}{9}\,\frac{a\hspace{0.03cm}b}{c}.
\label{eq:2s}
\end{equation}
Here, we believe that all the parameters under discussion are different from zero. Substitution of the preceding expression into (\ref{eq:2a}) leads to the equation
\[
\frac{16}{9}\,a^{2} + c^{\hspace{0.02cm}2} + \frac{40}{9}\,a\hspace{0.02cm}c = 0
\]
which connects unknown quantities $a$ and $c$. Considering the equation as that for $c$, we derive its two solutions:
\[
c_{1} = -\hspace{0.02cm}4\hspace{0.03cm}a, \qquad
c_{2} = -\frac{4}{9}\,a.
\]
Equation (\ref{eq:2s}) produces us two other relations for the parameter $d$:
\[
d_{1} = -\frac{4}{9}\,b,  \qquad
d_{2} = -\hspace{0.02cm}4\hspace{0.03cm}b.
\]
Further, substitution of the parameters $(c_1,\,d_1)$ into (\ref{eq:2p}) gives
\[
b_{1}^{\pm} = \pm\hspace{0.02cm}18\hspace{0.03cm}a, \qquad
d_{1}^{\pm} =  \mp\hspace{0.03cm}8\hspace{0.03cm}a,
\]
and substitution of the parameters $(c_2,\,d_2)$ into the same equation results in the relations
\[
b_{2}^{\pm} = \pm\hspace{0.03cm}\frac{2}{27}\,a, \qquad
d_{2}^{\pm} =  \mp\hspace{0.03cm}\frac{8}{27}\,a.
\]
Thus, equations (\ref{eq:2o})\,--\,(\ref{eq:2a}) admits four possible solutions (as functions of the parameter $a$) which are conveniently written in the form of the table:
\begin{equation}
\begin{array}{llll}
(\textrm{I}): \qquad \qquad \qquad    &c_{1} = -\hspace{0.02cm}4\hspace{0.03cm}a,\qquad\qquad
&b_{1}^{+} = 18\hspace{0.03cm}a, \qquad\qquad
&d_{1}^{+} =  -\hspace{0.02cm}8\hspace{0.03cm}a;  \\[2ex]
(\textrm{I\!\hspace{0.04cm}I}):\hspace{2cm}     &c_{1} = -\hspace{0.02cm}4\hspace{0.03cm}a,\qquad\qquad
&b_{1}^{-} = -\hspace{0.02cm}18\hspace{0.04cm}a, \quad
&d_{1}^{-} =  8\hspace{0.03cm}a; \\[2ex]
(\textrm{I\!\hspace{0.045cm}I\!\hspace{0.045cm}I}): \hspace{2cm}     &c_{2} = -\hspace{0.02cm}\displaystyle\frac{4}{9}\,a,\qquad
&b_{2}^{+} = \displaystyle\frac{2}{27}\,a, \quad
&d_{2}^{+} =  -\hspace{0.02cm}\displaystyle\frac{8}{27}\,a; \\[2ex]
(\textrm{IV}):\hspace{2cm}     &c_{2} = -\hspace{0.02cm}\displaystyle\frac{4}{9}\,a,\qquad
&b_{2}^{+} = -\hspace{0.02cm}\displaystyle\frac{2}{27}\,a, \quad
&d_{2}^{-} = \displaystyle\frac{8}{27}\,a.
\end{array}
\label{eq:2d}
\end{equation}
For determining a value of the parameter $a$, let us substitute the solution (I) into the remaining equation (\ref{eq:2i}). Then, we have
\begin{equation}
a_{\rm I}^{2} = -\frac{\!1}{64}\,\frac{1}{m^{2}} \quad\mbox{or}\quad
a_{\rm I} = \pm\, i\hspace{0.03cm}\frac{1}{8\hspace{0.02cm}m}.
\label{eq:2f}
\end{equation}
By virtue of the invariance of equation (\ref{eq:2i}) with respect to the replacement $(b,\,d) \rightarrow (-b,\,-d)$, the solution (II) in (\ref{eq:2d}) leads to the same values for the $a$. Further, for the solutions (III) and (IV), we get
\begin{equation}
a_{\rm I\!\hspace{0.02cm}I\!\hspace{0.02cm}I}^{2} = -\biggl(\frac{9}{8}\biggr)^{\!2}\frac{\!1}{m^{2}} \quad\mbox{or}\quad
a_{\rm I\!\hspace{0.02cm}I\!\hspace{0.02cm}I} = \pm\, i\hspace{0.03cm}\frac{9}{8\hspace{0.02cm}m}.
\label{eq:2g}
\end{equation}
\indent
Thus, the values of the parameters in the expansion of the matrix $A^{2}$, Eq.\,(\ref{eq:2u}), are defined in full. At the second step, it is necessary to determine the coefficients in the expansion of the original matrix $A$, Eq.\,(\ref{eq:2y}). For this purpose we make use of the following connection between the coefficients $(\alpha,\,\beta,\,\gamma,\,\delta)$ and $(a,\,b,\,c,\,d)$:
\begin{equation}
\begin{split}
&\alpha^{\hspace{0.01cm}2} - \frac{9}{16}\,\gamma^{\hspace{0.02cm}2} - \frac{45}{32}\,\delta^{\hspace{0.03cm}2} - \frac{9}{8}\,\beta\hspace{0.015cm}\delta = a, \\[1ex]
&\beta^{\hspace{0.02cm}2} + \frac{5}{2}\,\gamma^{2} + \frac{91}{16}\,\delta^{\hspace{0.03cm}2} + 2\hspace{0.02cm}\alpha\hspace{0.02cm}\gamma + 5\hspace{0.02cm}\beta\hspace{0.015cm}\delta = c,  \\[1ex]
&2\hspace{0.015cm}\alpha\hspace{0.02cm}\beta - \frac{9}{8}\hspace{0.03cm}\gamma\hspace{0.02cm}\delta = b,  \\[1ex]
&2\hspace{0.015cm}\alpha\hspace{0.02cm}\delta + 2\hspace{0.015cm}\beta\hspace{0.02cm}\gamma + 5\hspace{0.015cm}\gamma\hspace{0.02cm}\delta = d.
\end{split}
\label{eq:2h}
\end{equation}
Note that all the equations of this system (as opposed to (\ref{eq:2i})\,--\,(\ref{eq:2a})) are inhomogeneous. One may somewhat simplify the system. For definiteness, we consider on the right-hand side of (\ref{eq:2h}) the values of parameters $(a,\,b,\,c,\,d)$ for the solution (I) in (\ref{eq:2d}). Multiplying the third equation in (\ref{eq:2h}) by $4/9$ and summing it with the last one, we obtain
\[
\biggl(\frac{4}{9}\,\alpha + \gamma\biggr)\biggl(\frac{4}{9}\,\beta + \delta\biggr) = 0.
\]
It might be required vanishing each of the expressions in parentheses that enables one to reduce the number of unknown parameters by one half. However, as can be easily shown, in this case the system (\ref{eq:2h}) results in a contradiction. Therefore, we have to require vanishing only one of the expressions. For definiteness, we set
\begin{equation}
\gamma = -\frac{4}{9}\,\alpha,
\label{eq:2j}
\end{equation}
considering the parameters $\beta$ and $\delta$ as arbitrary ones. The relation (\ref{eq:2j}) enables us to reduce the system (\ref{eq:2h}) to three equations:
\begin{align}
&\frac{8}{9}\;\alpha^{2} - \frac{45}{32}\,\delta^{\hspace{0.03cm}2} - \frac{9}{8}\,\beta\hspace{0.015cm}\delta = a_{\hspace{0.02cm}{\rm I}},  \label{eq:2k}\\[1ex]
&\beta^{\hspace{0.02cm}2} - \frac{32}{81}\,\alpha^{2} + \frac{91}{16}\,\delta^{\hspace{0.03cm}2} + 5\hspace{0.015cm}\beta\hspace{0.015cm}\delta = -4\hspace{0.02cm}a_{\hspace{0.02cm}{\rm I}},  \label{eq:2l}\\[1ex]
&\frac{1}{9}\,\alpha\hspace{0.015cm}\beta + \frac{1}{36}\,\alpha\hspace{0.02cm}\delta = a_{\hspace{0.02cm}{\rm I}},
\label{eq:2z}
\end{align}
where the value of the parameter $a_{\hspace{0.02cm}{\rm I}}$ is equal to (\ref{eq:2f}).\\
\indent
 In the case of the solution (I\!\hspace{0.02cm}I\!\hspace{0.025cm}I) in (\ref{eq:2d}) the relation
\begin{equation}
\gamma = -\hspace{0.01cm}4\hspace{0.02cm}\alpha
\label{eq:2x}
\end{equation}
will be analog of the relation (\ref{eq:2j}), and the corresponding reduced system takes the form
\begin{align}
&8\hspace{0.03cm}\alpha^{2} + \frac{45}{32}\,\delta^{\hspace{0.03cm}2} + \frac{9}{8}\,\beta\hspace{0.015cm}\delta = -a_{\rm I\!\hspace{0.02cm}I\!\hspace{0.02cm}I},  \label{eq:2c}\\[1ex]
&\beta^{\hspace{0.02cm}2} + 32\hspace{0.03cm}\alpha^{2} + \frac{91}{16}\,\delta^{\hspace{0.02cm}2} + 5\hspace{0.015cm}\beta\hspace{0.015cm}\delta = -\frac{4}{9}\,
a_{\rm I\!\hspace{0.02cm}I\!\hspace{0.02cm}I},  \label{eq:2v}\\[1ex]
&2\hspace{0.02cm}\alpha\hspace{0.02cm}\beta + \frac{9}{2}\,\alpha\hspace{0.03cm}\delta = \frac{2}{27}\,a_{\rm I\!\hspace{0.02cm}I\!\hspace{0.02cm}I}  \label{eq:2b}
\end{align}
with the value $a_{\rm I\!\hspace{0.02cm}I\!\hspace{0.02cm}I}$ equal to (\ref{eq:2g}).\\
\indent
Straightforward solving the system (\ref{eq:2k})\,-- (\ref{eq:2z})  (as well as
(\ref{eq:2c})\,--\,(\ref{eq:2b})) is very cumbersome. Here, at the end of all algebraic manipulations, we arrive at the necessity of solving a quartic equation. We will follow a more simple way. In constructing solutions of these systems we involve additional algebraic equations for the required parameters ($\alpha,\,\beta,\,\gamma,\,\delta$), which follow from the second matrix equation for $A$, namely, from Eq.\,(\ref{eq:2w}). However, for an analysis of equation (\ref{eq:2w}) we need the rules of rearrangement of the matrix $\Omega$ with the matrices $\beta_{\mu}$. This, in turn, requires a knowledge of an explicit form of the matrix $\Omega$. Therefore, in the next section we consider the construction of this matrix and derive all the required commutation rules.

\section{\bf Explicit form of the $\Omega$ matrix}
\setcounter{equation}{0}

The matrix $\Omega$ must satisfy the fundamental relation (\ref{eq:2r}). In the construction of an explicit form this matrix we will essentially use the results of papers \cite{madhavarao_1946, venkatachaliengar_1954}. Let us introduce a new matrix $\theta$ setting by definition:
\begin{equation}
\theta \equiv P_{2} - P_{1},
\label{eq:3q}
\end{equation}
where the matrices $P_1$ and $P_2$ are defined in Appendix A, Eq.\,(\ref{ap:A6}). The minimal equation to which $\theta$ satisfies is:
\begin{equation}
\theta^{\hspace{0.03cm}3} - 2\hspace{0.02cm}\theta^{\hspace{0.02cm}2} -
15\hspace{0.02cm}\theta = 0
\label{eq:3w}
\end{equation}
and as a result we get
\begin{equation}
\theta^{\hspace{0.03cm}4} - 19\hspace{0.02cm}\theta^{\hspace{0.02cm}2} -
30\hspace{0.02cm}\theta = 0.
\label{eq:3e}
\end{equation}
The matrix $\theta$ is a central element of the algebra $A_{\xi}$. The second central element (as it is defined in (\ref{ap:A5})) can be presented as a polynomial in $\theta$ in the following form \cite{venkatachaliengar_1954}:
\begin{equation}
P_{4} - 2\hspace{0.015cm}P_{3} = \theta^{\hspace{0.02cm}2} - 2\hspace{0.03cm}\theta
-\frac{15}{2}\,I.
\label{eq:3r}
\end{equation}
It is not difficult to show that as a consequence of (\ref{eq:3w}) and (\ref{eq:3e}) the following relation
\[
\bigl(P_{4} - 2\hspace{0.015cm}P_{3}\bigr)^{2} = \biggl(\frac{15}{2}\biggr)^{\!\!2}I.
\]
holds.\\
\indent
We seek for the matrix $\Omega$ in the form of the decomposition
\begin{equation}
\Omega = \gamma_{5}\otimes\omega,
\label{eq:3t}
\end{equation}
where $\gamma_5=\gamma_1 \gamma_2 \gamma_3 \gamma_4$ and $\omega$ is the unknown matrix. From (\ref{eq:3t}), we have further
\[
\Omega^{\hspace{0.02cm}2} = I_{\gamma}\otimes\omega^{\hspace{0.02cm}2},
\qquad
\Omega^{\hspace{0.02cm}4} = I_{\gamma}\otimes\omega^{\hspace{0.02cm}4}\!,
\]
where $I_{\gamma}$ is the unity $4 \times 4$ matrix of Dirac's algebra. In choice of the presentation (\ref{eq:3t}) the characteristic equation (\ref{eq:2r}) turns into the equation for the matrix $\omega$:
\begin{equation}
\omega^{\hspace{0.02cm}4} - \frac{5}{2}\, \omega^{\hspace{0.02cm}2} +  \frac{9}{16}\,I_{\xi} = 0.
\label{eq:3y}
\end{equation}
Here, we have taken into account that $I=I_{\gamma} \otimes I_{\xi}$, where $I_{\xi}$ is the unity matrix of proper dimension of the $A_{\xi}$-\hspace{0.01cm}algebra. We will search for the matrix $\omega$ as a second-order polynomial in the central element $\theta$:
\begin{equation}
\omega = \mu\hspace{0.02cm}\theta^{\hspace{0.02cm}2} + \nu\hspace{0.02cm}\theta + \lambda\hspace{0.02cm}I_{\xi},
\label{eq:3u}
\end{equation}
where $\mu$, $\nu$ and $\lambda$ are the unknown parameters. Alternatively, the expansion (\ref{eq:3u}) can be written in the terms of matrices $P_1, P_2, P_3$ and $P_4$, in view of (\ref{eq:3q}) and (\ref{eq:3r}), as follows:
\begin{equation}
\omega = \mu\hspace{0.02cm}\bigl(P_{4} - 2\hspace{0.02cm}P_{3}\bigr) +
(2\hspace{0.02cm}\mu +\nu)\hspace{0.02cm}\bigl(P_{2} - P_{1}\bigr) +
\Bigl(\frac{15}{2}\,\mu + \lambda\Bigr)\hspace{0.02cm}I_{\xi}.
\label{eq:3i}
\end{equation}
\indent
Before proceeding with the calculation of the coefficients ($\mu$, $\nu$, $\lambda$) in the expression (\ref{eq:3u}), let us consider one essential for the subsequent discussion consequence of the decomposition (\ref{eq:3t}) and of the choice of the matrix $\omega$ in the form of (\ref{eq:3u}). By virtue of the fact that the matrix $\omega$ is made up of the elements of the center of $A_{\xi}$-\hspace{0.01cm}algebra, the following relation:
\[
[\hspace{0.03cm}\xi_{\mu},\hspace{0.02cm}\omega\hspace{0.02cm}] = 0
\]
is true. For the matrices $\beta_{\mu}$ given in the form of the direct product (\ref{eq:1o}) the following commutative rules
\begin{equation}
\{\hspace{0.02cm}\beta_{\mu},\hspace{0.02cm}\Omega\hspace{0.02cm}\} = 0,
\qquad
[\hspace{0.03cm}\beta_{\mu},\hspace{0.02cm}\Omega^{\hspace{0.02cm}2}\hspace{0.02cm}] = 0,
\qquad
\{\hspace{0.02cm}\beta_{\mu},\hspace{0.02cm}\Omega^{\hspace{0.02cm}3}\hspace{0.02cm}\} = 0
\label{eq:3o}
\end{equation}
will be a consequence of (\ref{eq:3t}) and the commutativity of $\xi_{\mu}$ and $\omega$. The first relation here is an analog of the corresponding relation in the Dirac theory, namely, $\{\gamma_{\mu}, \gamma_5 \}=0$. These commutative rules are much more simple in contrast to the corresponding ones in the DKP-theory (see Appendix A in \cite{markov_2015}). In the latter these rules are  too tangled that makes analysis of the matrix equations more tedious. On the other hand, the formalism for describing the $3/2$-spin particle developed here, is incomparably more cumbersome in contrast to the spin-1 case, and only the simple commutative rules (\ref{eq:3o}) enable us ultimately to solve the problem posed in the present work.\\
\indent
In much the same way as in the previous section we divide the procedure of calculating the unknown coefficients in (\ref{eq:3u}) into two steps. First, we define the square of the matrix $\omega$. Taking into account (\ref{eq:3w}) and (\ref{eq:3e}), we derive
\begin{equation}
\omega^{\hspace{0.02cm}2} = m\hspace{0.03cm}\theta^{\hspace{0.02cm}2} + n\hspace{0.03cm}\theta + l\hspace{0.015cm}I_{\xi},
\label{eq:3p}
\end{equation}
where
\begin{align}
&m\equiv 19\hspace{0.015cm}\mu^{2} + \nu^{2} + 4\hspace{0.02cm}\mu\hspace{0.015cm}\nu + 2\hspace{0.015cm}\mu\hspace{0.01cm}\lambda, \notag\\[0.5ex]
&n\equiv 30\hspace{0.015cm}\mu^{2} + 30\hspace{0.015cm}\mu\hspace{0.015cm}\nu + 2\hspace{0.02cm}\nu\lambda, \label{eq:3a}\\[0.5ex]
&l\equiv \lambda^{2}.\notag
\end{align}
\indent
Further we define a system of algebraic equations for the constants $m,\,n$ and $l$. Putting (\ref{eq:3p}) into (\ref{eq:3y}) and equating to zero the coefficients of the matrices $\theta^{2},\,\theta$ and $I_{\xi}$ yields
\begin{align}
&\theta^{2}: \qquad 19\hspace{0.02cm}m^{2} + n^{2} + 4\hspace{0.02cm}m\hspace{0.02cm}n + 2\hspace{0.02cm}m\hspace{0.02cm}l = \frac{5}{2}\,m, \notag\\
&\theta: \,\;\qquad 30\hspace{0.02cm}m^{2} + 30\hspace{0.02cm}m\hspace{0.02cm}n + 2\hspace{0.02cm}n\hspace{0.02cm}l = \frac{5}{2}\,n, \label{eq:3s}\\
&I_{\xi}: \;\qquad l^{\hspace{0.02cm}2} = \frac{5}{2}\,l -  \frac{9}{16}.\notag
\end{align}
From the last equation we can immediately define the parameter $l$. Here, we have two possibilities:
\[
l_{\rm I} = \frac{9}{4} \quad \mbox{and} \quad l_{\rm I\!\hspace{0.025cm}I} = \frac{1}{4}\,.
\]
For definiteness, we fix the first value, i.e. we set $l=l_{\rm I}$. In this case the remaining two equations in (\ref{eq:3s}) take the form
\begin{equation}
\left\{
\begin{split}
19\hspace{0.03cm}&m^{2} + n^{2} + 4\hspace{0.02cm}m\hspace{0.02cm}n + 2\hspace{0.02cm}m = 0,\\
&m^{2} + m\hspace{0.02cm}n + \frac{1}{15}\,n = 0.
\end{split}
\label{eq:3d}
\right.
\end{equation}
In order not to overburden the following considerations, the detailed analysis of solutions of this system is given in Appendix B. Here, we write out only the final result. The system (\ref{eq:3d}) has the following three solutions:
\begin{equation}
\begin{array}{llll}
(1) \quad\; &l_{\rm I} = \displaystyle\frac{9}{4}, \qquad\; &n_{1} = \displaystyle\frac{5}{12}, \qquad\; &m_{1}^{+} = -\hspace{0.02cm}\displaystyle\frac{1}{12}\,; \\[2ex]
(2) \quad\; &l_{\rm I} = \displaystyle\frac{9}{4}, \qquad\; &n_{2} = -\hspace{0.02cm}\displaystyle\frac{3}{20}, \qquad &m_{2}^{-} = -\hspace{0.02cm}\displaystyle\frac{1}{20}\,; \\[2ex]
(3) \quad\; &l_{\rm I} = \displaystyle\frac{9}{4}, \qquad\; &n = \displaystyle\frac{4}{15}, \qquad\; &m = -\hspace{0.02cm}\displaystyle\frac{2}{15}\,.
\end{array}
\label{eq:3f}
\end{equation}
\indent
Let us consider now a solution of the algebraic system (\ref{eq:3a}) for the parameters $(\mu,\,\nu,\,\lambda)$. To be specific, let us take as the parameters $(m, n, l)$ on the left-hand side of (\ref{eq:3a}) the first triple of numbers in (\ref{eq:3f}). Then the system (\ref{eq:3a}) takes the form
\begin{align}
&19\hspace{0.015cm}\mu^{2} + \nu^{2} + 4\hspace{0.015cm}\mu\hspace{0.015cm}\nu + 2\hspace{0.015cm}\mu\hspace{0.015cm}\lambda = -\frac{1}{12}, \notag\\[0.5ex]
&\mu^{2} + \mu\hspace{0.015cm}\nu + \frac{1}{15}\,\nu\lambda = \frac{1}{72}, \notag\\[0.5ex]
&\lambda^{2} = \frac{9}{4}.\notag
\end{align}
From the last equation we define two solutions: $\lambda_{\rm I,\, I\!\hspace{0.025cm}I}=\pm\,3/2$. Further, let us examine the case
\[
\lambda \equiv \lambda_{\rm I} = \frac{3}{2}\,.
\]
Then, the first two equations in the system above turn to
\begin{equation}
\left\{
\begin{split}
19\hspace{0.02cm}&\mu^{2} + \nu^{2} + 4\hspace{0.02cm}\mu\hspace{0.01cm}\nu + 3\hspace{0.015cm}\mu = -\frac{1}{12}\,, \\[0.5ex]
&\mu^{2} + \mu\hspace{0.02cm}\nu + \frac{1}{10}\,\nu = \frac{1}{72}\,.
\end{split}
\label{eq:3g}
\right.
\end{equation}
From the second equation here we define the parameter $\mu$ as a function of $\nu$
\begin{equation}
\mu^{\pm} = \frac{1}{2}\,\bigl(-\nu\hspace{0.02cm} \pm \sqrt{D}), \qquad
D \equiv \nu^{2} - \frac{2}{5}\,\nu + \frac{1}{18}
\label{eq:3h}
\end{equation}
and substitute it into the first equation. After simplification the first equation becomes
\[
\frac{17}{2}\,\nu^{2} - \frac{17}{5}\,\nu + \frac{25}{72} =
\biggl(\pm\frac{15}{2}\,\nu \hspace{0.02cm}\mp\hspace{0.02cm} \frac{3}{2}\biggr)\sqrt{D}.
\]
Having taken the square of the previous expression we finally define
\begin{equation}
\biggl(\frac{17}{2}\,\nu^{2} - \frac{17}{5}\,\nu + \frac{25}{72}\biggr)^{\!\!2}
=
\biggl(\frac{15}{2}\,\nu - \frac{3}{2}\biggr)^{\!\!2}\hspace{0.03cm}
\biggl(\nu^{2} - \frac{2}{5}\,\nu + \frac{1}{18}\biggr).
\label{eq:3j}
\end{equation}
The equation represents that of the fourth degree with respect to $\nu$. However, it can be easily solved if one notes that the unknown parameter $\nu$ enters into the left- and the right-hand side {\it only} in the combination
\begin{equation}
\nu^{2} - \frac{2}{5}\,\nu \equiv x.
\label{eq:3k}
\end{equation}
This enables us to reduce solving Eq.\,(\ref{eq:3j}) to successive solving two quadratic equations. The first of them (the equation for $x$), by virtue of (\ref{eq:3j}), has the form
\[
16\hspace{0.03cm}x^{2} + \frac{19}{36}\,x - \frac{\!23}{(72)^{2}} = 0
\]
and its two solutions are
\[
x_{+} = \frac{\!1}{(12)^2}, \qquad x_{-} = -\frac{\!23}{(24)^2}.
\]
Further we consider solutions of equation (\ref{eq:3k}) for a given value $x=x_{\pm}$, i.e.
\[
\nu^{2} - \frac{2}{5}\,\nu - x_{\pm} = 0.
\]
For every of two values of $x_{\pm}$ we define a set of the solutions $\nu_{1,\,2}^{\pm}$ of the preceding equation and in turn for each of four solutions $\nu_{1,\,2}^{\pm}$ in view of the relation (\ref{eq:3h}) we derive two sets of solutions for the parameter $\mu$. However, by a direct substitution of the obtained solutions into the initial system (\ref{eq:3g}) we check that only half of them obey this system. Thus, choosing the first triple of numbers in (\ref{eq:3f}) and setting $\lambda\equiv\lambda_{\rm I}=3/2$, one can write out the following set of  permissible coefficients in the expansion (\ref{eq:3u}):
\begin{align}
(1)\qquad\quad &\lambda_{\rm I} = \frac{3}{2},  &\nu_{1}^{+} &= \frac{5}{12}, &\mu^{+} &= -\,\frac{1}{12}\,; \label{eq:3l}\\[1ex]
(2)\qquad\quad  &\lambda_{\rm I} = \frac{3}{2}, &\nu_{2}^{+} &= -\hspace{0.02cm}\frac{1}{60}, &\mu^{-} &= -\,\frac{7}{60}\,;
\notag \\[1ex]
(3)\qquad\quad  &\lambda_{\rm I} = \frac{3}{2},  &\nu_{1}^{-} &= \frac{5}{24}, &\mu^{+} &= -\hspace{0.02cm}\frac{1}{24}\,;
\label{eq:3z}\\[1ex]
(4)\qquad\quad  &\lambda_{\rm I} = \frac{3}{2},  &\nu_{2}^{-} &= \frac{23}{120}, &\mu^{-} &= -\hspace{0.02cm}\frac{19}{120}\,.
\notag
\end{align}
One may perform a similar analysis of solutions of the system (\ref{eq:3a}) if we choose as the values of the parameters ($l, m, n$) the values from the second and third lines in (\ref{eq:3f}) and fix $\lambda_{\rm I\!\hspace{0.025cm}I}=-3/2$.\\
\indent
In this section we have constructed the representation of the matrix $\Omega$ in an explicit form and shown that there exists a finite number of variants of choosing the $\Omega$ (although it is possible that they relate among themselves through a certain symmetry transformation). However, here there is a question of principle: whether it is possible to write the matrix $\Omega$ which we defined in the form of the direct product (\ref{eq:3t}) solely in terms of the original $\beta_{\mu}$ matrices? Recall that the matrix $\omega$ in the decomposition can be given in the form of an expansion in the central elements (including the unity matrix) of the $A_{\xi}$ algebra, Eq.\,(\ref{eq:3i}).  It is not difficult to show that for the term with $P_4$ in the representation (\ref{eq:3i}) passage to the $\beta$\hspace{0.01cm}-\hspace{0.01cm}matrices is really possible. Indeed, by virtue of decomposition of the $\beta$\hspace{0.02cm}-\hspace{0.02cm}matrices
\[
\hspace{3cm}
\beta_{\mu} = \gamma_{\mu}\otimes\xi_{\mu} \quad (\mbox{no summation!})
\]
and the properties of the direct product, the following equality
\[
\hspace{4cm}
\beta_{\mu}\beta_{\nu}\beta_{\lambda}\beta_{\sigma} =
(\gamma_{\mu}\gamma_{\nu}\gamma_{\lambda}\gamma_{\sigma})\otimes
(\xi_{\mu}\hspace{0.02cm}\xi_{\nu}\hspace{0.02cm}\xi_{\lambda}\hspace{0.02cm}\xi_{\sigma})
\quad (\mbox{no summation!})
\]
is true. The contraction of this expression with the totally antisymmetric tensor
$\varepsilon_{\mu\nu\lambda\sigma}$ and allowance for the properties of the $\gamma$-\hspace{0.01cm}matrices give
\[
\epsilon_{\mu\nu\lambda\sigma}
(\gamma_{\mu}\gamma_{\nu}\gamma_{\lambda}\gamma_{\sigma})\otimes
(\xi_{\mu}\hspace{0.02cm}\xi_{\nu}\hspace{0.02cm}\xi_{\lambda}\hspace{0.02cm}\xi_{\sigma})
=
\epsilon_{1234}\hspace{0.03cm}(\gamma_{1}\gamma_{2}\gamma_{3}\gamma_{4})\otimes (\xi_{1}\hspace{0.02cm}\xi_{2}\hspace{0.02cm}\xi_{3}\hspace{0.02cm}\xi_{4}) +
\epsilon_{2134}\hspace{0.03cm}(\gamma_{2}\gamma_{1}\gamma_{3}\gamma_{4})\otimes (\xi_{2}\hspace{0.02cm}\xi_{1}\hspace{0.02cm}\xi_{3}\hspace{0.02cm}\xi_{4}) +\, \ldots
\]
\[
=
\gamma_{1}\gamma_{2}\gamma_{3}\gamma_{4}\otimes \sum\hspace{0.02cm} \xi_{\mu}\hspace{0.02cm}\xi_{\nu}\hspace{0.02cm}\xi_{\lambda}\hspace{0.02cm}\xi_{\sigma}.
\]
The indices is being unequal in the summation. Thus, we have the equality
\[
\epsilon_{\mu\nu\lambda\sigma}\beta_{\mu}\beta_{\nu}\beta_{\lambda}\beta_{\sigma}
\equiv
\gamma_{5}\otimes P_{4}
\]
and by doing so the term in (\ref{eq:3t}) (with consideration for (\ref{eq:3i})) containing $P_4$ is  uniquely expressed by the antisymmetrized product of the original $\beta$-matrices. Unfortunately, we can say nothing about similar representations for the terms with the matrices $P_1, P_2$ and $P_3$, and this problem is still an open one.

\section{\bf Solving matrix equation (\ref{eq:2w})}
\setcounter{equation}{0}

We proceed now to analysis of matrix equation (\ref{eq:2w}) for the required matrix $A$. The commutation rules for the matrices $\beta_{\mu}$ and $\Omega$ as they were defined in the previous section, Eq.\,(\ref{eq:3o}), and the characteristic equation (\ref{eq:2r}) will  play a decisive role in this analysis.\\
\indent
By virtue of the commutation rules (\ref{eq:3o}) we have the following relationship
\begin{equation}
\beta_{\mu}A = \bar{A}\hspace{0.02cm}\beta_{\mu},
\label{eq:4q}
\end{equation}
where the matrix $\bar{A}$ is given by
\[
\bar{A} = \alpha\hspace{0.015cm}I  - \beta\hspace{0.02cm}\Omega
+ \gamma\hspace{0.03cm}\Omega^{\hspace{0.02cm}2} - \delta\hspace{0.03cm}\Omega^{\hspace{0.02cm}3}.
\]
This expression is differ from the matrix $A$ by change in the sign of the terms with odd powers of $\Omega$. In view of (\ref{eq:4q}), the equation (\ref{eq:2w}) can be reduced to the following form:
\begin{equation}
\bigl(A\bar{A}\bigr)\bigl[\hspace{0.02cm}\bar{A}^{2} + \bigl(A\bar{A}\bigr) + A^{2}\hspace{0.02cm}\bigr]\beta_{\mu}
=
\frac{1}{m^{2}}\,\beta_{\mu}.
\label{eq:4w}
\end{equation}
The expression for the matrix $A^2$ is defined by (\ref{eq:2u}), and the expression for $\bar{A}^2$ is obtained from (\ref{eq:2u}) by a simple replacement of two parameters: $(b, d) \rightarrow (-b, -d)$. It is easy to convince in the validity of the latter from an analysis of the general relations between the coefficients $(\alpha, \beta, \gamma, \delta)$ and $(a, b, c, d)$, Eq.\,(\ref{eq:2h}). With that said, for the sum of $A^2$ and $\bar{A}^2$  we have
\[
\bar{A}^{2} + A^{2} = 2\hspace{0.015cm}a\hspace{0.01cm}I
+
2\hspace{0.02cm}c\hspace{0.03cm}\Omega^{\hspace{0.015cm}2}.
\]
Further, a product of the matrices $A$ and $\bar{A}$ with regard for the characteristic equation (\ref{eq:2r}) and also the first and third relations in the system (\ref{eq:2h}), gives the following equality:
\begin{equation}
\bigl(A\bar{A}\bigr) = \Bigl(\hspace{0.02cm}2\hspace{0.015cm}\alpha^{2} - \frac{9}{8}\,\gamma^{2} -  a \Bigr)I +
\bigl(
4\hspace{0.02cm}\alpha\hspace{0.03cm}\gamma + 5\hspace{0.02cm}\gamma^{2} - c\hspace{0.02cm}\bigr)\hspace{0.03cm}\Omega^{\hspace{0.015cm}2}.
\label{eq:4e}
\end{equation}
Summing the expressions obtained, we derive an explicit form of the expression in the square brackets in (\ref{eq:4w}). By virtue of the fact that the matrices $\beta_{\mu}$ are nonsingular\footnote{\,The matrices $\beta_{\mu}^{-1} = -\frac{16}{9}\,\beta_{\mu}^{3} + \frac{45}{9}\,I$ are the inverse matrices of $\beta_{\mu}$.}, they can be canceled on the left- and right-hand sides in (\ref{eq:4w}). This circumstance is qualitatively distinct from an analysis of the corresponding matrix equations in the DKP-case \cite{markov_2015}, where the matrices $\beta_{\mu}$ are singular. This has made analysis of the corresponding matrix equation rather difficult.\\
\indent
Multiplying the expression in the square brackets in (\ref{eq:4w}) by $(A \bar{A})$ and collecting similar terms with the unity matrix $I$ and with the squared matrix $\Omega$, we result in the following system of two algebraic equations:
\begin{equation}
\begin{split}
&I: \qquad\;\; \biggl(\hspace{0.02cm}2\hspace{0.02cm}\alpha^{2} - \frac{9}{8}\,\gamma^{2}\biggr)^{\!\!2} -
\frac{9}{16}\,\bigl(
4\hspace{0.02cm}\alpha\hspace{0.03cm}\gamma + 5\hspace{0.02cm}\gamma^{2}\hspace{0.02cm}\bigr)^{2} -
a^{2} + \frac{9}{16}\,c^{2} = \frac{1}{m^2}\, ,\\[1ex]
&\Omega^{2}: \qquad \frac{5}{2}\hspace{0.03cm}\bigl(
4\hspace{0.02cm}\alpha\hspace{0.03cm}\gamma + 5\hspace{0.02cm}\gamma^{2}\hspace{0.02cm}\bigr)^{2}
\hspace{0.02cm}+\hspace{0.02cm}
2\biggl(2\hspace{0.02cm}\alpha^{2} - \frac{9}{8}\,\gamma^{2}\biggr)
\bigl(4\hspace{0.02cm}\alpha\hspace{0.03cm}\gamma + 5\hspace{0.02cm}\gamma^{2}\bigr)
-
\frac{5}{2}\,c^{2} - 2\hspace{0.02cm}ac = 0.
\end{split}
\label{eq:4r}
\end{equation}
Let us analyze in detail the case when the relation (\ref{eq:2j}) between the parameters $\alpha$ and $\gamma$ is fixed. As the parameter $a$ we take the value $a_{\rm I}$, Eq.\,(\ref{eq:2f}), and as the $c$ we choose the value $c_1=-4\hspace{0.02cm}a$ in (\ref{eq:2d}). It is not difficult to verify that in such a choice of the parameters $\gamma,\,a$ and $c$ the system (\ref{eq:4r}) is consistent and turns into an identity then and only then the free parameter $\alpha$ satisfies the condition:
\begin{equation}
\alpha^{4} = - \biggl(\frac{9}{4}\biggr)^{\!\!4} a_{I}^{2} \,\equiv\,
\frac{1}{4}\biggl(\frac{9}{8}\biggr)^{\!\!4}\frac{\!1}{m^2}\,.
\label{eq:4t}
\end{equation}
\indent
Now we return to the reduced system of equations for the parameters $(\alpha,\hspace{0.02cm}\beta,\hspace{0.02cm}\delta)$, Eqs.\,(\ref{eq:2k})\,--\,(\ref{eq:2z}). For convenience of future reference, we write out the system once again:
\begin{align}
&\frac{8}{9}\,\alpha^{2} - \frac{45}{32}\,\delta^{\hspace{0.02cm}2} - \frac{9}{8}\,\beta\hspace{0.015cm}\delta = a_{\rm I},  \notag\\[1ex]
&\beta^{2} - \frac{32}{81}\,\alpha^{2} + \frac{91}{16}\,\delta^{\hspace{0.02cm}2} + 5\hspace{0.015cm}\beta\hspace{0.015cm}\delta = -4\hspace{0.03cm}a_{\rm I},  \label{eq:4y}\\[1ex]
&\frac{1}{9}\,\alpha\hspace{0.01cm}\beta + \frac{1}{36}\,\alpha\hspace{0.02cm}\delta = a_{\rm I}.
\notag
\end{align}
In view of the relation (\ref{eq:4t}), it is necessary to consider a value of the parameter $\alpha$ fixed, and thereby we have the overdetermined system for the remaining unknown quantities $\beta$ and $\delta$. From the last equation in the system (\ref{eq:4y}), we express the parameter $\delta$ in terms of $\beta$:
\begin{equation}
\delta  = -4\hspace{0.015cm}\beta + 36\,\frac{a_{I}}{\alpha}\,.
\label{eq:4u}
\end{equation}
Substituting $\delta$ into the first two equations, we obtain that they are consistent if and only if the parameters $\alpha$ and $a_{\rm I}$ are connected with each other by the relation (\ref{eq:4t}). It is convenient to rewrite an equation for the parameter $\beta$ in terms of a dimensionless variable $y\equiv\beta/\alpha$:
\begin{equation}
y^{2} - \frac{9^2}{4}\biggl(\frac{a_{\rm I}}{\alpha^2}\biggr) y
-
\biggl[\hspace{0.03cm}\frac{\!4}{9^2}\hspace{0.02cm} - \hspace{0.02cm} 5\,\frac{9^{\hspace{0.02cm}2}}{\!4}\biggl(\frac{a_{\rm I}^2}{\alpha^4}\biggr)
-
\frac{1}{18}\biggl(\frac{a_{\rm I}}{\alpha^2}\biggr)\biggr] = 0,
\label{eq:4i}
\end{equation}
where, by virtue of (\ref{eq:4t}), we have for the coefficients in (\ref{eq:4i})
\[
\biggl(\frac{a_{\rm I}^2}{\alpha^4}\biggr)  = - \biggl(\frac{4}{9}\biggr)^{\!\!4},
\qquad
\biggl(\frac{a_{\rm I}}{\alpha^2}\biggr) = \pm\hspace{0.03cm} i \hspace{0.03cm}\biggl(\frac{4}{9}\biggr)^{\!\!2}.
\]
Since the relation $a_{I}/\alpha^2$ admits two signs, we conclude that equation (\ref{eq:4i}) has four solutions for the parameter $\beta$ as a function of $\alpha$:
\begin{align}
&\beta_{1}^{(\pm)} = 2\hspace{0.02cm}\biggl[(\pm\hspace{0.02cm} i) +
\frac{\!\sqrt{2}}{3^3}\,\sqrt{\mp \hspace{0.02cm} i}\hspace{0.03cm}\, \biggr]\hspace{0.03cm}\alpha,  \notag\\[0.7ex]
&\beta_{2}^{(\pm)} = 2\hspace{0.02cm}\biggl[(\pm\hspace{0.02cm} i) -
\frac{\!\sqrt{2}}{3^3}\,\sqrt{\mp \hspace{0.02cm} i}\hspace{0.03cm}\, \biggr]\hspace{0.03cm}\alpha  \notag\hspace{0.02cm}.
\end{align}
By the symbol $\sqrt{\mp i}$ here we mean
\begin{equation}
\sqrt{\mp \hspace{0.02cm} i} = \frac{1}{\sqrt{2}}\,(1 \mp i ).
\label{eq:4o}
\end{equation}
The relation (\ref{eq:4u}) enables us to find the corresponding values for the parameter $\delta$ as a function of $\alpha$. By a direct substitution of the obtained solutions for $\beta$ and $\delta$ into the first two equations of the system (\ref{eq:4y}), we verify that all four possible variants in the choice of values for the parameters $\beta_{1,\,2}^{(\pm)}$ and $\delta_{1,\,2}^{(\pm)}$ reduce these equations to an identity.\\
\indent
Let us finally write out an explicit form of the coefficients $(\alpha,\hspace{0.02cm}\beta,\hspace{0.02cm}\gamma,\hspace{0.02cm}\delta)$ of the expansion for the matrix $A$, Eq.\,(\ref{eq:2y}), in choosing the relation $\gamma = -4\hspace{0.02cm}\alpha/9$:
\begin{equation}
\left\{
\begin{array}{l}
\beta_{1}^{(\pm)} = \biggl[\hspace{0.03cm}\displaystyle\frac{\!\!2}{3^3} \hspace{0.03cm}+\hspace{0.03cm} (\pm\hspace{0.02cm} i)
\biggl(2 - \displaystyle\frac{\!\!2}{3^3}\biggr) \biggr]\hspace{0.03cm}\alpha,   \\[3ex]
\delta_{1}^{(\pm)} = \biggl[-\displaystyle\frac{\!\!8}{3^3} \hspace{0.03cm}+\hspace{0.03cm} (\pm\hspace{0.02cm} i)
\biggl(-\displaystyle\frac{\!\!8}{3^2} + \displaystyle\frac{\!\!8}{3^3}\biggr) \biggr]\hspace{0.03cm}
\alpha,
\end{array}
\right.
\qquad
\left\{
\begin{array}{l}
\beta_{2}^{(\pm)} = \biggl[-\displaystyle\frac{\!\!2}{3^3} \hspace{0.03cm}+\hspace{0.03cm} (\pm\hspace{0.02cm} i)
\biggl(2 + \displaystyle\frac{\!\!2}{3^3}\biggr) \biggr]\hspace{0.03cm}\alpha,   \\[3ex]
\delta_{2}^{(\pm)} = \biggl[\hspace{0.03cm}\displaystyle\frac{\!\!8}{3^3} \hspace{0.03cm}+\hspace{0.03cm} (\pm\hspace{0.02cm} i)
\biggl(-\displaystyle\frac{\!\!8}{3^2} - \displaystyle\frac{\!\!8}{3^3}\biggr) \biggr]\hspace{0.03cm}
\alpha, \end{array}
\right.
\label{eq:4p}
\end{equation}
where the parameter $\alpha$ is fixed by the condition (\ref{eq:4t}). When we wrote out these solutions we had taken into account (\ref{eq:4o}).\\
\indent
If we fix the relation (\ref{eq:2x}), i.e. we set $\gamma=-4\hspace{0.02cm}\alpha$, then one should analyze the reduced system of equations (\ref{eq:2c})\,--\,(\ref{eq:2b}) with the value $a = a_{\rm I\!\hspace{0.02cm}I\!\hspace{0.02cm}I}$, Eq.\,(\ref{eq:2g}). Similar reasoning leads to other possible values of the parameters in the expansion (\ref{eq:2y}):
\[
\left\{
\begin{array}{l}
\beta_{1}^{(\pm)} = \biggl[\hspace{0.03cm}2\cdot3^2 \hspace{0.03cm}+\hspace{0.03cm} (\mp\hspace{0.02cm} i)
\biggl(\displaystyle\frac{2}{3} - 2\cdot3^2\biggr)\! \biggr]\hspace{0.03cm}\alpha,   \\[3ex]
\delta_{1}^{(\pm)} = \biggl[\hspace{0.03cm}-8 \hspace{0.03cm}+\hspace{0.03cm} (\mp\hspace{0.02cm} i)
\biggl(-\displaystyle\frac{8}{3} + 8\biggr)\! \biggr]\hspace{0.03cm}\alpha,\end{array}
\right.
\qquad
\left\{
\begin{array}{l}
\beta_{2}^{(\pm)} = \biggl[\hspace{0.03cm}-2\cdot3^2 \hspace{0.03cm}+\hspace{0.03cm} (\mp\hspace{0.02cm} i)
\biggl(\displaystyle\frac{2}{3} + 2\cdot3^2\biggr)\! \biggr]\hspace{0.03cm}\alpha,   \\[3ex]
\delta_{2}^{(\pm)} = \biggl[\hspace{0.03cm}8 \hspace{0.03cm}+\hspace{0.03cm} (\mp\hspace{0.02cm} i) \biggl(-\displaystyle\frac{8}{3} - 8\biggr)\! \biggr]\hspace{0.03cm}\alpha,   \end{array}
\right.
\]
where in turn the parameter $\alpha$ is fixed by the condition
\[
\alpha^{4} = \frac{\!1}{4^7}\,\frac{\!1}{m^2}\,.
\]
\indent
Thus we have completely solved two first matrix equations (\ref{eq:2q}) and (\ref{eq:2w}) and thereby defined an explicit form of the matrix $A$. As it was discussed in the comment following Eq.\,(\ref{eq:2e}), three remaining equations must either be identically fulfilled or lead to a contradiction. Let us consider the third matrix equation (\ref{eq:2e}).  With allowance for the relation (\ref{eq:4q}), the equation under consideration can be resulted in the following form:
\[
\bigl[\hspace{0.02cm}\bigl(A\bar{A}\bigr)A^{2} + \bigl(A\bar{A}\bigr)^{2} + A^{2}\bigl(A\bar{A}\bigr)\hspace{0.02cm}\bigr]\bigl\{\beta_{\mu},\beta_{\nu}\bigr\}
=
-\,\frac{1}{m^2}\,\bigl\{\beta_{\mu},\beta_{\nu}\bigr\}.
\]
By virtue of nonsingularity of the matrices $\beta_{\mu}$, the anticommutator on the left- and right-hand sides can be omitted. Taking into account Eq.\,(\ref{eq:4w}) (without the $\beta_{\mu}$ matrices), we rewrite the equation above in the equivalent form
\begin{equation}
\bigl(A\bar{A}\bigr)\bigl[\hspace{0.02cm}A^{2} - \bar{A}^{2}\hspace{0.02cm}\bigr]
=
-\hspace{0.03cm}\frac{\!2}{m^2}\,I.
\label{eq:4a}
\end{equation}
This equation in contrast to (\ref{eq:4w}) contains already the difference of $A^2$ and $\bar{A}^2$, which is equal to
\[
A^{2} - \bar{A}^{2} = 2\hspace{0.015cm}b\hspace{0.03cm}\Omega + 2\hspace{0.015cm}d\hspace{0.03cm}\Omega^{\hspace{0.02cm}3}.
\]
Multiplication of this expression by the matrix $(A \bar{A})$, as it was defined by Eq.\,(\ref{eq:4e}), gives us the expression containing only {\it odd} powers of the matrix $\Omega$ that thereby leads to the contradiction with the right-hand side in (\ref{eq:4a}). It is clear that the ``na\"{\i}ve'' representation of the fourth root as was defined on the left-hand side of expression (\ref{eq:1c}) is unsuitable. This concerns not only the specific choice of the right-hand side as was presented in (\ref{eq:1c}), but also a choice of the right-hand side in the form of the multi-mass fourth-order Klein-Gordon-Fock operator (\ref{eq:1a}) or the usual single-mass second-order Klein-Gordon-Fock operator (\ref{eq:1s}). Here, we need to develop more subtle approach to solving the problem in hand.

\section{The $\eta_{\mu}$ matrices}
\setcounter{equation}{0}

In the previous section we have shown that a straightforward approach to the construction of the fourth root of the fourth-order wave equation in the form of (\ref{eq:1c}) leads to contradiction. Here, it is necessary to involve some additional considerations of algebraic character. In this section, we attempt to outline a general approach to the stated problem. In our research we will follow the same line of reasoning suggested for the spin-1 case \cite{markov_2015}.\\
\indent
Let us introduce a new set of matrices $\eta_{\mu}$ instead of the original matrices $\beta_{\mu}$, that would satisfy the following condition:
\begin{equation}
A\hspace{0.02cm}\eta_{\mu} = {\rm w}\hspace{0.02cm}\eta_{\mu}\hspace{0.02cm}A,
\label{eq:5q}
\end{equation}
where $ {\rm w}$ is some complex number, and the matrix $A$ obeys equations (\ref{eq:2q}) and (\ref{eq:2w}). Let us return to the expression (\ref{eq:1c}), where now on the left-hand side, instead of the matrices $\beta_{\mu}$, we take $\eta_{\mu}$. Expanding the four power of the operator expression and taking into account the rule of the rearrangement (\ref{eq:5q}) and Eq.\,(\ref{eq:2q}), we derive
\begin{equation}
\bigl[\hspace{0.03cm}A\hspace{0.02cm}(\hspace{0.02cm}\eta\cdot\partial + m\hspace{0.02cm}I)\bigr]^{4} =
\label{eq:5w}
\end{equation}
\[
=
-\frac{1}{m^2}\,{\rm w}^{6}\hspace{0.02cm}(\hspace{0.02cm}\eta\cdot\partial\hspace{0.02cm})^{4}
-\frac{1}{m}\,{\rm w}^{3}\hspace{0.02cm}\varepsilon
({\rm w})(\hspace{0.02cm}\eta\cdot\partial\hspace{0.02cm})^{3}
- {\rm w}\hspace{0.02cm}\bigl[\hspace{0.02cm}\varepsilon({\rm w}) + {\rm w}^{2} +
{\rm w}^{4}\hspace{0.02cm}\bigr](\hspace{0.02cm}\eta\cdot\partial\hspace{0.02cm})^{2}
\]
\[
-\, m\hspace{0.025cm}\varepsilon({\rm w})(\hspace{0.02cm}\eta\cdot\partial\hspace{0.02cm})
- m^{2}I.
\]
Here, we have introduced the function
\[
\varepsilon({\rm w}) = 1 + {\rm w} + {\rm w}^{2} + {\rm w}^{3} \equiv
({\rm w} - q)({\rm w} - q^{2})({\rm w} - q^{3}),
\]
where $q$ is a primitive root of the equation ${\rm w}^4-1=0$. It is evident that if we set the complex number ${\rm w}$ equal to $q$ (or $q^3$), then the right-hand side of (\ref{eq:5w}) is reduced to
\begin{equation}
\frac{1}{m^2}\,(\hspace{0.02cm}\eta\cdot\partial\hspace{0.02cm})^{4} - m^{2}I.
\label{eq:5e}
\end{equation}
Further, we could reproduce the right-hand side of the relation\footnote{\,Formally, we can consider also the limit ${\rm w} \rightarrow q^2$, although $q^2$ is not a primitive root. In this case we obtain expression (\ref{eq:1x}) with the corresponding replacement $\beta_{\mu}\rightarrow\eta_{\mu}$, i.e. analog of the square of the second order Dirac equation.} (\ref{eq:1c}) if the matrices $\eta_{\mu}$ obeyed the identity of the form (\ref{eq:1l})(with the replacement $\beta_{\mu}$ by  $\eta_{\mu}$).\\
\indent
Let us now turn to the construction of an explicit form of the matrices $\eta_{\mu}$. To this end, let us introduce the following {\it deformed} commutator
\begin{equation}
[\hspace{0.03cm}A,\beta_{\mu}\hspace{0.03cm}]_{\hspace{0.01cm}z}
 \equiv A\hspace{0.02cm}\beta_{\mu} - z\hspace{0.015cm}\beta_{\mu}\hspace{0.02cm}A,
\label{eq:5ee}
\end{equation}
where $z$ is an arbitrary complex number. We rearrange the matrix $A$ to the left
\begin{equation}
[\hspace{0.03cm}A,\beta_{\mu}\hspace{0.03cm}]_{\hspace{0.01cm}z}
 =
A\bigl(\hspace{0.02cm}\beta_{\mu} + z\hspace{0.025cm}m^{2\!}\hspace{0.01cm}A^{3\!}
\hspace{0.01cm}\beta_{\mu}\hspace{0.02cm}A\bigr)
\equiv
A\bigl(\hspace{0.02cm}\beta_{\mu} + z\hspace{0.025cm}m^{2\!}A^{2}(A\bar{A})
\beta_{\mu}\bigr).
\label{eq:5r}
\end{equation}
Here, we have taken into account an explicit form of the reciprocal matrix $A^{-1}=-m^2A^3$ and the property (\ref{eq:4q}). On the other hand, we can rearrange the same matrix to the right
\begin{equation}
[\hspace{0.03cm}A,\beta_{\mu}\hspace{0.03cm}]_{\hspace{0.01cm}z}
 =
\bigl(-m^{2\!}A\hspace{0.02cm}\beta_{\mu}\hspace{0.02cm}A^{3} -
z\hspace{0.02cm}\beta_{\mu}\bigr)A
\equiv
-\bigl(\hspace{0.02cm}m^{2\!}\hspace{0.025cm}(A\bar{A})\bar{A}^{2}\hspace{0.01cm}\beta_{\mu} +
z\hspace{0.015cm}\beta_{\mu}\bigr)A.
\label{eq:5t}
\end{equation}
\indent
The expressions for the matrices $A^2$ and $(A \bar{A})$ in the general form are defined by
Eqs.\,(\ref{eq:2u}) and (\ref{eq:4e}), correspondingly (recall that $\bar{A}^2$ is obtained from $A^2$ by means of the simple replacement of the coefficients $(b, d)$ by $(-b, -d)$). A product of these matrices determines $A^2 (A \bar{A})$ and $(A \bar{A}) \bar{A}^2$ which enter into the right-hand side of the expressions (\ref{eq:5r}) and (\ref{eq:5t}). To be specific, let us substitute the values of parameters, corresponding to the solution $({\rm I})$ in the general list of solutions (\ref{eq:2d}), into $A^2 (A \bar{A})$ and $(A \bar{A}) \bar{A}^2$ and fix the relation (\ref{eq:2j}). Then the most right-hand side of Eq.\,(\ref{eq:5r}) takes the following form
\begin{equation}
A\biggl[\biggl(1 - \frac{1}{8}\,z\biggr)\beta_{\mu}
\hspace{0.03cm}\pm\hspace{0.03cm} i\hspace{0.02cm}z\hspace{0.035cm}\frac{9}{4}\,\Omega\hspace{0.03cm}\beta_{\mu}
\hspace{0.03cm}
+
\hspace{0.03cm} z\hspace{0.025cm}\frac{1}{2}\,
\Omega^{\hspace{0.02cm}2\!}\hspace{0.02cm}\beta_{\mu} \hspace{0.03cm}
\mp
\hspace{0.03cm}i\hspace{0.02cm}z\hspace{0.03cm}\Omega^{\hspace{0.02cm}3\!}
\hspace{0.02cm}\beta_{\mu}
\biggr]
\label{eq:5y}
\end{equation}
and similarly for the most right-hand side of Eq.\,(\ref{eq:5t}) we have
\begin{equation}
\biggl[\biggl(-z + \frac{1}{8}\biggr)\beta_{\mu}
\hspace{0.03cm}\pm\hspace{0.03cm}
i\,\frac{9}{4}\,\Omega\hspace{0.03cm}\hspace{0.02cm}\beta_{\mu}
\hspace{0.03cm}
-
\hspace{0.03cm} \frac{1}{2}\,
\Omega^{\hspace{0.02cm}2\!}\hspace{0.03cm}\beta_{\mu}
\hspace{0.03cm}\mp\hspace{0.03cm}
i\hspace{0.03cm}\Omega^{\hspace{0.02cm}3\!}\hspace{0.03cm}\beta_{\mu}
\biggr]A
\label{eq:5u}
\end{equation}
\[
\equiv
-\hspace{0.01cm}z\biggl[\biggl(1- \frac{1}{8\hspace{0.02cm}z}\biggr)\beta_{\mu}
\hspace{0.03cm}\mp\hspace{0.03cm} i\hspace{0.03cm}\frac{9}{4\hspace{0.015cm}z}\;\Omega\hspace{0.035cm}\beta_{\mu}
\hspace{0.03cm}
+
\hspace{0.03cm} \frac{1}{2z}\,
\Omega^{\hspace{0.02cm}2\!}\hspace{0.03cm}\beta_{\mu} \hspace{0.03cm}\pm\hspace{0.03cm}
i\hspace{0.03cm}\frac{1}{z}\;\Omega^{\hspace{0.02cm}3\!}\hspace{0.03cm}\beta_{\mu}
\biggr]A.
\]
Ambiguity of the choice of the signs in front of the terms with odd powers of $\Omega$ in (\ref{eq:5y}) and (\ref{eq:5u}) is connected with the ambiguity of the square root of $\alpha^{4\!}$, Eq.\,(\ref{eq:4t}). It is precisely this choice of the signs in these expressions that is connected with ``synchronization'' of a similar choice of the signs in the coefficients of the matrix $A$ (the symbol $(\pm)$ in the notations of the coefficients $\beta$ and $\delta$ in (\ref{eq:4p})). It is not difficult to obtain also the expressions analogous to (\ref{eq:5y}) and (\ref{eq:5u}) and for the solution $({\rm I\!\hspace{0.03cm}I\!
\hspace{0.03cm}I})$ in the general list (\ref{eq:2d}) with fixing the relation (\ref{eq:2x}).\\
\indent
We introduce by definition the following set of matrices $\eta_{\mu}^{(\pm)}(z)$ depending on a complex number $z$, playing a fundamental role in the subsequent discussion
\begin{equation}
\eta_{\mu}^{(\pm)}(z) \equiv
\biggl(1 - \frac{1}{8}\,z\biggr)\beta_{\mu}
\hspace{0.03cm}\pm\hspace{0.03cm} i\hspace{0.02cm}z\hspace{0.03cm}\frac{9}{4}\;\Omega\hspace{0.03cm}\beta_{\mu}
\hspace{0.03cm}
+
\hspace{0.03cm} z\hspace{0.02cm}\frac{1}{2}\,
\Omega^{\hspace{0.02cm}2\!}\hspace{0.03cm}\beta_{\mu}
\hspace{0.03cm}\mp\hspace{0.03cm}
i\hspace{0.02cm}z\hspace{0.03cm}\Omega^{\hspace{0.02cm}3\!}\hspace{0.03cm}\beta_{\mu}.
\label{eq:5i}
\end{equation}
We equate the expressions (\ref{eq:5y}) and (\ref{eq:5u}), taking successively as the complex number $z$ at first the primitive root $q$ and then $q^{3}$. As a result, we obtain the following expressions:
\begin{align}
&A\hspace{0.02cm}\eta_{\mu}^{(\pm)}(q) = q^{3}\hspace{0.01cm}\eta_{\mu}^{(\pm)}(q)\hspace{0.02cm}A
- 2\hspace{0.02cm}\bigl(\hspace{0.02cm}\Pi_{1/2}\beta_{\mu}\bigr)A, \label{eq:5o}\\[0.8ex]
&A\hspace{0.02cm}\eta_{\mu}^{(\pm)}(q^{3}) = q\hspace{0.04cm}\eta_{\mu}^{(\pm)}(q^3)\hspace{0.02cm}A
-\! 2\hspace{0.02cm}\bigl(\hspace{0.02cm}\Pi_{1/2}\beta_{\mu}\bigr)A, \label{eq:5p}
\end{align}
where we have introduced the notation
\begin{equation}
\Pi_{1/2}\equiv\frac{1}{2}\,\biggl(\Omega^{\hspace{0.02cm}2} - \frac{1}{4}\,I\biggr).
\label{eq:5a}
\end{equation}
The expressions (\ref{eq:5o}) and (\ref{eq:5p}) are transformed into each other with respect to the replacement $q\rightleftarrows q^3$. For definiteness, as the basic relation we take (\ref{eq:5o}). Notice in addition that by virtue of hermitian character of the matrix $\Omega$ (and the matrices $\beta_{\mu}$), with allowance for  the commutation rules (\ref{eq:3o}) and the property $q^{\ast}=q^3$,  the matrices $\eta_{\mu}^{(\pm)}(q)$ and $\eta_{\mu}^{(\pm)}(q^3)$ are connected with each other by the hermitian conjugation:
\[
\bigl[\eta_{\mu}^{(\pm)}(q)\bigr]^{\dagger} = \eta_{\mu}^{(\pm)}(q^{3}).
\]
The asterisk and dagger denote the complex and hermitian conjugations, correspondingly.\\
\indent
Comparing (\ref{eq:5o}) and (\ref{eq:5q}), we see that on the right-hand side of (\ref{eq:5o}) we have the ``redundant'' term: $-2\hspace{0.02cm}(\Pi_{1/2} \beta_{\mu}) A$, which does not enable us to give the (\ref{eq:5q}) form to (\ref{eq:5o}) in choosing $\eta_{\mu}=\eta_{\mu}^{(\pm)}(q)$ and ${\rm w} = q^3$. Note that this circumstance in a qualitative sense distinguishes the present consideration from a similar one in the case of the DKP theory \cite{markov_2015}. In the latter we obtained at once the relation of the form (\ref{eq:5q}) (Eq.\,(\ref{eq:4u}) in \cite{markov_2015}). One can suggest that this circumstance is closely related to the presence of spin-1/2 component in the general construction of a formalism for the spin-3/2 case.\\
\indent
Let us analyzed the expression (\ref{eq:5o}) in more detail. To this end we note that the matrix $\Pi_{1/2}$, Eq.\,(\ref{eq:5a}), by virtue of the characteristic equation (\ref{eq:2r}), is idempotent
\[
\bigl(\Pi_{1/2}\bigr)^{2} = \Pi_{1/2}.
\]
We want to construct a matrix $\Pi_{3/2}$ that satisfies the condition
\[
\Pi_{1/2}\Pi_{3/2} = \Pi_{3/2}\Pi_{1/2} = 0.
\]
It is not difficult to verify that the matrix has the following structure:
\[
\Pi_{3/2}^{(\sigma)} = \sigma\biggl(\Omega^{\hspace{0.02cm}3} \hspace{0.02cm}-\hspace{0.02cm} \frac{9}{4}\,\Omega\biggr),
\]
where $\sigma$ is an arbitrary number parameter. However, this matrix in contrast to $\Pi_{1/2}$ is not idempotent, since
\[
\bigl(\Pi_{3/2}^{(\sigma)}\bigr)^{2} = \sigma^{2}\bigl(\hspace{0.02cm}I - \Pi_{1/2}\bigr).
\]
The cube of this matrix is equal to
\[
\bigl(\Pi_{3/2}^{(\sigma)}\bigr)^{3} = \sigma^{2\,}\Pi_{3/2}^{(\sigma)}.
\]
We will require that this matrix be {\it tripotent}, then
\[
\sigma^{2} = 1\quad \mbox{or} \quad \sigma = \pm\hspace{0.02cm} 1.
\]
Thus, we have a set of three matrices
\[
\Pi_{1/2}^{\phantom{(\pm)}}\equiv\frac{1}{2}\,\biggl(\Omega^{\hspace{0.02cm}2} - \frac{1}{4}\,I\biggr),
\qquad
\Pi_{3/2}^{(\pm)} = \pm\biggl(\Omega^{\hspace{0.02cm}3} - \frac{9}{4}\;\Omega\biggr)
\]
possessing the properties
\begin{equation}
\begin{array}{ll}
\bigl(\Pi_{1/2}^{\phantom{(\pm)}}\bigr)^{2} = \Pi_{1/2}^{\phantom{(\pm)}},
&\qquad \bigl(\Pi_{3/2}^{(\pm)}\bigr)^{2} = I - \Pi_{1/2}^{\phantom{(\pm)}},  \\[3ex]
\bigl(\Pi_{3/2}^{(\pm)}\bigr)^{3} = \Pi_{3/2}^{(\pm)},
&\qquad \Pi_{1/2}^{\phantom{(\pm)}}\hspace{0.02cm}\Pi_{3/2}^{(\pm)} = \Pi_{3/2}^{(\pm)}\hspace{0.02cm}\Pi_{1/2}^{\phantom{(\pm)}} = 0.
\end{array}
\label{eq:5s}
\end{equation}
This set of matrices does not generate a system of the projectors by virtue of tripotent character of the matrices $\Pi_{3/2}^{(\pm)}$. Besides, their sum does not give us the unity matrix. Nevertheless, one can introduce a new set of three matrices possessing all the properties of projectors. For this purpose, we write out an explicit form of the matrices $\eta_{\mu}^{(\pm)}(q)$ in terms of $\Pi_{1/2}$ and $\Pi_{3/2}^{(\pm)}$. It follows from the original expression (\ref{eq:5i}) at $z = q$ that
\begin{equation}
\eta_{\mu}^{(\pm)}(q) = \beta_{\mu} + q\hspace{0.03cm}\Pi_{1/2}^{\phantom{(\pm)}}\beta_{\mu}
-
i\hspace{0.015cm}q\hspace{0.03cm}\Pi_{3/2}^{(\pm)\,}\beta_{\mu}.
\label{eq:5d}
\end{equation}
We may replace the first term on the right-hand side of (\ref{eq:5d}) by
$\bigl[\hspace{0.02cm}\Pi_{1/2}^{\phantom{(\pm)}} + (\Pi_{3/2}^{(\pm)})^2\hspace{0.02cm}\bigr]\hspace{0.02cm}\beta_{\mu}$. This is legitimate in view of one of the properties in (\ref{eq:5s}). Then, instead of (\ref{eq:5d}), we have
\begin{equation}
\eta_{\mu}^{(\pm)}(q) = (1 + q)\hspace{0.03cm}\Pi_{1/2}^{\phantom{(\pm)}}\beta_{\mu}
+
\left[\bigl(\Pi_{3/2}^{(\pm)}\bigr)^{2} - \hspace{0.02cm}
i\hspace{0.015cm}q\hspace{0.035cm}\Pi_{3/2}^{(\pm)}\hspace{0.03cm}\right]\!\beta_{\mu}.
\label{eq:5f}
\end{equation}
Let us introduce the notation
\begin{equation}
{\cal P}_{3/2}^{(\pm)}(q)
\equiv
\frac{1}{2}\left[\bigl(\Pi_{3/2}^{(\pm)}\bigr)^{2} -
i\hspace{0.015cm}q\hspace{0.035cm}\Pi_{3/2}^{(\pm)}\hspace{0.03cm}\right]\!.
\label{eq:5ff}
\end{equation}
By a direct calculation with the use of the properties (\ref{eq:5s}), it is not difficult to verify that the following relation:
\[
\bigl({\cal P}_{3/2}^{(\pm)}(q)\bigr)^{2} = {\cal P}_{3/2}^{(\pm)}(q)
\]
holds. In what follows, for the uniformity of notations, we also put ${\cal P}_{1/2}^{\phantom{(\pm)}}\! \equiv\! \Pi_{1/2}^{\phantom{(\pm)}}$. A set of the matrices
$({\cal P}_{1/2}^{\phantom{(\pm)}},\hspace{0.02cm} {\cal P}_{3/2}^{(\pm)}(q))$ satisfies the standard relations of the usual algebra of projectors:
\begin{equation}
\begin{array}{ll}
\bigl({\cal P}_{1/2}^{\phantom{(\pm)}}\bigr)^{2} = {\cal P}_{1/2}^{\phantom{(\pm)}},
&\qquad \bigl({\cal P}_{3/2}^{(\pm)}(q)\bigr)^{2} = {\cal P}_{3/2}^{(\pm)}(q),  \\[3ex]
{\cal P}_{3/2}^{(\pm)}(q){\cal P}_{3/2}^{(\mp)}(q) = 0,
&\qquad {\cal P}_{1/2}^{\phantom{(\pm)}}{\cal P}_{3/2}^{(\pm)}(q)
=
{\cal P}_{3/2}^{(\pm)}(q){\cal P}_{1/2}^{\phantom{(\pm)}} = 0
\end{array}
\label{eq:5g}
\end{equation}
and the completeness relation\footnote{\,We intentionally use the symbols $(\pm)$ or $(\mp)$ in all the expressions without concretizing the signs $(+)$ or $(-)$. This is very convenient, since this allows us to not care of right arrangement of the signs in formulae, when the choice of the signs becomes important.}:
\begin{equation}
{\cal P}_{1/2}^{\phantom{(\pm)}} + {\cal P}_{3/2}^{(\pm)}(q) + {\cal P}_{3/2}^{(\mp)}(q) = I.
\label{eq:5gg}
\end{equation}
The rules of rearrangement with the matrices $\beta_{\mu}$
\begin{equation}
{\cal P}_{1/2}^{\phantom{(\pm)}}\beta_{\mu} = \beta_{\mu}{\cal P}_{1/2}^{\phantom{(\pm)}} ,
\qquad
{\cal P}_{3/2}^{(\pm)}\beta_{\mu}(q) = \beta_{\mu} {\cal P}_{3/2}^{(\mp)}(q)
\label{eq:5h}
\end{equation}
and also the property
\begin{equation}
{\cal P}_{3/2}^{(\pm)}(q) = {\cal P}_{3/2}^{(\mp)}(q^{3})
\label{eq:5j}
\end{equation}
will be useful for the subsequent discussion.\\
\indent
Further we introduce by definition the following projected $\beta$\hspace{0.0cm}-\hspace{0.01cm}matrices:
\begin{equation}
\eta_{\mu}^{(1/2)} \equiv {\cal P}_{1/2}^{\phantom{(\pm)}}\beta_{\mu},
\qquad
\eta_{\mu}^{(\pm\hspace{0.02cm} 3/2)\!}(q) \equiv
{\cal P}_{3/2}^{(\pm)}(q)\beta_{\mu}.
\label{eq:5k}
\end{equation}
The projector ${\cal P}_{1/2}^{\phantom{(\pm)}}$ does not depend on the primitive root $q$ and, therefore, we do not write any argument of the matrix $\eta_{\mu}^{(1/2)}$. It is evident that the following relations:
\begin{equation}
\begin{array}{lll}
{\cal P}_{1/2}^{\phantom{(\pm)}}\eta_{\mu}^{(1/2)} = \eta_{\mu}^{(1/2)},
&\quad {\cal P}_{3/2}^{(\pm)}(q)\hspace{0.02cm} \eta_{\mu}^{(1/2)} = 0,
\qquad {\cal P}_{3/2}^{(\pm)}(q)\hspace{0.02cm} \eta_{\mu}^{(\mp\hspace{0.02cm} 3/2)\!}(q)
= 0, \\[3ex]
{\cal P}_{1/2}^{\phantom{(\pm)}}\eta_{\mu}^{(\pm\hspace{0.02cm} 3/2)\!}(q)  = 0,
&\quad {\cal P}_{3/2}^{(\pm)}(q)\hspace{0.02cm} \eta_{\mu}^{(\pm\hspace{0.02cm} 3/2)\!}(q)
=
\eta_{\mu}^{(\pm\hspace{0.02cm} 3/2)\!}(q)
\end{array}
\label{eq:5l}
\end{equation}
are true and besides
\[
\eta_{\mu}^{(1/2)}\hspace{0.01cm}\eta_{\mu}^{(\pm\hspace{0.02cm} 3/2)\!}(q) =
\eta_{\mu}^{(\pm\hspace{0.025cm} 3/2)\!}(q)\hspace{0.02cm}\eta_{\mu}^{(1/2)} = 0.
\]
The matrices $\eta_{\mu}^{(\pm)}(q)$, Eq.\,(\ref{eq:5f}), can be rewritten in the following form:
\begin{equation}
\eta_{\mu}^{(\pm)}(q) = (1 + q)\hspace{0.02cm}\eta_{\mu}^{(1/2)} + 2\hspace{0.02cm} \eta_{\mu}^{(\pm\hspace{0.02cm} 3/2)\!}(q).
\label{eq:5z}
\end{equation}
\indent
Now we turn to the expression (\ref{eq:5o}) and substitute the matrix (\ref{eq:5z}) into it. Multiplying Eq.\,(\ref{eq:5o}) by the projector ${\cal P}_{3/2}^{(\pm)}(q)$ on the left and taking into account the commutativity of the matrices ${\cal P}_{3/2}^{(\pm)}(q)$ and $A$, and the properties (\ref{eq:5l}), we obtain
\begin{equation}
A\hspace{0.02cm}\eta_{\mu}^{(\pm\hspace{0.02cm} 3/2)}(q) =
q^{3}\hspace{0.025cm}\eta_{\mu}^{(\pm \hspace{0.02cm}3/2)}(q)\hspace{0.02cm}A.
\label{eq:5x}
\end{equation}
We see that the structure of this expression is exactly the same as that of (\ref{eq:5q}), where by the matrices $\eta_{\mu}$ it is necessary to mean the projected matrices $\eta_{\mu}^{(\pm\hspace{0.02cm} 3/2)\!}(q)$ and as the complex number ${\rm w}$ the primitive root $q^{3}$ should be taken. For (\ref{eq:5p}) we will have a similar expression
\begin{equation}
A\hspace{0.02cm}\eta_{\mu}^{(\pm\hspace{0.02cm} 3/2)}(q^{3}) =
q\hspace{0.03cm}\eta_{\mu}^{(\pm \hspace{0.02cm}3/2)}(q^{3})\hspace{0.02cm}A,
\label{eq:5c}
\end{equation}
where $\eta_{\mu}^{(\pm\hspace{0.02cm} 3/2)}(q^{3}) \equiv {\cal P}_{3/2}^{(\pm)}(q^3) \beta_{\mu}$. If one takes into account the property (\ref{eq:5j}), then the relation
\[
\eta_{\mu}^{(\pm\hspace{0.02cm} 3/2)}(q^{3}) =
\eta_{\mu}^{(\mp\hspace{0.02cm} 3/2)}(q)
\]
holds and, therefore, the equality (\ref{eq:5c}) can be rewritten in the other equivalent form
\begin{equation}
A\hspace{0.02cm}\eta_{\mu}^{(\mp\hspace{0.02cm} 3/2)}(q) =
q\hspace{0.03cm}\eta_{\mu}^{(\mp\hspace{0.02cm} 3/2)}(q)\hspace{0.02cm}A.
\label{eq:5v}
\end{equation}
This expression will be needed in the subsequent consideration. One can verify the validity of the important formulae (\ref{eq:5x}) and (\ref{eq:5c}) by straightforward calculations.\\
\indent
It remains to consider a similar relation with the projector ${\cal P}_{1/2}^{\phantom{(\pm)}}$. Multiplying Eq.\,(\ref{eq:5o}) by the projector  ${\cal P}_{1/2}^{\phantom{(\pm)}}$ on the left, we find
\[
(1 + q)A\hspace{0.02cm}\eta_{\mu}^{(1/2)} = q^{3}(1 + q)\hspace{0.02cm}\eta_{\mu}^{(1/2)\!}A -
2\hspace{0.02cm}\eta_{\mu}^{(1/2)\!}A.
\]
Dividing this expression by $(1+q)$ and taking into account that
\[
q^{3} - \frac{2}{1 + q} = q^{2},
\]
we finally obtain
\[
A\hspace{0.02cm}\eta_{\mu}^{(1/2)} = q^{2}\hspace{0.03cm}\eta_{\mu}^{(1/2)\!}A.
\]
The structure of this expression also coincides with (\ref{eq:5q}), only the matrices $\eta_{\mu}$ should be identified now with $\eta_{\mu}^{(1/2)}$ and as the complex number ${\rm w}$ it is necessary to take $q^2\,(=-1)$.

\section{Commutation relations for the $\eta_{\mu}^{(\pm\hspace{0.02cm} 3/2)\!}(q)$ matrices}
\setcounter{equation}{0}

In this section we define two commutation relations for the matrices $\eta_{\mu}^{(\pm\hspace{0.02cm} 3/2)}(q)$. Our first step is to consider the commutator of two $\eta_{\mu}^{(\pm\hspace{0.02cm} 3/2)}(q)$ matrices. For this purpose we examine a product of these two matrices. By virtue of the definition (\ref{eq:5k}) and properties (\ref{eq:5g}), (\ref{eq:5h}) we have
\begin{equation}
\eta_{\mu}^{(\pm\hspace{0.02cm} 3/2)}(q)\hspace{0.02cm}
\eta_{\nu}^{(\pm\hspace{0.02cm} 3/2)}(q)
=
\bigl({\cal P}_{3/2}^{(\pm)}(q)\beta_{\mu}\bigr)
\bigl({\cal P}_{3/2}^{(\pm)}(q)\beta_{\nu}\bigr)
=
{\cal P}_{3/2}^{(\pm)}(q){\cal P}_{3/2}^{(\mp)}(q)
\beta_{\mu}\hspace{0.02cm} \beta_{\nu} = 0.
\label{eq:6q}
\end{equation}
Thus, the product of two matrices $\eta_{\mu}^{(\pm\hspace{0.02cm} 3/2)}(q)$ with the same set of signs $(\pm)$ is nilpotent and, therefore, the usual definition of the commutator with the matrices of interest is identically vanishing. For obtaining a nontrivial expression we make use of an approach suggested in our paper for the spin-1 case \cite{markov_2015}.\\
\indent
First of all we rewrite the matrix $\eta_{\mu}^{(\pm)}(z)$ depending on an arbitrary complex number $z$, Eq.\,(\ref{eq:5i}), in terms of the matrices $\eta_{\mu}^{(1/2)}$ and $\eta_{\mu}^{(\pm\hspace{0.02cm} 3/2)}(q)$:
\begin{align}
\eta_{\mu}^{(\pm)}(z) &= \beta_{\mu} + z\hspace{0.025cm}\Pi_{1/2}^{\phantom{(\pm)}}\beta_{\mu} - i\hspace{0.015cm}z\hspace{0.025cm}\Pi_{3/2}^{(\pm)\,}\beta_{\mu} \notag
\\[1ex]
&=
(z + 1)\hspace{0.03cm}\Pi_{1/2}^{\phantom{(\pm)}}\beta_{\mu} +
\Bigl[\hspace{0.03cm}\bigl(\Pi_{3/2}^{(\pm)}\bigr)^{2} -
i\hspace{0.005cm}z\hspace{0.015cm}\Pi_{3/2}^{(\pm)\,}\Bigr]\beta_{\mu}
\notag \\[1ex]
&= (z - q^{2})\hspace{0.03cm}{\cal P}_{1/2}^{\phantom{(\pm)}}\beta_{\mu}
+
\Bigl[\hspace{0.03cm}(1 - z\hspace{0.02cm}q)\hspace{0.02cm}{\cal P}_{3/2}^{(\pm)}(q) +
(1 + z\hspace{0.02cm}q)\hspace{0.02cm}{\cal P}_{3/2}^{(\mp)}(q)\Bigr]\beta_{\mu}\notag
\\[1ex]
&\equiv
 (z - q^{2})\hspace{0.03cm}\eta_{\mu}^{(1/2)}
+
\bigl[\hspace{0.03cm}(1 - z\hspace{0.02cm}q)
\hspace{0.02cm}\eta_{\mu}^{(\pm\hspace{0.02cm} 3/2)}(q) +
(1 + z\hspace{0.02cm}q)\hspace{0.02cm}\eta_{\mu}^{(\mp\hspace{0.02cm} 3/2)}(q)\bigr].
\notag
\end{align}
In this general expression our concern is only with the part associated with the matrices $\eta_{\mu}^{(\pm\hspace{0.02cm} 3/2)}(q)$ which we separate as follows:
\begin{equation}
\eta_{\mu}^{(\pm\hspace{0.02cm} 3/2)}(z) \equiv
\bigl[\hspace{0.02cm}{\cal P}_{3/2}^{(\pm)}(q) + {\cal P}_{3/2}^{(\mp)}(q)\bigr]\hspace{0.02cm}
\eta_{\mu}^{(\pm)}(z)
=
(1 - z\hspace{0.02cm}q)
\hspace{0.02cm}\eta_{\mu}^{(\pm\hspace{0.02cm} 3/2)}(q) +
(1 + z\hspace{0.02cm}q)\hspace{0.02cm}\eta_{\mu}^{(\mp\hspace{0.02cm} 3/2)}(q).
\label{eq:6w}
\end{equation}
Further, we can present the matrices $\eta_{\mu}^{(\pm\hspace{0.02cm} 3/2)}(z)$ in the form of an expansion in terms of $\delta \equiv z - q\hspace{0.02cm}$:
\begin{equation}
\eta_{\mu}^{(\pm\hspace{0.02cm} 3/2)}(z) = 2\hspace{0.02cm}
\eta_{\mu}^{(\pm\hspace{0.02cm} 3/2)}(q)
+ \delta\hspace{0.015cm}\eta_{\mu}^{\hspace{0.02cm}\prime
\hspace{0.015cm}(\pm\hspace{0.02cm} 3/2)}(q),
\label{eq:6r}
\end{equation}
where the matrices $\eta_{\mu}^{\hspace{0.02cm}\prime\hspace{0.015cm}(\pm\hspace{0.02cm} 3/2)}(q)$ have the form
\begin{equation}
\left.
\eta_{\mu}^{\hspace{0.02cm}\prime\hspace{0.015cm}(\pm\hspace{0.02cm} 3/2)}(q)
\equiv \frac{d\hspace{0.02cm}\eta_{\mu}^{(\pm\hspace{0.02cm} 3/2)}(z)}
{d\hspace{0.02cm}z}\right|_{z\hspace{0.02cm} =\hspace{0.02cm} q}\! =
q^{3}\eta_{\mu}^{(\pm\hspace{0.02cm} 3/2)}(q) +
q\hspace{0.02cm}\eta_{\mu}^{(\mp\hspace{0.02cm} 3/2)}(q).
\label{eq:6t}
\end{equation}
\indent
Let us consider a product of two matrices $\eta_{\mu}^{(\pm\hspace{0.02cm} 3/2)}(z)$. In the limit $\delta \rightarrow 0$ and with allowance for (\ref{eq:6q}), (\ref{eq:6r}) and (\ref{eq:6t}), we get
\begin{align}
&\hspace{2cm}\eta_{\mu}^{(\pm\hspace{0.02cm} 3/2)}(q + \delta)\hspace{0.02cm}
\eta_{\nu}^{(\pm\hspace{0.02cm} 3/2)}(q + \delta) \label{eq:6y} \\[1.3ex]
&=
2\hspace{0.01cm}\delta\bigl[\hspace{0.02cm}\eta_{\mu}^{\hspace{0.02cm}\prime\hspace{0.015cm} (\pm\hspace{0.02cm} 3/2)}(q)
\hspace{0.02cm}\eta_{\nu}^{(\pm\hspace{0.02cm} 3/2)}(q) +
\eta_{\mu}^{(\pm\hspace{0.02cm} 3/2)}(q)\hspace{0.02cm}
\eta_{\nu}^{\hspace{0.02cm}\prime\hspace{0.015cm} (\pm\hspace{0.02cm} 3/2)}(q)\bigr]
+
{\cal O}\hspace{0.01cm}(\delta^{\hspace{0.02cm}2})
\notag \\[1.3ex]
&=
2\hspace{0.01cm}\delta\hspace{0.01cm}q\bigl[\hspace{0.02cm}\eta_{\mu}^{\hspace{0.015cm} (\mp\hspace{0.02cm}3/2)}(q)
\hspace{0.02cm}\eta_{\nu}^{(\pm\hspace{0.02cm} 3/2)}(q) +
\eta_{\mu}^{(\pm\hspace{0.02cm} 3/2)}(q)\hspace{0.02cm}
\eta_{\nu}^{\hspace{0.015cm}(\mp\hspace{0.02cm} 3/2)}(q)\bigr]
+
{\cal O}\hspace{0.01cm}(\delta^{\hspace{0.02cm}2}).
\notag
\end{align}
Taking into account the aforesaid, as the commutation relation for the $\eta_{\mu}^{(\pm\hspace{0.02cm} 3/2)}$ matrices we take the following expression:
\begin{equation}
S_{\mu\nu}(q) \equiv
\lim_{z\rightarrow\hspace{0.03cm} q}\hspace{0.01cm}
\frac{1}{\,\epsilon(z)}\,\bigl[\hspace{0.03cm}
\eta_{\mu}^{(\pm\hspace{0.02cm} 3/2)}(z),
\eta_{\nu}^{(\pm\hspace{0.02cm} 3/2)}(z)\hspace{0.02cm}\bigr]
\label{eq:6u}
\end{equation}
\[
=
\bigl[\hspace{0.03cm}\eta_{\mu}^{(\pm\hspace{0.02cm} 3/2)}(q),\eta_{\nu}^{(\mp\hspace{0.02cm} 3/2)}(q)\hspace{0.02cm}\bigr]
+
\bigl[\hspace{0.03cm}\eta_{\mu}^{(\mp\hspace{0.02cm} 3/2)}(q),\eta_{\nu}^{(\pm\hspace{0.02cm} 3/2)}(q)\hspace{0.02cm}\bigr]
\equiv
S_{\mu\nu}^{\rm (I)}(q) + S_{\mu\nu}^{\rm (I\!\hspace{0.025cm}I)}(q).
\]
Here, we have introduced by definition the following function:
\begin{equation}
\epsilon(z) = (z - q)(z - q^{3})
\label{eq:6e}
\end{equation}
and considered that $\lim_{z \rightarrow q}\epsilon^{-1}(z)\sim 1/(2q \delta)$. The multiplier $\epsilon^{-1}(z) $ in (\ref{eq:6u}) exactly compensates the appropriate factor in front of the square brackets in (\ref{eq:6y}). We note that the spin-tensor (\ref{eq:6u}) can be presented in another equivalent form
\begin{equation}
S_{\mu\nu}(q) = {\cal P}_{3/2}^{(\pm)}(q)S_{\mu\nu}(q)
+
{\cal P}_{3/2}^{(\mp)}(q)S_{\mu\nu}(q),
\label{eq:6ee}
\end{equation}
where
\begin{equation}
{\cal P}_{3/2}^{(\pm)}(q)S_{\mu\nu}(q) =
\eta_{\mu}^{(\pm\hspace{0.02cm} 3/2)}(q)\hspace{0.02cm}\eta_{\nu}^{(\mp\hspace{0.02cm} 3/2)}(q)
-
\eta_{\nu}^{(\pm\hspace{0.02cm} 3/2)}(q)\hspace{0.02cm}\eta_{\mu}^{(\mp\hspace{0.02cm} 3/2)}(q)
\label{eq:6eee}
\end{equation}
and correspondingly
\begin{equation}
{\cal P}_{3/2}^{(\mp)}(q)S_{\mu\nu}(q) =
\eta_{\mu}^{(\mp\hspace{0.02cm} 3/2)}(q)\hspace{0.02cm}\eta_{\nu}^{(\pm\hspace{0.02cm} 3/2)}(q)
-
\eta_{\nu}^{(\mp\hspace{0.02cm} 3/2)}(q)\hspace{0.02cm}\eta_{\mu}^{(\pm\hspace{0.02cm} 3/2)}(q).
\label{eq:6eeee}
\end{equation}
In spite of the fact that the right-hand sides of the projections (\ref{eq:6eee}) and (\ref{eq:6eeee}) do not represent the matrix commutators, nevertheless each of the expressions is antisymmetric {\it by itself} upon interchange of the indices $\mu \leftrightarrow \nu$, whereas in the initial definition (\ref{eq:6u}) two commutators $S_{\mu\nu}^{\rm (I)}(q)$ and $S_{\mu\nu}^{\rm (I\!\hspace{0.025cm}I)}(q)$ on the right-hand side transforms to each other with the opposite sign upon interchange $\mu \leftrightarrow \nu$ by virtue of the relation
\[
S_{\mu\nu}^{\rm (I\!\hspace{0.025cm}I)}(q) = - S_{\nu\mu}^{\rm (I)}(q).
\]
Thus, the expressions (\ref{eq:6eee}) and (\ref{eq:6eeee}) represent two completely independent spin structures in this consideration.
\\
\indent
Let us consider the double commutation relation with the $\eta_{\mu}^{(\pm\hspace{0.02cm} 3/2)}(q)$
matrices. By using (\ref{eq:6u}), we have
\begin{equation}
\lim_{z\rightarrow\hspace{0.03cm} q}\hspace{0.01cm}
\frac{1}{\,\epsilon(z)}\,\bigl[\bigl[\hspace{0.03cm}\eta_{\mu}^{(\pm\hspace{0.02cm} 3/2)}(z),\eta_{\nu}^{(\pm\hspace{0.02cm} 3/2)}(z)\hspace{0.02cm}\bigr],
\eta_{\lambda}^{(\pm\hspace{0.02cm} 3/2)}(z)\hspace{0.02cm}\bigr]
\label{eq:6i}
\end{equation}
\[
=
\bigl[\bigl[\hspace{0.03cm}\eta_{\mu}^{(\pm\hspace{0.02cm} 3/2)}(q),\eta_{\nu}^{(\mp\hspace{0.02cm} 3/2)}(q)\hspace{0.02cm}\bigr],
\eta_{\lambda}^{(\pm\hspace{0.02cm} 3/2)}(q)\hspace{0.02cm}\bigr]
+
\bigl[\bigl[\hspace{0.03cm}
\eta_{\mu}^{(\mp\hspace{0.02cm} 3/2)}(q),\eta_{\nu}^{(\pm\hspace{0.02cm} 3/2)}(q)\hspace{0.02cm}\bigr],
\eta_{\lambda}^{(\pm\hspace{0.02cm} 3/2)}(q)\hspace{0.02cm}\bigr].
\]
We will analyze the right-hand side of this expression. With this in mind we recall that the original matrices $\beta_{\mu}$ satisfy the trilinear relation
\[
[\hspace{0.03cm}[\hspace{0.03cm}\beta_{\mu},\beta_{\nu}],\beta_{\lambda}\hspace{0.02cm}]
= \beta_{\mu}\hspace{0.02cm}\delta_{\nu\lambda} - \beta_{\nu}\hspace{0.02cm}\delta_{\mu\lambda}.
\]
As already mentioned in Introduction, in the Bhabha theory of the higher spin particles, it is postulated that this relation must valid for all spins. Let us multiply this relation by $({\cal P}_{3/2}^{(\pm)}(q))^3$ from the left. Taking into account the properties (\ref{eq:5g}), (\ref{eq:5h}), and the definition (\ref{eq:5k}), we find
\[
\bigl[\bigl[\hspace{0.03cm}\eta_{\mu}^{(\pm\hspace{0.02cm}3/2)}(q),
\eta_{\nu}^{(\mp\hspace{0.02cm} 3/2)}(q)\hspace{0.02cm}\bigr],
\eta_{\lambda}^{(\pm\hspace{0.02cm} 3/2)}(q)\hspace{0.02cm}\bigr]
+
\bigl[\bigl[\hspace{0.03cm}
\eta_{\mu}^{(\mp\hspace{0.02cm} 3/2)}(q),\eta_{\nu}^{(\pm\hspace{0.02cm} 3/2)}(q)\hspace{0.02cm}\bigr],
\eta_{\lambda}^{(\pm\hspace{0.02cm} 3/2)}(q)\hspace{0.02cm}\bigr]
\]
\[
=
\eta_{\mu}^{(\pm\hspace{0.02cm}3/2)}(q)\hspace{0.02cm}\delta_{\nu\lambda} - \eta_{\nu}^{(\pm\hspace{0.02cm}3/2)}(q)\hspace{0.02cm}\delta_{\mu\lambda}
\]
and thus the limit (\ref{eq:6i}) takes the final form:
\[
\lim_{z\rightarrow\hspace{0.03cm} q}\hspace{0.01cm}
\frac{1}{\,\epsilon(z)}\,\bigl[\bigl[\hspace{0.03cm}\eta_{\mu}^{(\pm\hspace{0.02cm} 3/2)}(z),\eta_{\nu}^{(\pm\hspace{0.02cm} 3/2)}(z)\hspace{0.02cm}\bigr],
\eta_{\lambda}^{(\pm\hspace{0.02cm} 3/2)}(z)\hspace{0.02cm}\bigr]
=
\eta_{\mu}^{(\pm\hspace{0.02cm}3/2)}(q)
\hspace{0.02cm}\delta_{\nu\lambda} - \eta_{\nu}^{(\pm\hspace{0.02cm}3/2)}(q)\hspace{0.02cm}\delta_{\mu\lambda}.
\]
This relation will assure us the relativistic covariance of the following wave equation (in the limit $z \rightarrow q$):
\begin{equation}
A\biggl[\frac{\!1}{\,\epsilon^{1/2}(z)}\,\eta_{\mu}^{(\pm\hspace{0.02cm}3/2)}(z)
\hspace{0.03cm}\partial_{\mu}
+
\Bigl(\hspace{0.02cm}{\cal P}_{3/2}^{(\pm)}(q) +
{\cal P}_{3/2}^{(\mp)}(q)\!\hspace{0.02cm}\Bigr)\hspace{0.02cm}m\biggr]\psi(x;z) = 0.
\label{eq:6o}
\end{equation}
A careful analysis of this equation will be considered in the next section. Note only that in the notation of the wave function $\psi$ we have explicitly separated out the dependence on the deformation parameter $z$, and in the mass term, instead of the unity matrix $I$, we have entered a sum of projectors which single out in $\psi$ only the part connected with the spin\hspace{0.01cm}-\hspace{0.01cm}$3/2$ sector.\\
\indent
Finally, let us consider the question of four-linear algebra to which the matrices $\eta_{\mu}^{(\pm\hspace{0.02cm}3/2)}(z)$ have to satisfy. In other words, what is analog of the algebra (\ref{eq:1i}) for these matrices? As a preliminary step, we consider the following limit:\\
\begin{align}
\lim_{z\rightarrow\hspace{0.03cm} q}\hspace{0.01cm}
\frac{1}{\,\epsilon^{2}(z)}\,\Bigl[\hspace{0.03cm}&\eta_{\mu}^{(\pm\hspace{0.02cm}3/2)}(z)
\Bigl(\hspace{0.02cm}
\eta_{\nu}^{(\pm\hspace{0.02cm} 3/2)}(z)\hspace{0.02cm}
\eta_{\lambda}^{(\pm\hspace{0.02cm} 3/2)}(z)\hspace{0.02cm}
\eta_{\sigma}^{(\pm\hspace{0.02cm} 3/2)}(z)
+
\bigl(\nu \rightleftarrows \sigma\bigr)\!\Bigr)
\notag \\[1ex]
&+\Bigl(\hspace{0.02cm}\eta_{\nu}^{(\pm\hspace{0.02cm} 3/2)}(z)\hspace{0.02cm}
\eta_{\lambda}^{(\pm\hspace{0.02cm} 3/2)}(z)\eta_{\sigma}^{(\pm\hspace{0.02cm} 3/2)}(z)
+
\bigl(\nu \rightleftarrows \sigma\bigr)\!\Bigr)\hspace{0.02cm}\eta_{\mu}^{(\pm\hspace{0.02cm}3/2)}(z)\Bigr]
\notag \\[1ex]
= \frac{1}{4}\,\Bigl[\hspace{0.03cm}&\eta_{\mu}^{(\pm\hspace{0.02cm}3/2)}(q)
\Bigl(\hspace{0.02cm}
\eta_{\nu}^{(\mp\hspace{0.02cm} 3/2)}(q)\hspace{0.02cm}
\eta_{\lambda}^{(\pm\hspace{0.02cm} 3/2)}(q)\hspace{0.02cm}
\eta_{\sigma}^{(\mp\hspace{0.02cm} 3/2)}(q)
+
\bigl(\nu \rightleftarrows \sigma\bigr)\!\Bigr)
\label{eq:6p} \\[1ex]
&+\Bigl(\hspace{0.02cm}\eta_{\nu}^{(\pm\hspace{0.02cm} 3/2)}(q)\hspace{0.02cm}
\eta_{\lambda}^{(\mp\hspace{0.02cm} 3/2)}(q)
\eta_{\sigma}^{(\pm\hspace{0.02cm} 3/2)}(q)
+
\bigl(\nu \rightleftarrows \sigma\bigr)\!\Bigr)\hspace{0.02cm}
\eta_{\mu}^{(\mp\hspace{0.02cm}3/2)}(q)\Bigr]
\notag \\[1ex]
+\hspace{0.02cm} \frac{1}{4}\,\Bigl[\hspace{0.03cm}&\eta_{\mu}^{(\mp\hspace{0.02cm}3/2)}(q)
\Bigl(\hspace{0.02cm}
\eta_{\nu}^{(\pm\hspace{0.02cm} 3/2)}(q)\hspace{0.02cm}
\eta_{\lambda}^{(\mp\hspace{0.02cm} 3/2)}(q)\hspace{0.02cm}
\eta_{\sigma}^{(\pm\hspace{0.02cm} 3/2)}(q)
+
\bigl(\nu \rightleftarrows \sigma\bigr)\!\Bigr)
\notag \\[1ex]
&+\Bigl(\hspace{0.02cm}\eta_{\nu}^{(\mp\hspace{0.02cm} 3/2)}(q)\hspace{0.02cm}
\eta_{\lambda}^{(\pm\hspace{0.02cm} 3/2)}(q)
\eta_{\sigma}^{(\mp\hspace{0.02cm} 3/2)}(q)
+
\bigl(\nu \rightleftarrows \sigma\bigr)\!\Bigr)\hspace{0.02cm}
\eta_{\mu}^{(\pm\hspace{0.02cm}3/2)}(q)\Bigr].
\notag
\end{align}
The expression in the square brackets on the left-hand side coincides in fact with the expression on the left-hand side of the original algebra (\ref{eq:1i}) with the replacement of $\beta_{\mu}$ by $\eta_{\mu}^{(\pm\hspace{0.02cm}3/2)}(z)$. On the right-hand side of (\ref{eq:6p}) we have two independent groups of the terms, which can be separated out when we multiply the right-hand side by the projectors ${\cal P}_{3/2}^{(\pm)}(q)$ and ${\cal P}_{3/2}^{(\mp)}(q)$, correspondingly. Let us analyze the first group of terms. For this purpose we multiply the expression (\ref{eq:1i}) from the left by the matrix $\bigl[{\cal P}_{3/2}^{(\pm)}(q)\bigr]^4$. Taking into account the property of rearrangement of the matrices $\beta_{\mu}$ and projectors ${\cal P}_{3/2}^{(\pm)}(q)$, Eq.\,(\ref{eq:5h}), the definition of the matrices $\eta_{\mu}^{(\pm 3/2)}(q)$, and the property of idempotency of the projector ${\cal P}_{3/2}^{(\pm)}$, we find the required relation
\begin{align}
2\hspace{0.02cm}\Bigl[\hspace{0.03cm}&\eta_{\mu}^{(\pm\hspace{0.02cm}3/2)}(q)
\Bigl(\hspace{0.02cm}
\eta_{\nu}^{(\mp\hspace{0.02cm} 3/2)}(q)\hspace{0.02cm}
\eta_{\lambda}^{(\pm\hspace{0.02cm} 3/2)}(q)\hspace{0.02cm}
\eta_{\sigma}^{(\mp\hspace{0.02cm} 3/2)}(q)
+
\bigl(\nu \rightleftarrows \sigma\bigr)\!\Bigr)
\notag\\[1ex]
&+\Bigl(\hspace{0.02cm}\eta_{\nu}^{(\pm\hspace{0.02cm} 3/2)}(q)\hspace{0.02cm}
\eta_{\lambda}^{(\mp\hspace{0.02cm} 3/2)}(q)
\eta_{\sigma}^{(\pm\hspace{0.02cm} 3/2)}(q)
+
\bigl(\nu \rightleftarrows \sigma\bigr)\!\Bigr)\hspace{0.02cm}
\eta_{\mu}^{(\mp\hspace{0.02cm}3/2)}(q)\Bigr]
\label{eq:6a} \\[1ex]
= 3\hspace{0.02cm}\Bigl(\hspace{0.02cm}\Bigl[\hspace{0.03cm}&\eta_{\mu}^{(\pm\hspace{0.02cm}3/2)}(q)
\hspace{0.02cm}\eta_{\nu}^{(\mp\hspace{0.02cm} 3/2)}(q)
+
\eta_{\nu}^{(\pm\hspace{0.02cm}3/2)}(q)\hspace{0.02cm}
\eta_{\mu}^{(\mp\hspace{0.02cm} 3/2)}(q)\Bigr]\hspace{0.015cm}\delta_{\lambda\sigma}
+
\bigl(\nu \rightleftarrows \sigma\bigr)\!\Bigr)
\notag \\[1ex]
+\Bigl(\hspace{0.02cm}\Bigl[\hspace{0.03cm}&\eta_{\lambda}^{(\pm\hspace{0.02cm}3/2)}(q)
\hspace{0.02cm}\eta_{\sigma}^{(\mp\hspace{0.02cm} 3/2)}(q)
+
\eta_{\sigma}^{(\pm\hspace{0.02cm}3/2)}(q)
\hspace{0.02cm}\eta_{\lambda}^{(\mp\hspace{0.02cm} 3/2)}(q)\Bigr]\hspace{0.015cm}\delta_{\mu\nu}
+
\bigl(\nu \rightleftarrows \sigma\bigr)\!\Bigr)
\notag
\end{align}
\begin{align}
+\hspace{0.03cm} \Bigl[\hspace{0.03cm}&\eta_{\nu}^{(\pm\hspace{0.02cm}3/2)}(q)
\hspace{0.02cm}\eta_{\sigma}^{(\mp\hspace{0.02cm} 3/2)}(q)
+
\eta_{\sigma}^{(\pm\hspace{0.02cm}3/2)}(q)
\hspace{0.02cm}\eta_{\nu}^{(\mp\hspace{0.02cm} 3/2)}(q)
\Bigr]\hspace{0.015cm}\delta_{\mu\lambda}
\notag \\[1ex]
+\hspace{0.03cm} \Bigl[\hspace{0.03cm}&\eta_{\mu}^{(\pm\hspace{0.02cm}3/2)}(q)
\hspace{0.02cm}\eta_{\lambda}^{(\mp\hspace{0.02cm} 3/2)}(q)
+
\eta_{\lambda}^{(\pm\hspace{0.02cm}3/2)}(q)
\hspace{0.02cm}\eta_{\mu}^{(\mp\hspace{0.02cm} 3/2)}(q)
\Bigr]\hspace{0.015cm}\delta_{\nu\sigma}
\notag \\[1ex]
-\,&\frac{3}{2}\,\bigl(\hspace{0.02cm} \delta_{\mu\nu}\delta_{\lambda\sigma}
+
\delta_{\mu\lambda}\delta_{\nu\sigma}
+
\delta_{\nu\lambda}\delta_{\mu\sigma}\bigr)\hspace{0.02cm}
{\cal P}_{3/2}^{(\pm)}(q).
\notag
\end{align}
For the second group of the terms on the right-hand side of (\ref{eq:6p}), the completely similar relation only with the replacement $(\pm\hspace{0.02cm}3/2) \rightleftarrows (\mp\hspace{0.02cm}3/2)$ holds. Substituting the relation (\ref{eq:6a}) and relation with the replacement given just above into the right-hand side of (\ref{eq:6p}), we derive the final expression which we take as the desired four-linear algebra for the matrices $\eta_{\mu}^{(\pm\hspace{0.02cm}3/2)}(z)$. It is a direct analog of the original algebra (\ref{eq:1i}).\\
\indent
In closing, we would like to analyze in more detail the structure of the projectors ${\cal P}_{3/2}^{(\pm)}(q)$ and correspondingly the structure of the matrices $\eta_{\mu}^{(\pm 3/2)}(q)$ associated with them. By virtue of the definition (\ref{eq:5ff}) we write these projectors in an expanded form:
\begin{equation}
{\cal P}_{3/2}^{(\pm)}(q) = \frac{1}{2}\,
\Bigl[\hspace{0.02cm}I_{\gamma}
\otimes\!\hspace{0.01cm}\Bigl(-\frac{1}{2}\,\omega^{\hspace{0.02cm}2} + \frac{9}{8}\,I_{\xi}\Bigr)
\,\mp\,i\hspace{0.02cm}q\hspace{0.02cm}\gamma_{5}
\otimes\!\hspace{0.01cm}\Bigl(\omega^{\hspace{0.02cm}3} - \frac{9}{4}\;\omega\Bigr)\Bigr].
\label{eq:6s}
\end{equation}
We recall that $I_{\gamma}$ and $I_{\xi}$ are the unity matrices of the Dirac and $A_{\xi}$ algebras correspondingly. Let us introduce the notation
\[
\pi_{1/2}\equiv\frac{1}{2}\,\Bigl(\omega^{\hspace{0.02cm}2} - \frac{1}{4}\,I_{\xi}\Bigr),
\qquad
\pi_{3/2}^{(\pm)} = \omega^{\hspace{0.02cm}3} - \frac{9}{4}\;\omega.
\]
It is evident that these matrices satisfy the relations similar to (\ref{eq:5s}). Further we turn from the matrices $(I_{\gamma},\hspace{0.01cm}\gamma_5)$ to the chiral projection operators  $(P_{L},\hspace{0.01cm}P_{R})$:
\begin{equation}
\left\{
\begin{array}{l}
I_{\gamma} = P_{R} + P_{L}, \\[1ex]
\gamma_{5} = P_{R} - P_{L},
\end{array}
\right.
\qquad
\left\{
\begin{array}{l}
P_{L} = \displaystyle\frac{1}{2}\,\bigl( I_{\gamma} - \gamma_{5}\bigr),   \\[1.5ex]
P_{R} = \displaystyle\frac{1}{2}\,\bigl( I_{\gamma} + \gamma_{5}\bigr).
\end{array}
\right.
\label{eq:6d}
\end{equation}
Substituting (\ref{eq:6d}) into (\ref{eq:6s}) and collecting the terms similar to $P_{L}$ and $P_{R}$, we rewrite the projectors ${\cal P}_{3/2}^{(\pm)}(q)$ in the following equivalent form:
\begin{equation}
\begin{split}
&{\cal P}_{3/2}^{(+)}(q) = P_{R}\otimes\!\hspace{0.03cm}\pi_{3/2}^{(+)}(q) + P_{L}\otimes\!\hspace{0.03cm}\pi_{3/2}^{(-)}(q), \\[1.5ex]
&{\cal P}_{3/2}^{(-)}(q) = P_{R}\otimes\!\hspace{0.03cm}\pi_{3/2}^{(-)}(q) + P_{L}\otimes\!\hspace{0.03cm}\pi_{3/2}^{(+)}(q),
\end{split}
\label{eq:6f}
\end{equation}
where
\[
\pi_{3/2}^{(\pm)}(q) \equiv \frac{1}{2}\,\bigl(-\pi_{1/2} + I_{\xi} \mp
i\hspace{0.02cm}q\hspace{0.02cm}\pi_{3/2}\bigr).
\]
By direct calculations, it is not difficult to verify the validity of the following properties for the matrices $\pi_{3/2}^{(\pm)}(q)$:
\begin{equation}
\pi_{3/2}^{(\pm)}(q)\hspace{0.02cm} \pi_{3/2}^{(\mp)}(q) = 0,
\qquad
\bigl(\pi_{3/2}^{(\pm)}(q)\bigr)^{2} = \pi_{3/2}^{(\pm)}(q).
\label{eq:6g}
\end{equation}
Thus, these matrices represent the projectors in subspace generated only by the matrices of the $A_{\xi}$-algebra. We have intentionally written separately the expressions for the projectors ${\cal P}_{3/2}^{(+)}(q)$ and ${\cal P}_{3/2}^{(-)}(q)$ in (\ref{eq:6f}). The expressions (\ref{eq:6f}) possess a remarkable feature: by virtue of the first property in (\ref{eq:6g}) and also the properties of the chiral projector operators
\[
P_{L}P_{R} = P_{R}P_{L} =0,
\]
the four terms on the right-hand sides of (\ref{eq:6f}) are orthogonal among themselves!\\
\indent
Further, by using the definition (\ref{eq:5k}) and decomposition (\ref{eq:1o}), the basic matrices $\eta_{\mu}^{(\pm 3/2)}(q)$ can be presented in a more descriptive form:
\begin{align}
&\eta_{\mu}^{(+\hspace{0.02cm}3/2)}(q) = \bigl(P_{R}\hspace{0.02cm}\gamma_{\mu}\bigr)
\!\hspace{0.02cm}\otimes\!\hspace{0.02cm}\bigl(\pi_{3/2}^{(+)}(q)\hspace{0.03cm}\xi_{\mu} \bigr)
+
\bigl(P_{L}\hspace{0.02cm}\gamma_{\mu}\bigr)
\!\hspace{0.02cm}\otimes\!\hspace{0.02cm}\bigl(\pi_{3/2}^{(-)}(q)\hspace{0.03cm}\xi_{\mu} \bigr),
\notag \\[1ex]
&\eta_{\mu}^{(-\hspace{0.02cm}3/2)}(q) = \bigl(P_{R}\hspace{0.02cm}\gamma_{\mu}\bigr)
\!\hspace{0.02cm}\otimes\!\hspace{0.02cm}\bigl(\pi_{3/2}^{(-)}(q)\hspace{0.03cm}\xi_{\mu} \bigr)
+
\bigl(P_{L}\hspace{0.02cm}\gamma_{\mu}\bigr)
\!\hspace{0.02cm}\otimes\!\hspace{0.02cm}\bigl(\pi_{3/2}^{(+)}(q)\hspace{0.03cm}\xi_{\mu} \bigr).
\notag
\end{align}
The existence of two projectors ${\cal P}_{3/2}^{(+)}(q)$ and ${\cal P}_{3/2}^{(-)}(q)$ for the same spin-3/2 sector possible indicates that in the system under consideration there exists another additional internal degree of freedom\footnote{\,The signs $\pm$ do not imply the projections of the spin on any chosen direction as might appear at first sight.} (and a quantum number associated with it). This degree of freedom arises by virtue of introducing an additional algebraic object, namely a system of the roots of unity containing  two primitive ones $q$ and $q^3$ in the spin-3/2 case. This is a consequence of the property (\ref{eq:5j}). If one denotes the projector  ${\cal P}_{3/2}^{(+)}(q)$ as
\[
{\cal P}_{3/2}^{(+)}(q) \equiv {\cal P}_{3/2}(q),
\]
then the projector ${\cal P}_{3/2}^{(-)}(q)$ will represent the same projector ${\cal P}_{3/2}$ taken only for another value of the primitive root, i.e.
\[
{\cal P}_{3/2}^{(-)}(q) = {\cal P}_{3/2}(q^{3}).
\]
It is appropriate at this point to mention the papers of Indian mathematician Alladi Ramakrishnan (see, for example \cite{ramakrishnan_1971}), where the original approach to interpretation of various internal quantum numbers of particles in terms of the generalized Clifford algebra was suggested. As well known \cite{jagannathan_2010}, one of the crucial moments of this theory is the use of the primitive roots of the unity.

\section{The general structure of a solution of the first-order differential equation (\ref{eq:6o})}
\setcounter{equation}{0}

We analyze now the general structure of a solution of the wave equation (\ref{eq:6o}) which we present as follows:
\begin{equation}
\hat{\cal L}^{(3/2)}(z,\partial)\hspace{0.02cm}\psi(x;z) = 0.
\label{eq:7q}
\end{equation}
Here, we have introduced a short-hand notation for the first order differential operator
\begin{equation}
\hat{\cal L}^{(3/2)}(z;\partial)\equiv
A\biggl[\frac{\!1}{\,\epsilon^{1/2}(z)}\,\eta_{\mu}^{(\pm\hspace{0.02cm}3/2)}(z)
\hspace{0.03cm}\partial_{\mu}
+
\Bigl(\hspace{0.02cm}{\cal P}_{3/2}^{(\pm)}(q) +
{\cal P}_{3/2}^{(\mp)}(q)\!\hspace{0.02cm}\Bigr)\hspace{0.02cm}m\hspace{0.02cm}\biggr]
\label{eq:7w}
\end{equation}
\begin{align}
= A\biggl[\hspace{0.02cm}2\,\frac{\!1}{\,\delta^{1/2}}\,\frac{1}{\rho^{1/2}}\,
\eta_{\mu}^{(\pm\hspace{0.02cm}3/2)}(q)
\hspace{0.03cm}\partial_{\mu}
+\,
\delta^{1/2}\,\frac{1}{\rho^{1/2}}\,&\bigl[q^{3}\eta_{\mu}^{(\pm\hspace{0.02cm} 3/2)}(q) +
q\hspace{0.02cm}\eta_{\mu}^{(\mp\hspace{0.02cm} 3/2)}(q)\bigr]
\hspace{0.02cm}\partial_{\mu}
\notag \\[1ex]
&+\!
\Bigl(\hspace{0.02cm}{\cal P}_{3/2}^{(\pm)}(q) +
{\cal P}_{3/2}^{(\mp)}(q)\!\hspace{0.02cm}\Bigr)\hspace{0.02cm}m\hspace{0.02cm}\biggr],
\notag
\end{align}
where $\varrho\equiv q - q^{3}$. In the second line we have taken into account the expansion (\ref{eq:6r}), (\ref{eq:6t}).\\
\indent
The solution of Eq.\,(\ref{eq:7q}) can be unambiguously presented in the following form:
\begin{equation}
\psi(x;z) = \bigl[\hat{\cal L}^{(3/2)}(z;\partial)\bigr]^{\!\hspace{0.03cm}3}\varphi(x;z),
\label{eq:7e}
\end{equation}
where in turn the function $\varphi(x;z)$ is a solution of the fourth-order wave equation
\begin{equation}
\bigl[\hat{\cal L}^{(3/2)}(z;\partial)\bigr]^{4}\varphi(x;z) = 0.
\label{eq:7r}
\end{equation}
Our first step is to define an expansion of the cube of the operator  $\hat{\cal L}^{(3/2)}(z; \partial)$ in terms of $\delta^{1/2}$. Taking into account (\ref{eq:7w}) and the definition of the function $\varepsilon(z)$, Eq.\,(\ref{eq:6e}), we have the starting expression\\
\begin{align}
&\bigl[\hat{\cal L}^{(3/2)}(z;\partial)\bigr]^{3} =
\frac{1}{\delta^{3/2}}\,\frac{1}{\varrho^{3/2}}
\Bigr[A\hspace{0.03cm}\eta_{\mu}^{(\pm\hspace{0.02cm}3/2)}(z) A\hspace{0.03cm}\eta_{\nu}^{(\pm\hspace{0.02cm}3/2)}(z)
A\hspace{0.03cm}\eta_{\lambda}^{(\pm\hspace{0.02cm}3/2)}(z)
\Bigr]\partial_{\mu}\hspace{0.03cm}\partial_{\nu}
\hspace{0.03cm}\partial_{\lambda} \notag \\[1ex]
&+ m\hspace{0.03cm}\frac{1}{\delta}\,\frac{1}{\varrho}
\Bigl[\hspace{0.03cm}A\hspace{0.03cm}\eta_{\mu}^{(\pm\hspace{0.02cm}3/2)}(z) A\hspace{0.03cm}\eta_{\nu}^{(\pm\hspace{0.02cm}3/2)}(z)A
\!+\!
A\hspace{0.03cm}\eta_{\mu}^{(\pm\hspace{0.02cm}3/2)}(z) A^{2}\hspace{0.03cm}\eta_{\nu}^{(\pm\hspace{0.02cm}3/2)}(z)
\!+\!
A^{2}\hspace{0.03cm}\eta_{\mu}^{(\pm\hspace{0.02cm}3/2)}(z) A\hspace{0.03cm}\eta_{\nu}^{(\pm\hspace{0.02cm}3/2)}(z)
\Bigr]\partial_{\mu}\hspace{0.03cm}\partial_{\nu} \notag \\[1ex]
&+ m^{2}\frac{1}{\delta^{1/2}}\,\frac{1}{\varrho^{1/2}}
\Bigl[\hspace{0.03cm}A\hspace{0.03cm}\eta_{\mu}^{(\pm\hspace{0.02cm}3/2)}(z) A^{2}
+
A^{2}\hspace{0.03cm}\eta_{\mu}^{(\pm\hspace{0.02cm}3/2)}(z) A
+
A^{3}\hspace{0.03cm}\eta_{\mu}^{(\pm\hspace{0.02cm}3/2)}(z)
\Bigr]\partial_{\mu} \notag\\[1ex]
&+ m^{3\!}A^{3}\Bigl(\hspace{0.02cm}{\cal P}_{3/2}^{(\pm)}(q) +
{\cal P}_{3/2}^{(\mp)}(q)\!\hspace{0.02cm}\Bigr).
\notag
\end{align}
The expansion of this expression is rather cumbersome, albeit simple in structure. Here, it is necessary to use subsequently the expansion of the $\eta_{\mu}^{(\pm3/2)}$ matrices, Eqs.\,(\ref{eq:6r}) and (\ref{eq:6t}), the rules of rearrangement (\ref{eq:5x}) and (\ref{eq:5v}), and the property of nilpotency (\ref{eq:6q}). As a result the first four terms of the expansion of the operator $\bigl[\hat{\cal L}^{(3/2)}(z; \partial)\bigr]^{3}$ in powers of $\delta^{1/2}$ have the following form:\\
\begin{align}
\bigl[&\hat{\cal L}^{(3/2)}(z;\partial)\bigr]^{3} =
\frac{1}{\delta^{1/2}}\,\frac{\!2}{\varrho^{1/2}}\,A^{3}\biggl\{\hspace{0.02cm}2\hspace{0.02cm}
\frac{1}{\varrho}\;q^{2}
\bigl[\hspace{0.03cm}\eta_{\mu}^{(\pm\hspace{0.02cm}3/2)}(q) \hspace{0.03cm}\eta_{\nu}^{(\mp\hspace{0.02cm}3/2)}(q)
\hspace{0.03cm}\eta_{\lambda}^{(\pm\hspace{0.02cm}3/2)}(q)
\bigr]\hspace{0.01cm}\partial_{\mu}\hspace{0.03cm}\partial_{\nu}
\hspace{0.03cm}\partial_{\lambda}
+
q\hspace{0.025cm}m^{2}\eta_{\mu}^{(\pm\hspace{0.02cm}3/2)}(q)\hspace{0.01cm}\partial_{\mu}
\biggr\} \notag \\[1ex]
&+ m\hspace{0.02cm}A^{3}\biggl\{\hspace{0.02cm}\frac{2}{\varrho}\,
\bigl[\hspace{0.03cm}(2 - q)\hspace{0.02cm}\eta_{\mu}^{(\mp\hspace{0.02cm}3/2)}(q) \hspace{0.03cm}\eta_{\nu}^{(\pm\hspace{0.02cm}3/2)}(q)
-
(2 + q)\hspace{0.02cm}\eta_{\mu}^{(\pm\hspace{0.02cm}3/2)}(q) \hspace{0.03cm}\eta_{\nu}^{(\mp\hspace{0.02cm}3/2)}(q)
\bigr]\hspace{0.01cm}\partial_{\mu}\hspace{0.03cm}\partial_{\nu}  \notag \\[1ex]
&+ m^{2}\Bigl(\hspace{0.02cm}{\cal P}_{3/2}^{(\pm)}(q) +
{\cal P}_{3/2}^{(\mp)}(q)\!\hspace{0.02cm}\Bigr)
\biggr\} \label{eq:7t} \\[1ex]
&+ \delta^{1/2}\,\frac{\!1}{\varrho^{1/2}}\,A^{3}\biggl\{\hspace{0.02cm}\frac{2}{\varrho}\,
\bigl[\hspace{0.03cm}2\hspace{0.025cm}q\hspace{0.02cm}\eta_{\mu}^{(\pm\hspace{0.02cm}3/2)}(q) \hspace{0.03cm}\eta_{\nu}^{(\mp\hspace{0.02cm}3/2)}(q)
\hspace{0.03cm}\eta_{\lambda}^{(\pm\hspace{0.02cm}3/2)}(q)
+
q\hspace{0.03cm}\eta_{\mu}^{(\mp\hspace{0.02cm}3/2)}(q) \hspace{0.03cm}\eta_{\nu}^{(\pm\hspace{0.02cm}3/2)}(q)
\hspace{0.03cm}\eta_{\lambda}^{(\mp\hspace{0.02cm}3/2)}(q)
\bigr]\hspace{0.01cm}\partial_{\mu}\hspace{0.03cm}\partial_{\nu}
\hspace{0.03cm}\partial_{\lambda}  \notag \\[1ex]
&+ m^{2}\bigl[\hspace{0.03cm}\eta_{\mu}^{(\pm\hspace{0.02cm}3/2)}(q)
+ \eta_{\mu}^{(\mp\hspace{0.02cm}3/2)}(q) \bigr]\hspace{0.01cm}\partial_{\mu}\biggr\}
\notag \\[1ex]
&+ \delta\hspace{0.025cm}m\hspace{0.02cm}A^{3}\hspace{0.02cm}\frac{1}{\varrho}\,
\Bigl\{\hspace{0.01cm}(2\hspace{0.01cm}q - 1)\hspace{0.02cm}\eta_{\mu}^{(\pm\hspace{0.02cm}3/2)}(q) \hspace{0.03cm}\eta_{\nu}^{(\mp\hspace{0.02cm}3/2)}(q)
-
(2\hspace{0.01cm}q + 1)\hspace{0.02cm}\eta_{\mu}^{(\mp\hspace{0.02cm}3/2)}(q) \hspace{0.03cm}\eta_{\nu}^{(\pm\hspace{0.02cm}3/2)}(q)\!
\Bigr\}\,\partial_{\mu}\hspace{0.01cm}\partial_{\nu}
+ {\cal O}(\delta^{3/2}). \notag
\end{align}
It is naturally to make the assumption that the solution $\varphi(x; z)$ is regular at $z=q$ and it can be presented in the form of a formal series expansion in positive powers of $\delta^{1/2}$:
\begin{equation}
\varphi(x;z) =  \varphi_{0}(x) +\delta^{\hspace{0.02cm}1/2}\hspace{0.02cm}\varphi_{\frac{1}{2}}(x) + \delta\hspace{0.02cm}\varphi_{1}(x) + \delta^{\hspace{0.02cm}3/2}\hspace{0.02cm}\varphi_{\frac{3}{2}}(x)\, + \,\ldots\,.
\label{eq:7y}
\end{equation}
Substituting the expansions (\ref{eq:7t}) and (\ref{eq:7y}) into the relation (\ref{eq:7e}) and collecting terms of the same power in $\delta^{1/2}$, we obtain
\begin{equation}
\psi(x;z) = \frac{1}{\delta^{\hspace{0.01cm}1/2}}\,\psi_{-\frac{1}{2}}(x) + \psi_{0}(x) +\delta^{\hspace{0.01cm}1/2}\psi_{\frac{1}{2}}(x) +\, \ldots\,,
\label{eq:7u}
\end{equation}
where
\begin{align}
&\psi_{-\frac{1}{2}}(x) = 2\hspace{0.02cm}
\frac{1}{\varrho^{1/2}}\,A^{3}\biggl\{\hspace{0.02cm}
\frac{2}{\varrho}\,q^{2}
\bigl[\hspace{0.03cm}\eta_{\mu}^{(\pm\hspace{0.02cm}3/2)}(q) \hspace{0.03cm}\eta_{\nu}^{(\mp\hspace{0.02cm}3/2)}(q)
\hspace{0.03cm}\eta_{\lambda}^{(\pm\hspace{0.02cm}3/2)}(q)
\bigr]\hspace{0.01cm}\partial_{\mu}\hspace{0.03cm}\partial_{\nu}
\hspace{0.03cm}\partial_{\lambda}
+
q\hspace{0.03cm}m^{2}\eta_{\mu}^{(\pm\hspace{0.02cm}3/2)}(q)\hspace{0.02cm}\partial_{\mu}
\biggr\}\hspace{0.02cm}\varphi_{0}(x),
\label{eq:7i}
\end{align}
\begin{align}
&\psi_{0}(x) =  m\hspace{0.02cm}A^{3}\biggl\{\hspace{0.02cm}\frac{2}{\varrho}\,
\bigl[\hspace{0.03cm}(2 - q)\hspace{0.02cm}\eta_{\mu}^{(\mp\hspace{0.02cm}3/2)}(q) \hspace{0.03cm}\eta_{\nu}^{(\pm\hspace{0.02cm}3/2)}(q)
-
(2 + q)\hspace{0.02cm}\eta_{\mu}^{(\pm\hspace{0.02cm}3/2)}(q) \hspace{0.03cm}\eta_{\nu}^{(\mp\hspace{0.02cm}3/2)}(q)
\bigr]\hspace{0.01cm}\partial_{\mu}\hspace{0.03cm}\partial_{\nu}  \notag \\[1ex]
&+ m^{2}\Bigl(\hspace{0.02cm}{\cal P}_{3/2}^{(\pm)}(q) +
{\cal P}_{3/2}^{(\mp)}(q)\!\hspace{0.02cm}\Bigr)
\biggr\}\hspace{0.03cm}\varphi_{0}(x)  \label{eq:7o} \\[1ex]
&+2\hspace{0.02cm}\frac{1}{\varrho^{1/2}}\,A^{3}\biggl\{\hspace{0.02cm}
\frac{2}{\varrho}\;q^{2}
\bigl[\hspace{0.03cm}\eta_{\mu}^{(\pm\hspace{0.02cm}3/2)}(q) \hspace{0.03cm}\eta_{\nu}^{(\mp\hspace{0.02cm}3/2)}(q)
\hspace{0.03cm}\eta_{\lambda}^{(\pm\hspace{0.02cm}3/2)}(q)
\bigr]\hspace{0.01cm}\partial_{\mu}\hspace{0.03cm}\partial_{\nu}
\hspace{0.03cm}\partial_{\lambda}
+
q\hspace{0.03cm}m^{2}\eta_{\mu}^{(\pm\hspace{0.02cm}3/2)}(q)\hspace{0.01cm}\partial_{\mu}
\biggr\}\hspace{0.03cm}\varphi_{\frac{1}{2}}(x), \notag
\end{align}
and
\begin{align}
&\psi_{\frac{1}{2}}(x) = \frac{1}{\varrho^{1/2}}\,A^{3}\biggl\{\hspace{0.02cm}\frac{2}{\varrho}\,
\bigl[\hspace{0.03cm}2\hspace{0.02cm}q\hspace{0.02cm}\eta_{\mu}^{(\pm\hspace{0.02cm}3/2)}(q) \hspace{0.03cm}\eta_{\nu}^{(\mp\hspace{0.02cm}3/2)}(q)
\hspace{0.03cm}\eta_{\lambda}^{(\pm\hspace{0.02cm}3/2)}(q)
+
q\hspace{0.02cm}\eta_{\mu}^{(\mp\hspace{0.02cm}3/2)}(q) \hspace{0.03cm}\eta_{\nu}^{(\pm\hspace{0.02cm}3/2)}(q)
\hspace{0.03cm}\eta_{\lambda}^{(\mp\hspace{0.02cm}3/2)}(q)
\bigr]\hspace{0.01cm}\partial_{\mu}\hspace{0.03cm}\partial_{\nu}
\hspace{0.03cm}\partial_{\lambda}  \notag \\[1ex]
&+ m^{2}\bigl[\hspace{0.03cm}\eta_{\mu}^{(\pm\hspace{0.02cm}3/2)}(q)
+ \eta_{\mu}^{(\mp\hspace{0.02cm}3/2)}(q) \bigr]\partial_{\mu}\biggr\}
\hspace{0.02cm}\varphi_{0}(x) \notag \\[2ex]
&+
m\hspace{0.02cm}A^{3}\biggl\{\hspace{0.02cm}\frac{2}{\varrho}\,
\bigl[\hspace{0.03cm}(2 - q)\hspace{0.02cm}\eta_{\mu}^{(\mp\hspace{0.02cm}3/2)}(q) \hspace{0.03cm}\eta_{\nu}^{(\pm\hspace{0.02cm}3/2)}(q)
-
(2 + q)\hspace{0.02cm}\eta_{\mu}^{(\pm\hspace{0.02cm}3/2)}(q) \hspace{0.03cm}\eta_{\nu}^{(\mp\hspace{0.02cm}3/2)}(q)
\bigr]\hspace{0.01cm}\partial_{\mu}\hspace{0.03cm}\partial_{\nu}  \notag \\[1ex]
&+ m^{2}\Bigl(\hspace{0.02cm}{\cal P}_{3/2}^{(\pm)}(q) +
{\cal P}_{3/2}^{(\mp)}(q)\!\hspace{0.02cm}\Bigr)
\biggr\}\hspace{0.03cm}\varphi_{\frac{1}{2}}(x)  \notag \\[1ex]
&+2\hspace{0.02cm}\frac{1}{\varrho^{1/2}}\,A^{3}\biggl\{\hspace{0.02cm}
\frac{2}{\varrho}\,q^{2}
\bigl[\hspace{0.03cm}\eta_{\mu}^{(\pm\hspace{0.02cm}3/2)}(q) \hspace{0.03cm}\eta_{\nu}^{(\mp\hspace{0.02cm}3/2)}(q)
\hspace{0.03cm}\eta_{\lambda}^{(\pm\hspace{0.02cm}3/2)}(q)
\bigr]\hspace{0.01cm}\partial_{\mu}\hspace{0.03cm}\partial_{\nu}
\hspace{0.03cm}\partial_{\lambda}
+
q\hspace{0.03cm}m^{2}\eta_{\mu}^{(\pm\hspace{0.02cm}3/2)}(q)\hspace{0.02cm}\partial_{\mu}
\biggr\}\hspace{0.03cm}\varphi_{1}(x), \quad \mbox{etc.}  \notag
\end{align}
Thus, the wave function $\psi(x; z) $ in the general case is not a regular function of the parameter $z$ in the limit $z \rightarrow q$ and behaves as $1/ \delta^{1/2}$. Let us analyze more closely the first term $\psi_{-\frac{1}{2}}(x)$ of the expansion (\ref{eq:7u}). By virtue of the completeness condition (\ref{eq:5gg}) we may present the function $\varphi_0 (x)$ as follows
\begin{equation}
\varphi_{0}(x) = \varphi_{0}^{(1/2)}(x) + \varphi_{0}^{(\pm\hspace{0.02cm}3/2)}(x;q) +
\varphi_{0}^{(\mp\hspace{0.02cm}3/2)}(x;q),
\label{eq:7p}
\end{equation}
where the projected wave functions are
\[
\varphi_{0}^{(1/2)}(x)\equiv {\cal P}_{1/2}^{\phantom{(\pm)}}\varphi_{0}(x), \qquad
\varphi_{0}^{(\pm\hspace{0.02cm}3/2)}(x;q) \equiv {\cal P}_{3/2}^{(\pm)}(q)\varphi_{0}(x).
\]
Further, by virtue of the definition of the matrices $\eta_{\mu}^{(\pm\hspace{0.02cm} 3/2)\!}(q)$ and the second property in (\ref{eq:5h}), we will have the evident relations
\begin{equation}
\eta_{\mu}^{(\pm\hspace{0.02cm}3/2)}(q)\hspace{0.01cm} \varphi_{0}^{(1/2)}(x) = 0,
\qquad
\eta_{\mu}^{(\pm\hspace{0.02cm}3/2)}(q)\hspace{0.01cm}
\varphi_{0}^{(\pm\hspace{0.02cm}3/2)}(x;q) = 0
\label{eq:7a}
\end{equation}
and meanwhile
\[
\eta_{\mu}^{(\pm\hspace{0.02cm}3/2)}(q)\hspace{0.01cm}
\varphi_{0}^{(\mp\hspace{0.02cm}3/2)}(x;q) \neq 0.
\]
Thus, on the right-hand side of the expression (\ref{eq:7i}) in the decomposition (\ref{eq:7p}) only one of the projected parts, namely $\varphi_0^{(\mp3/2)}(x; q)$, survives. This implies in particular that the singular contribution in the expansion (\ref{eq:7u}) can be dropped out, if we simply set
\begin{equation}
\varphi_{0}^{(\mp\hspace{0.02cm}3/2)}(x;q) \equiv 0.
\label{eq:7s}
\end{equation}
In this case, in the limit $z \rightarrow q$, the first nonzero contribution in the wave function $\psi(x; q)$ on the strength of the expression (\ref{eq:7o}) will have the form
\begin{equation}
\psi_{0}(x) = -\hspace{0.02cm} m\hspace{0.02cm}A^{3}\biggl\{\hspace{0.02cm}\frac{2}{\varrho}\,
(2 + q)\bigl[\hspace{0.03cm}\eta_{\mu}^{(\pm\hspace{0.02cm}3/2)}(q) \hspace{0.03cm}\eta_{\nu}^{(\mp\hspace{0.02cm}3/2)}(q)
\bigr]\hspace{0.01cm}\partial_{\mu}\hspace{0.03cm}\partial_{\nu} - m^{2}I
\biggr\}\hspace{0.03cm}\varphi_{0}^{(\pm\hspace{0.02cm}3/2)}(x;q)
\label{eq:7d}
\end{equation}
\[
+\,2\hspace{0.02cm}\frac{1}{\varrho^{1/2}}\,A^{3}\biggl\{\hspace{0.02cm}
\frac{2}{\varrho}\,q^{2}
\bigl[\hspace{0.03cm}\eta_{\mu}^{(\pm\hspace{0.02cm}3/2)}(q) \hspace{0.03cm}\eta_{\nu}^{(\mp\hspace{0.02cm}3/2)}(q)
\hspace{0.03cm}\eta_{\lambda}^{(\pm\hspace{0.02cm}3/2)}(q)
\bigr]\hspace{0.01cm}\partial_{\mu}\hspace{0.03cm}\partial_{\nu}
\hspace{0.03cm}\partial_{\lambda}
+
q\hspace{0.025cm}m^{2}\eta_{\mu}^{(\pm\hspace{0.02cm}3/2)}(q)\hspace{0.02cm}\partial_{\mu}
\biggr\}\hspace{0.03cm}\varphi_{\frac{1}{2}}^{(\mp\hspace{0.02cm}3/2)}(x;q)
\]
and the corresponding first ``correction'' is
\begin{align}
\psi_{\frac{1}{2}}(x) &=\frac{1}{\varrho^{1/2}}\,A^{3}\biggl\{\hspace{0.02cm}
\frac{2}{\varrho}\,q
\bigl[\hspace{0.03cm}\eta_{\mu}^{(\mp\hspace{0.02cm}3/2)}(q) \hspace{0.03cm}\eta_{\nu}^{(\pm\hspace{0.02cm}3/2)}(q)
\hspace{0.03cm}\eta_{\lambda}^{(\mp\hspace{0.02cm}3/2)}(q)
\bigr]\hspace{0.01cm}\partial_{\mu}\hspace{0.03cm}\partial_{\nu}
\hspace{0.03cm}\partial_{\lambda}
+
m^{2}\eta_{\mu}^{(\mp\hspace{0.02cm}3/2)}(q)\hspace{0.02cm}\partial_{\mu}
\biggr\}\hspace{0.03cm}\varphi_{0}^{(\pm\hspace{0.02cm}3/2)}(x;q) \notag \\
&+ m\hspace{0.02cm}A^{3}\biggl\{\hspace{0.02cm}\frac{2}{\varrho}\hspace{0.02cm}
\bigl[\hspace{0.03cm}(2 - q)\hspace{0.02cm}\eta_{\mu}^{(\mp\hspace{0.02cm}3/2)}(q) \hspace{0.03cm}\eta_{\nu}^{(\pm\hspace{0.02cm}3/2)}(q)
-
(2 + q)\hspace{0.02cm}\eta_{\mu}^{(\pm\hspace{0.02cm}3/2)}(q) \hspace{0.03cm}\eta_{\nu}^{(\mp\hspace{0.02cm}3/2)}(q)
\bigr]\hspace{0.01cm}\partial_{\mu}\hspace{0.03cm}\partial_{\nu}  \notag \\[1ex]
&+ m^{2}\Bigl(\hspace{0.02cm}{\cal P}_{3/2}^{(\pm)}(q) +
{\cal P}_{3/2}^{(\mp)}(q)\!\hspace{0.02cm}\Bigr)
\biggr\}\bigl(\varphi_{\frac{1}{2}}^{(\pm\hspace{0.02cm}3/2)}(x;q)
+ \varphi_{\frac{1}{2}}^{(\mp\hspace{0.02cm}3/2)}(x;q)\bigr) \notag \\[1ex]
&+2\hspace{0.02cm}\frac{1}{\varrho^{1/2}}\,A^{3}\biggl\{\hspace{0.02cm}
\frac{2}{\varrho}\;q^{2}
\bigl[\hspace{0.03cm}\eta_{\mu}^{(\pm\hspace{0.02cm}3/2)}(q) \hspace{0.03cm}\eta_{\nu}^{(\mp\hspace{0.02cm}3/2)}(q)
\hspace{0.03cm}\eta_{\lambda}^{(\pm\hspace{0.02cm}3/2)}(q)
\bigr]\hspace{0.01cm}\partial_{\mu}\hspace{0.03cm}\partial_{\nu}
\hspace{0.03cm}\partial_{\lambda}
+
q\hspace{0.03cm}m^{2}\eta_{\mu}^{(\pm\hspace{0.02cm}3/2)}(q)\hspace{0.02cm}\partial_{\mu}
\biggr\}\hspace{0.03cm}\varphi_{1}^{(\mp\hspace{0.02cm}3/2)}(x;q). \notag
\end{align}
Note that although the condition (\ref{eq:7s}) enables us to remove the singular contribution in the expansion (\ref{eq:7u}), nevertheless the ``memory'' of this term remains. Thus, for instance, in the case of the function $\psi_0(x)$, Eq.\,(\ref{eq:7d}), this becomes apparent in the presence of the contribution including the first correction $\varphi_{\frac{1}{2}}^{(\mp3/2)}(x; q)$ in the expansion (\ref{eq:7y}). In other words, for defining a finite value of the function $\psi(x; z)$ in the limit $z \rightarrow q$ we need to know not only a limiting value $\varphi_{0}(x)$ of the wave function $\varphi(x; z)$, but as well the following term of the expansion (\ref{eq:7y}), i.e. $\varphi_{\frac{1}{2}}(x)$. We will analyze the fourth-order equation (\ref{eq:7r}) just below and obtain the equations (more exactly, a self-consistent system of equations) for the functions $\varphi_0^{(\pm3/2)}(x; q)$ and $\varphi_{\frac{1}{2}}^{(\mp3/2)}(x; q)$.\\
\indent
It is necessary to say a few words about the matrix $A^3$ which enters into all the expressions above. Recall that the parameter $\alpha$ in the representation (\ref{eq:2y}) has to satisfy the condition (\ref{eq:4t}). From the latter follows that the $\alpha$ can assume four possible values, two of these are pure real and further two are pure imaginary. If as the parameter $\alpha$ we take only pure imaginary values so that $\alpha^{\ast} = -\alpha$, then the following relation will be true:
\[
A^{3} = \frac{1}{m}\,A^{\dagger}.
\]
The proof of this relation which is not fairly obvious, is given in Appendix C.\\
\indent
The differential equations to which the functions $\varphi_{0}(x),\,\varphi_{\frac{1}{2}}(x),\,\ldots$ must satisfy, are defined by the appropriate expansion of the operator $\hat{\cal L}^{(3/2)}(z;\partial)$ to  the fourth power. With allowance for the expressions (\ref{eq:7w}) and (\ref{eq:7t}), we get
\vspace{-0.3cm}
\begin{flushleft}
the \textit{singular} contribution:
\end{flushleft}
\vspace{-0.7cm}
%
%
\begin{equation}
\;\delta^{-1/2}\!:\hspace{0.45cm}    8\hspace{0.03cm}\frac{1}{m}\,\frac{1}{\varrho^{3/2}}\,(1 + q)\bigl[\hspace{0.03cm}\eta_{\mu}^{(\pm\hspace{0.02cm}3/2)}(q) \hspace{0.03cm}\eta_{\nu}^{(\mp\hspace{0.02cm}3/2)}(q)
\hspace{0.03cm}\eta_{\lambda}^{(\pm\hspace{0.02cm}3/2)}(q)
\bigr]\hspace{0.01cm}\partial_{\mu}\hspace{0.03cm}\partial_{\nu}
\hspace{0.03cm}\partial_{\lambda}
\label{eq:7f}
\end{equation}
and
\vspace{-0.3cm}
\begin{flushleft}
the \textit{regular} contributions:
\end{flushleft}
\vspace{-0.7cm}
\begin{align}
%
%
\delta^{\hspace{0.02cm}0}:\hspace{0.45cm}
&-4\hspace{0.03cm}\frac{1}{m^2}\,\frac{1}{\varrho^{2}}\,\bigl[
\hspace{0.03cm}\eta_{\mu}^{(\pm\hspace{0.02cm}3/2)}(q)
\hspace{0.03cm}\eta_{\nu}^{(\mp\hspace{0.02cm}3/2)}(q)
\hspace{0.03cm}\eta_{\lambda}^{(\pm\hspace{0.02cm}3/2)}(q)
\hspace{0.03cm}\eta_{\sigma}^{(\mp\hspace{0.02cm}3/2)}(q)
\label{eq:7g}\\[1ex]
&\qquad\quad+
\hspace{0.03cm}\eta_{\mu}^{(\mp\hspace{0.02cm}3/2)}(q)
\hspace{0.03cm}\eta_{\nu}^{(\pm\hspace{0.02cm}3/2)}(q)
\hspace{0.03cm}\eta_{\lambda}^{(\mp\hspace{0.02cm}3/2)}(q)
\hspace{0.03cm}\eta_{\sigma}^{(\pm\hspace{0.02cm}3/2)}(q)
\bigr]\hspace{0.01cm}\partial_{\mu}\hspace{0.03cm}\partial_{\nu}
\hspace{0.03cm}\partial_{\lambda}\hspace{0.03cm}\partial_{\sigma}
\notag\\[1ex]
&\qquad\quad+ 4\hspace{0.02cm}\frac{1}{\rho}\,
\bigl[\hspace{0.03cm}(1 + q)\hspace{0.02cm}\eta_{\mu}^{(\pm\hspace{0.02cm}3/2)}(q) \hspace{0.03cm}\eta_{\nu}^{(\mp\hspace{0.02cm}3/2)}(q)
-
(1 - q)\hspace{0.02cm}\eta_{\mu}^{(\mp\hspace{0.02cm}3/2)}(q) \hspace{0.03cm}\eta_{\nu}^{(\pm\hspace{0.02cm}3/2)}(q)
\bigr]\hspace{0.01cm}\partial_{\mu}\hspace{0.03cm}\partial_{\nu}
\notag \\[1ex]
&\qquad\quad- m^{2}\Bigl(\hspace{0.02cm}{\cal P}_{3/2}^{(\pm)}(q) +
{\cal P}_{3/2}^{(\mp)}(q)\!\hspace{0.02cm}\Bigr), \notag
\end{align}
%
%
\begin{align}
\delta^{\hspace{0.02cm}1/2}:\hspace{0.5cm}  &-4\hspace{0.03cm}\frac{1}{m}\,\frac{1}{\varrho^{3/2}}\,\biggl\{\hspace{0.01cm}
(1 + q)\bigl[\hspace{0.03cm}\eta_{\mu}^{(\mp\hspace{0.02cm}3/2)}(q) \hspace{0.03cm}\eta_{\nu}^{(\pm\hspace{0.02cm}3/2)}(q)
\hspace{0.03cm}\eta_{\lambda}^{(\mp\hspace{0.02cm}3/2)}(q)
\bigr]\hspace{2.5cm} \label{eq:7h}\\[1ex]
&\qquad\quad\quad
-2\hspace{0.02cm}
(1 - q)\bigl[\hspace{0.03cm}\eta_{\mu}^{(\pm\hspace{0.02cm}3/2)}(q) \hspace{0.03cm}\eta_{\nu}^{(\mp\hspace{0.02cm}3/2)}(q)
\hspace{0.03cm}\eta_{\lambda}^{(\pm\hspace{0.02cm}3/2)}(q)
\bigr]\biggr\}
\hspace{0.03cm}\partial_{\mu}\hspace{0.03cm}\partial_{\nu}
\hspace{0.03cm}\partial_{\lambda}, \notag
\end{align}
and so on. Substituting the expansions of the operator $[\hspace{0.02cm}\hat{\cal L}^{(3/2)}(z;\partial)]^{4}$ and of the function $\varphi(x;z)$ into (\ref{eq:7r}), we obtain the required equations for the functions $\varphi_{0}(x),\,\varphi_{\frac{1}{2}}(x),\,\ldots$\,. As this takes place, it is necessary to take into account the decomposition (\ref{eq:7p}) and properties (\ref{eq:7a}).\\
\indent
One can get rid of the equation connected with the singular contribution (\ref{eq:7f}) if one demands the fulfilment of the condition (\ref{eq:7s}). In this case we have the following equation to leading order in $\delta^{1/2}$:
\begin{align}
\delta^{\hspace{0.02cm}0}:\hspace{0.45cm}
\biggl\{\!&-\!\frac{1}{m^2}\,\frac{1}{\varrho^{2}}\hspace{0.03cm}\bigl[
\hspace{0.03cm}\eta_{\mu}^{(\pm\hspace{0.02cm}3/2)}(q)
\hspace{0.03cm}\eta_{\nu}^{(\mp\hspace{0.02cm}3/2)}(q)
\hspace{0.03cm}\eta_{\lambda}^{(\pm\hspace{0.02cm}3/2)}(q)
\hspace{0.03cm}\eta_{\sigma}^{(\mp\hspace{0.02cm}3/2)}(q)
\bigr]\hspace{0.01cm}\partial_{\mu}\hspace{0.03cm}\partial_{\nu}
\hspace{0.03cm}\partial_{\lambda}\hspace{0.03cm}\partial_{\sigma}
\notag \\[1ex]
&+ \frac{1}{\rho}\,
(1 + q)\hspace{0.02cm}\bigl[\hspace{0.03cm}\eta_{\mu}^{(\pm\hspace{0.02cm}3/2)}(q) \hspace{0.03cm}\eta_{\nu}^{(\mp\hspace{0.02cm}3/2)}(q)
\bigr]\hspace{0.01cm}\partial_{\mu}\hspace{0.03cm}\partial_{\nu}
- \frac{1}{4}\,m^{2}I\biggr\}\hspace{0.03cm}\varphi_{0}^{(\pm\hspace{0.02cm}3/2)}(x;q)
\label{eq:7j}\\[1ex]
&= - 2\hspace{0.03cm}\frac{1}{m}\,\frac{1}{\varrho^{3/2}}
(1 + q)\bigl[\hspace{0.03cm}\eta_{\mu}^{(\pm\hspace{0.02cm}3/2)}(q) \hspace{0.03cm}\eta_{\nu}^{(\mp\hspace{0.02cm}3/2)}(q)
\hspace{0.03cm}\eta_{\lambda}^{(\pm\hspace{0.02cm}3/2)}(q)
\bigr]
\hspace{0.01cm}\partial_{\mu}\hspace{0.03cm}\partial_{\nu}
\hspace{0.03cm}\partial_{\lambda}\,\varphi_{\frac{1}{2}}^{(\mp\hspace{0.02cm}3/2)}(x;q). \notag
\end{align}
We see that the equation for the function $\varphi_{0}^{(\pm\hspace{0.02cm}3/2)}(x; q)$ is  nonclosed. To understand whether  it is possible to set in a consistent fashion the correction $\varphi_{\frac{1}{2}}^{(\mp\hspace{0.02cm}3/2)}(x; q)$  identically equal to zero, it is necessary to write out the next-to-leading order equation (i.e. proportional to $\delta^{1/2}$). The desired equation easily follows from the expressions (\ref{eq:7f})\,--\,(\ref{eq:7h}), decomposition (\ref{eq:7p}), and expansion (\ref{eq:7y}). Actually this equation will represent a sum of two independent equations which we may separate by the projections on the ${\cal P}_{3/2}^{(\mp)}(q)$ and ${\cal P}_{3/2}^{(\pm)}(q)$ sectors. The first of them has the form
\begin{align}
\delta^{\hspace{0.02cm}1/2}:\hspace{0.45cm}
\biggl\{\!&-\!\frac{1}{m^2}\,\frac{1}{\varrho^{2}}\,\bigl[
\hspace{0.03cm}\eta_{\mu}^{(\mp\hspace{0.02cm}3/2)}(q)
\hspace{0.03cm}\eta_{\nu}^{(\pm\hspace{0.02cm}3/2)}(q)
\hspace{0.03cm}\eta_{\lambda}^{(\mp\hspace{0.02cm}3/2)}(q)
\hspace{0.03cm}\eta_{\sigma}^{(\pm\hspace{0.02cm}3/2)}(q)
\bigr]\hspace{0.01cm}\partial_{\mu}\hspace{0.03cm}\partial_{\nu}
\hspace{0.03cm}\partial_{\lambda}\hspace{0.03cm}\partial_{\sigma}
\notag \\[1ex]
&- \frac{1}{\rho}\,
(1 - q)\hspace{0.02cm}\bigl[\hspace{0.03cm}\eta_{\mu}^{(\mp\hspace{0.02cm}3/2)}(q) \hspace{0.03cm}\eta_{\nu}^{(\pm\hspace{0.02cm}3/2)}(q)
\bigr]\hspace{0.01cm}\partial_{\mu}\hspace{0.03cm}\partial_{\nu}
- \frac{1}{4}\,m^{2}I\biggr\}\hspace{0.03cm}\varphi_{\frac{1}{2}}^{(\mp\hspace{0.02cm}3/2)}(x;q)
\label{eq:7k}\\[1ex]
&= \frac{1}{m}\,\frac{1}{\varrho^{3/2}}
(1 + q)\bigl[\hspace{0.03cm}\eta_{\mu}^{(\mp\hspace{0.02cm}3/2)}(q) \hspace{0.03cm}\eta_{\nu}^{(\pm\hspace{0.02cm}3/2)}(q)
\hspace{0.03cm}\eta_{\lambda}^{(\mp\hspace{0.02cm}3/2)}(q)
\bigr]
\hspace{0.01cm}\partial_{\mu}\hspace{0.03cm}\partial_{\nu}
\hspace{0.03cm}\partial_{\lambda}\,\varphi_{0}^{(\pm\hspace{0.02cm}3/2)}(x;q)
\notag
\end{align}
and, correspondingly, the second one with the use of the condition (\ref{eq:7s}) is
\begin{align}
\delta^{\hspace{0.02cm}1/2}:\hspace{0.45cm}
\biggl\{\!&-\!\frac{1}{m^2}\,\frac{1}{\varrho^{2}}\,\bigl[
\hspace{0.03cm}\eta_{\mu}^{(\pm\hspace{0.02cm}3/2)}(q)
\hspace{0.03cm}\eta_{\nu}^{(\mp\hspace{0.02cm}3/2)}(q)
\hspace{0.03cm}\eta_{\lambda}^{(\pm\hspace{0.02cm}3/2)}(q)
\hspace{0.03cm}\eta_{\sigma}^{(\mp\hspace{0.02cm}3/2)}(q)
\bigr]\hspace{0.01cm}\partial_{\mu}\hspace{0.03cm}\partial_{\nu}
\hspace{0.03cm}\partial_{\lambda}\hspace{0.03cm}\partial_{\sigma}
\notag \\[1ex]
&+ \frac{1}{\rho}\,
(1 + q)\hspace{0.02cm}\bigl[\hspace{0.03cm}\eta_{\mu}^{(\pm\hspace{0.02cm}3/2)}(q) \hspace{0.03cm}\eta_{\nu}^{(\mp\hspace{0.02cm}3/2)}(q)
\bigr]\hspace{0.01cm}\partial_{\mu}\hspace{0.03cm}\partial_{\nu}
- \frac{1}{4}\,m^{2}I\biggr\}\hspace{0.03cm}\varphi_{\frac{1}{2}}^{(\pm\hspace{0.02cm}3/2)}(x;q)
\label{eq:7l}\\[1ex]
&= - 2\hspace{0.03cm}\frac{1}{m}\,\frac{1}{\varrho^{3/2}}
(1 + q)\bigl[\hspace{0.03cm}\eta_{\mu}^{(\pm\hspace{0.02cm}3/2)}(q) \hspace{0.03cm}\eta_{\nu}^{(\mp\hspace{0.02cm}3/2)}(q)
\hspace{0.03cm}\eta_{\lambda}^{(\pm\hspace{0.02cm}3/2)}(q)
\bigr]
\hspace{0.01cm}\partial_{\mu}\hspace{0.03cm}\partial_{\nu}
\hspace{0.03cm}\partial_{\lambda}\,\varphi_{1}^{(\mp\hspace{0.02cm}3/2)}(x;q).
\notag
\end{align}
From the expressions (\ref{eq:7j}) and (\ref{eq:7k}) we see that they generate a self-consistent system of equations for the functions $\varphi_{0}^{(\pm\hspace{0.02cm}3/2)}(x; q)$ and $\varphi_{\frac{1}{2}}^{(\mp\hspace{0.02cm}3/2)}(x; q)$. It is precisely these functions that determine the leading-order wave function $\psi_{0}(x)$, Eq.\,(\ref{eq:7d}). Quite apparently, if we put the function $\varphi_{\frac{1}{2}}^{(\mp\hspace{0.02cm}3/2)}(x; q)$ identically equal to zero, then this results in the trivial  degeneration of all the system. The remaining equation (\ref{eq:7l}) is connected with the other functions in the expansion (\ref{eq:7y}). Thus, the only restriction (\ref{eq:7s}) we have imposed by hand on the formalism under consideration, leads to a completely self-consistent calculation scheme of the wave function $\psi(x; z)$ obeying the first order wave equation (\ref{eq:7q}). This wave function is regular in the limit $z \rightarrow q$.\\
\indent
We can present the differential-matrix operator of fourth order in $\partial_{\mu}$ on the left-hand side of Eqs.\,(\ref{eq:7j}) and (\ref{eq:7k}) in the form similar to the expression (\ref{eq:1l}). For this purpose, we use the four-linear relation for the matrices $\eta_{\mu}^{(\pm\hspace{0.02cm} 3/2)}(q)$, Eq.\,(\ref{eq:6a}). Let us contract the algebraic relation (\ref{eq:6a}) with $\partial_{\mu}\hspace{0.03cm}\partial_{\nu}
\hspace{0.03cm}\partial_{\lambda}\hspace{0.03cm}\partial_{\sigma}$. As a result we will have:
\begin{equation}
\begin{split}
\bigl[\hspace{0.03cm}&\eta_{\mu}^{(\pm\hspace{0.02cm}3/2)}(q)\hspace{0.02cm}
\eta_{\nu}^{(\mp\hspace{0.02cm} 3/2)}(q)\hspace{0.02cm}
\eta_{\lambda}^{(\pm\hspace{0.02cm} 3/2)}(q)\hspace{0.02cm}
\eta_{\sigma}^{(\mp\hspace{0.02cm} 3/2)}(q)\bigr]
\hspace{0.01cm}\partial_{\mu}\hspace{0.03cm}\partial_{\nu}
\hspace{0.03cm}\partial_{\lambda}\hspace{0.03cm}\partial_{\sigma}
\\[2ex]
=\;
&\frac{5}{2}\,\Bigl(\bigl[\hspace{0.03cm}\eta_{\mu}^{(\pm\hspace{0.02cm}3/2)}(q)
\hspace{0.02cm}
\eta_{\nu}^{(\mp\hspace{0.02cm}3/2)}(q)\hspace{0.02cm}\bigr]\hspace{0.01cm}
\partial_{\mu}\hspace{0.03cm}\partial_{\nu}\Bigr)\hspace{0.03cm}
\Box
\hspace{0.03cm}-\hspace{0.03cm}
\frac{9}{16}\,{\Box}^{\hspace{0.03cm}2}\hspace{0.04cm}{\cal P}_{3/2}^{(\pm)}(q).
\end{split}
\label{eq:7z}
\end{equation}
A similar relation holds also for the matrix-differential operator of fourth-order in $\partial_{\mu}$ in equation (\ref{eq:7k}) with the replacement $(\pm\hspace{0.02cm}3/2) \rightleftarrows
 (\mp\hspace{0.02cm} 3/2)$.\\
\indent
At the end of the present section a remark on the previous equations should be made. We see inevitable appearance of the contributions of third order in derivatives $\partial_{\mu}$ in the equations (\ref{eq:7j})\,--\,(\ref{eq:7l}) for the terms of the expansion of the function $\varphi(x; z)$.
In this connection, it is worth noting that the equation of third order in derivatives in the theory of a massive particle with the spin 3/2 (in addition to the usual Klein-Gordon-Fock equation) was introduced by Joos \cite{joos_1962}, Weinberg \cite{weinberg_1964}, and Weaver {\it et al.} \cite{weaver_1964} within the boost technique and then was analyzed by Shay {\it et al.} \cite{shay_1965}, Tung \cite{tung_1966, tung_1967}, Nelson and Good \cite{nelson_1970}, Good \cite{good_1989}, and Napsuciale \cite{napsuciale_2003}. Sometimes this equation is referred to as the {\it Weinberg equation}. One of the purposes of its considering is the reduction of the number of components of the wave function. In our case, however, the contribution with derivatives of the third order get involved in the basic fourth order wave equation by a very distinctive manner generating the self-consistent system of equations.

\section{Interacting case}
\setcounter{equation}{0}

In the interaction free case we have derived a self-consistent system of equations (\ref{eq:7j}) and (\ref{eq:7k}) for the functions $\varphi_0^{(\pm3/2)}(x;q)$ and $\varphi_{\frac{1}{2}}^{(\mp3/2)}(x;q)$. Let us consider the question of a modification of the fourth-order wave operator (\ref{eq:7z}) in the presence of an external electromagnetic field. We introduce the interaction via the minimal substitution:
\[
\partial_{\mu} \rightarrow D_{\mu}\equiv\partial_{\mu} + i\hspace{0.02cm}eA_{\mu}(x).
\]
With an external gauge field in the system the left-hand side of (\ref{eq:7z}) takes the form
\begin{equation}
\bigl[\hspace{0.03cm}\eta_{\mu_{1}}^{(\pm\hspace{0.02cm}3/2)}(q)\hspace{0.02cm}
\eta_{\mu_{2}}^{(\mp\hspace{0.02cm} 3/2)}(q)\hspace{0.02cm}
\eta_{\mu_{3}}^{(\pm\hspace{0.02cm} 3/2)}(q)\hspace{0.02cm}
\eta_{\mu_{4}}^{(\mp\hspace{0.02cm} 3/2)}(q)\bigr]
\hspace{0.01cm}D_{\mu_{1}}D_{\mu_{2}}D_{\mu_{3}}D_{\mu_{4}}.
\label{eq:8w}
\end{equation}
For analysis of the expression (\ref{eq:8w}) we make use of the following identity for a product of four covariant derivatives
\begin{align}
&D_{\mu_{1}}D_{\mu_{2}}D_{\mu_{3}}D_{\mu_{4}} = \frac{1}{4!} \Bigl(\bigl\{\!D_{\mu_{1}},D_{\mu_{2}},D_{\mu_{3}},D_{\mu_{4}}\!\hspace{0.015cm}\bigr\}
\label{eq:8e} \\[1ex]
&+
12\hspace{0.015cm}i\hspace{0.01cm}eD_{\mu_{1}}D_{\mu_{2}}F_{\mu_{3}\mu_{4}}
+
4\hspace{0.015cm}i\hspace{0.005cm}eD_{\mu_{1}}D_{\mu_{3}}F_{\mu_{2}\mu_{4}}
+
2\hspace{0.015cm}i\hspace{0.005cm}e\bigl(D_{\mu_{2}}D_{\mu_{3}}F_{\mu_{1}\mu_{4}}
+
D_{\mu_{2}}D_{\mu_{4}}F_{\mu_{1}\mu_{3}}
+
D_{\mu_{3}}D_{\mu_{4}}F_{\mu_{1}\mu_{2}}\bigr) \notag \\[1ex]
&+
6\hspace{0.015cm}i\hspace{0.005cm}e\bigl(F_{\mu_{1}\mu_{2}}D_{\mu_{3}}D_{\mu_{4}}
+
F_{\mu_{1}\mu_{3}}D_{\mu_{2}}D_{\mu_{4}}
+
F_{\mu_{1}\mu_{4}}D_{\mu_{2}}D_{\mu_{3}}\bigr) \notag \\[1ex]
&+
8\hspace{0.015cm}i\hspace{0.005cm}eD_{\mu_{1}}F_{\mu_{2}\mu_{3}}D_{\mu_{4}}
+
2\hspace{0.015cm}i\hspace{0.005cm}e\sum_{({\cal P})} D_{\mu_{2}}F_{\hat{\mu}_{1}\mu_{3}}D_{\mu_{4}} \notag \\[-1ex]
&+ 4\hspace{0.015cm}e^{2}\bigl(F_{\mu_{1}\mu_{2}}F_{\mu_{3}\mu_{4}}
+ F_{\mu_{1}\mu_{3}}F_{\mu_{2}\mu_{4}} + F_{\mu_{1}\mu_{4}}F_{\mu_{2}\mu_{3}}\bigr)
\Bigr), \notag
\end{align}
where we have designated by the symbol $\bigl\{\!D_{\mu_{1}},\,\dots,\,D_{\mu_{4}}\!\hspace{0.02cm}\bigr\}$ a product of four $D$\hspace{0.02cm}-\hspace{0.02cm}operators completely symmetrized over the vector indices
$\mu_{i},\;i = 1,\ldots,4$:
\begin{equation}
\bigl\{\!D_{\mu_{1}},D_{\mu_{2}},D_{\mu_{3}},D_{\mu_{4}}\!\hspace{0.02cm}\bigr\}
\equiv
\sum_{({\cal P})}D_{\mu_{1}}D_{\mu_{2}}D_{\mu_{3}}D_{\mu_{4}}.
\label{eq:8r}
\end{equation}
In the last but one term on the right-hand side of (\ref{eq:8e}) the permutation $({\cal P})$ is performed over the free indices $\mu_2, \mu_3$ and $\mu_4$ with the exception of $\hat{\mu}_1$. The Abelian strength tensor $F_{\mu_{1}\mu_{2}}(x)$ is defined as follows:
\[
[\hspace{0.02cm}D_{\mu_{1}},D_{\mu_{2}}\hspace{0.02cm}] = i\hspace{0.02cm}e\hspace{0.005cm}F_{\mu_{1}\mu_{2}}(x).
\]
The proof of the identity (\ref{eq:8e}) is given in Appendix \ref{appendix_D}.\\
\indent
Our first step is to consider the contribution in (\ref{eq:8w}) due to the symmetrized part (\ref{eq:8r}). In view of a total symmetry over permutation of the indices, we have a chain of the equalities
\begin{align}
&\quad \bigl[\hspace{0.03cm}
\eta_{\mu_{1}}^{(\pm\hspace{0.02cm}3/2)}(q)\hspace{0.02cm} \eta_{\mu_{2}}^{(\mp\hspace{0.02cm}3/2)}(q)
\eta_{\mu_{3}}^{(\pm\hspace{0.02cm}3/2)}(q)\hspace{0.02cm} \eta_{\mu_{4}}^{(\mp\hspace{0.02cm}3/2)}(q)
\bigr]\hspace{0.015cm}
\bigl\{\!D_{\mu_{1}},D_{\mu_{2}},D_{\mu_{3}},D_{\mu_{4}}\!\hspace{0.02cm}\bigr\}
\label{eq:8t}\\[1.5ex]
&=\frac{1}{4!} \sum_{({\cal P})}\,\bigl[\hspace{0.03cm}
\eta_{\mu_{1}}^{(\pm\hspace{0.02cm}3/2)}(q)\hspace{0.02cm} \eta_{\mu_{2}}^{(\mp\hspace{0.02cm}3/2)}(q)
\eta_{\mu_{3}}^{(\pm\hspace{0.02cm}3/2)}(q)\hspace{0.02cm} \eta_{\mu_{4}}^{(\mp\hspace{0.02cm}3/2)}(q)
\bigr]\hspace{0.015cm}
\bigl\{\!D_{\mu_{1}},D_{\mu_{2}},D_{\mu_{3}},D_{\mu_{4}}\!\hspace{0.02cm}\bigr\}
\notag\\[0.6ex]
&=\frac{1}{4!}\sum_{({\cal P})} \biggl(\frac{5}{2}\,\bigl[\hspace{0.03cm}
\eta_{\mu_{1}}^{(\pm\hspace{0.02cm}3/2)}(q)\hspace{0.02cm} \eta_{\mu_{2}}^{(\mp\hspace{0.02cm}3/2)}(q)
\bigr]\delta_{\mu_{3}\mu_{4}}
+
\frac{9}{16}\,\delta_{\mu_{1}\mu_{2}}\delta_{\mu_{3}\mu_{4}}
\biggr)\hspace{0.015cm}
\bigl\{\!D_{\mu_{1}},D_{\mu_{2}},D_{\mu_{3}},D_{\mu_{4}}\!\hspace{0.02cm}\bigr\} \notag\\[0.6ex]
&=\biggl(\frac{5}{2}\,\bigl[\hspace{0.03cm}
\eta_{\mu_{1}}^{(\pm\hspace{0.02cm}3/2)}(q)\hspace{0.02cm} \eta_{\mu_{2}}^{(\mp\hspace{0.02cm}3/2)}(q)
\bigr] \delta_{\mu_{3}\mu_{4}}
+
\frac{9}{16}\,\delta_{\mu_{1}\mu_{2}}\delta_{\mu_{3}\mu_{4}}
\biggr)
\bigl\{\!D_{\mu_{1}},D_{\mu_{2}},D_{\mu_{3}},D_{\mu_{4}}\!\hspace{0.02cm}\bigr\}.\notag
\end{align}
In deriving this expression we have used the fact that the matrices $\eta_{\mu}^{(\mp\hspace{0.02cm}3/2)}(q)$ formally satisfy the relation like that for the $\beta$-matrices, Eq.\,(\ref{eq:1i}), and thereby we can use the completely symmetrized version  (\ref{eq:1g}) of the Bhabha-Madhavarao algebra written down for the $\eta$\hspace{0.02cm}-matrices.\\
\indent
Let us analyzed the contraction of the expression in parentheses in the last line of Eq.\,(\ref{eq:8t}) with the completely symmetrized product $\{D_{\mu_1}, D_{\mu_2}, D_{\mu_3}, D_{\mu_4}\}$. For this purpose it is convenient to present the latter product in the form of an expansion in the symmetrized product of three $D$\hspace{0.02cm}-\hspace{0.01cm}operators
\begin{equation}
\bigl\{\!D_{\mu_{1}},D_{\mu_{2}},D_{\mu_{3}},D_{\mu_{4}}\!\hspace{0.02cm}\bigr\}
\label{eq:8y}
\end{equation}
\[
=
D_{\mu_{1}}\bigl\{\!D_{\mu_{2}},D_{\mu_{3}},D_{\mu_{4}}\!\hspace{0.02cm}\bigr\}
+
D_{\mu_{2}}\bigl\{\!D_{\mu_{1}},D_{\mu_{3}},D_{\mu_{4}}\!\hspace{0.02cm}\bigr\}
+
D_{\mu_{3}}\bigl\{\!D_{\mu_{1}},D_{\mu_{2}},D_{\mu_{4}}\!\hspace{0.02cm}\bigr\}
+
D_{\mu_{4}}\bigl\{\!D_{\mu_{1}},D_{\mu_{2}},D_{\mu_{3}}\!\hspace{0.02cm}\bigr\}.
\]
The product $\{D_{\mu_1}, D_{\mu_2}, D_{\mu_3}\}$ was extensively used in the spin-1 case (Eq.\,(6.4) in \cite{markov_2015}).\\
\indent
At first we give consideration to the contraction of the term $\delta_{\mu_1\mu_2} \delta_{\mu_3\hspace{0.02cm}\mu_4}$ with (\ref{eq:8y}). The contraction of the product of two Kronecker deltas with the first term on the right-hand side of (\ref{eq:8y}) gives
\[
D_{\mu_{1}}\bigl\{\!D_{\mu_{2}},D_{\mu_{3}},D_{\mu_{4}}\!\hspace{0.02cm}\bigr\}
\hspace{0.03cm}\delta_{\mu_{1}\mu_{2}}\delta_{\mu_{3}\mu_{4}}
=
6\hspace{0.015cm}D^{4} - 2\hspace{0.015cm}i\hspace{0.01cm}e
D_{\mu}F_{\mu\nu}D_{\nu} - i\hspace{0.01cm}e D_{\mu}D_{\nu}F_{\mu\nu}.
\]
A similar contraction with the remaining terms in (\ref{eq:8y}) results in the same expression and, thus, we have
\begin{equation}
\bigl\{\!D_{\mu_{1}},D_{\mu_{2}},D_{\mu_{3}},D_{\mu_{4}}\!\hspace{0.02cm}\bigr\}
\hspace{0.03cm}\delta_{\mu_{1}\mu_{2}}\delta_{\mu_{3}\mu_{4}}
=
24\hspace{0.015cm}D^{4} - 8\hspace{0.015cm}i\hspace{0.015cm}e
D_{\mu}F_{\mu\nu}D_{\nu} - 4\hspace{0.02cm} i\hspace{0.01cm}e D_{\mu}D_{\nu}F_{\mu\nu}.
\label{eq:8u}
\end{equation}
\indent
We proceed now to the consideration of the contraction with the term containing the matrices $\eta_{\mu}^{(\pm\hspace{0.02cm}3/2)}(q)$, namely,
\[
\bigl[\hspace{0.03cm}
\eta_{\mu_{1}}^{(\pm\hspace{0.02cm}3/2)}(q)\hspace{0.02cm} \eta_{\mu_{2}}^{(\mp\hspace{0.02cm}3/2)}(q)
\bigr]
\delta_{\mu_{3}\mu_{4}}\hspace{0.02cm}
\bigl\{\!D_{\mu_{1}},D_{\mu_{2}},D_{\mu_{3}},D_{\mu_{4}}\!\hspace{0.02cm}\bigr\}.
\]
Unfortunately, the expressions here are already more cumbersome and tangled. We will give only the final result. The contraction with the first term on the right-hand side of (\ref{eq:8y}) can be put in the following form:
\begin{equation}
6\hspace{0.02cm}(\eta^{(\pm\hspace{0.02cm}3/2)\!}\cdot D)
(\eta^{(\mp\hspace{0.02cm}3/2)\!}\cdot D) D^{2}
-
4\hspace{0.015cm} i\hspace{0.015cm}e\hspace{0.02cm} (\eta^{(\pm\hspace{0.02cm}3/2)\!}\cdot D)
(\eta^{(\mp\hspace{0.02cm}3/2)\!}\cdot F\cdot D)
+
2\hspace{0.015cm} i\hspace{0.015cm}e\hspace{0.02cm} (\eta^{(\pm\hspace{0.02cm}3/2)\!}\cdot D)
(D\cdot F\cdot\eta^{(\mp\hspace{0.02cm}3/2)}).
\label{eq:8i}
\end{equation}
Here, for compactness of writing, we have introduced the notations: $(\eta\cdot D)\equiv\eta_{\mu}D_{\mu}\hspace{0.02cm};\, (\eta\cdot F\cdot D) \equiv \eta_{\mu} F_{\mu \nu} D_{\nu}$. Also, in the interest of brevity we have suppressed the $q$\hspace{0.01cm}-\hspace{0.01cm}dependence of the matrices $\eta_{\mu}^{(\mp\hspace{0.02cm}3/2)}(q)$. A similar contraction with the second term in (\ref{eq:8y}) is
\begin{equation}
6\hspace{0.02cm}(\eta^{(\pm\hspace{0.02cm}3/2)\!}\cdot D)
(\eta^{(\mp\hspace{0.02cm}3/2)\!}\cdot D) D^{2}
-\; 6\hspace{0.015cm} i\hspace{0.015cm}e\hspace{0.02cm}(\eta^{(\pm\hspace{0.02cm}3/2)\!} \cdot F
\cdot\eta^{(\mp\hspace{0.02cm}3/2)}) D^2
-
4\hspace{0.015cm} i\hspace{0.015cm}e\hspace{0.015cm}D_{\mu}
(\eta^{(\pm\hspace{0.02cm}3/2)\!}\cdot F\cdot D)\hspace{0.02cm}\eta^{(\mp\hspace{0.02cm}3/2)\!}_{\mu}
\label{eq:8o}
\end{equation}
\[
+\,
2\hspace{0.015cm}i\hspace{0.015cm}e\hspace{0.015cm}D_{\mu}
(D\cdot F\cdot\eta^{(\pm\hspace{0.02cm}3/2)})\hspace{0.02cm}\eta^{(\mp\hspace{0.02cm}3/2)\!}_{\mu}.
\]
Further, the contraction with the third term on the right-hand side of (\ref{eq:8y}) leads to
\[
2\hspace{0.02cm}(\eta^{(\pm\hspace{0.02cm}3/2)\!}\cdot D)
(\eta^{(\mp\hspace{0.02cm}3/2)\!}\cdot D) D^{2}
-
2\hspace{0.015cm}i\hspace{0.015cm}e\hspace{0.015cm} (\eta^{(\pm\hspace{0.02cm}3/2)\!}\cdot D)
(\eta^{(\mp\hspace{0.02cm}3/2)\!}\cdot F\cdot D)
-
2\hspace{0.015cm}i\hspace{0.015cm}e\hspace{0.015cm}(\eta^{(\pm\hspace{0.02cm}3/2)\!}\cdot F\cdot D)  (\eta^{(\mp\hspace{0.02cm}3/2)\!}\cdot D)
\]
\begin{equation}
-\; 2\hspace{0.02cm}e^{2} (\eta^{(\pm\hspace{0.02cm}3/2)\!}\cdot F\cdot F\cdot \eta^{(\mp\hspace{0.02cm}3/2)})
-
2\hspace{0.015cm} i\hspace{0.02cm}e\hspace{0.015cm} D^2
(\eta^{(\pm\hspace{0.02cm}3/2)\!}\cdot F\cdot\eta^{(\mp\hspace{0.02cm}3/2)})
+
4\hspace{0.015cm} D^2 (\eta^{(\pm\hspace{0.02cm}3/2)\!}\cdot D)
(\eta^{(\mp\hspace{0.02cm}3/2)\!}\cdot D)
\label{eq:8p}
\end{equation}
\[
-\, i\hspace{0.015cm}e\hspace{0.015cm} D_{\mu} (\eta^{(\pm\hspace{0.02cm}3/2)\!}\cdot F\cdot\eta^{(\mp\hspace{0.02cm}3/2)}) D_{\mu}
-
i\hspace{0.015cm}e\hspace{0.015cm}(D\cdot F\cdot\eta^{(\pm\hspace{0.02cm}3/2)})  (\eta^{(\mp\hspace{0.02cm}3/2)\!}\cdot D)
-
i\hspace{0.015cm}e\hspace{0.015cm} D_{\mu} F_{\mu\nu}\hspace{0.02cm} (D\cdot
\eta^{(\pm\hspace{0.02cm}3/2)})\hspace{0.03cm}\eta^{(\mp\hspace{0.02cm}3/2)\!}_{\nu}.
\]
Finally, the contraction with the fourth term has  the following form:
\begin{equation}
\begin{split}
4\hspace{0.02cm}&(\eta^{(\pm\hspace{0.02cm}3/2)\!}\cdot D)
(\eta^{(\mp\hspace{0.02cm}3/2)\!}\cdot D) D^{2}
-
4\hspace{0.015cm} i\hspace{0.015cm}e\hspace{0.015cm}(\eta^{(\pm\hspace{0.02cm}3/2)\!}\cdot D)
(\eta^{(\mp\hspace{0.02cm}3/2)\!}\cdot F\cdot D)
-
4\hspace{0.015cm}i\hspace{0.015cm}e\hspace{0.015cm}
(\eta^{(\pm\hspace{0.02cm}3/2)\!}\cdot F\cdot D) (\eta^{(\mp\hspace{0.02cm}3/2)\!}\cdot D)
\\[1ex]
&- 3\hspace{0.02cm}e^{2} (\eta^{(\pm\hspace{0.02cm}3/2)\!}\cdot F\cdot F\cdot \eta^{(\mp\hspace{0.02cm}3/2)})
+
2\hspace{0.015cm} D^2\hspace{0.01cm} (\eta^{(\pm\hspace{0.02cm}3/2)\!}\cdot D)
(\eta^{(\mp\hspace{0.02cm}3/2)\!}\cdot D)
-
 i\hspace{0.015cm}e\hspace{0.015cm} D^2\hspace{0.01cm} (\eta^{(\pm\hspace{0.02cm}3/2)\!} \cdot F
\cdot\eta^{(\mp\hspace{0.02cm}3/2)})
\\[1ex]
&- 2\hspace{0.015cm} i\hspace{0.02cm}e\hspace{0.015cm}D_{\mu} (\eta^{(\pm\hspace{0.02cm}3/2)\!} \cdot F\cdot\eta^{(\mp\hspace{0.02cm}3/2)})D_{\mu}
+ i\hspace{0.015cm}e\hspace{0.02cm}(\eta^{(\pm\hspace{0.02cm}3/2)\!}\cdot D)
(D\cdot F\cdot\eta^{(\mp\hspace{0.02cm}3/2)})
+
i\hspace{0.015cm}e\hspace{0.015cm} D_{\mu} D_{\nu} F_{\mu\lambda} \eta^{(\pm\hspace{0.02cm}3/2)\!}_{\lambda}\eta^{(\mp\hspace{0.02cm}3/2)\!}_{\nu}.
\end{split}
\label{eq:8a}
\end{equation}
\indent
The expressions (\ref{eq:8u})\,--\,(\ref{eq:8a}) completely define the contraction (\ref{eq:8t}) with the totally symmetrized product of the four covariant derivatives. Further, we need to consider a similar contraction with the remaining terms on the right-hand side of the identity (\ref{eq:8e}). By virtue of awkwardness of the expressions we give their final form in Appendix \ref{appendix_E}. We note only that here the contractions of the strength tensor $F_{\mu\nu}(x)$ with the spin structure $S_{\mu\nu}(q)$ will take place, as it was defined by the expression (\ref{eq:6u}).

\section{The Fock-Schwinger proper-time representation}
\setcounter{equation}{0}

In our paper \cite{markov_2015} we have discussed in detail a fundamental difficulty connected with the construction of the path integral representation for the spin-1 massive particle propagator interacting with a background gauge field within the standard Duffin-Kemmer-Petiau theory. It has been pointed out that this difficulty is closely related to noncommutativity of the DKP operator in the presence of an electromagnetic field
\[
L_{\rm D\hspace{0.03cm}\!K\hspace{0.02cm}\!P}(D) = \beta_{\mu}D_{\mu} + m\hspace{0.02cm} I,
\]
and the proper divisor
\[
d_{\rm D\hspace{0.03cm}\!K\hspace{0.02cm}\!P}(D) = \frac{1}{m}\,(\hspace{0.02cm}D^{2} - m^{2}\hspace{0.02cm})I  + \beta_{\mu}\hspace{0.02cm}D_{\mu}
-\frac{1}{m}\,\beta_{\mu}\beta_{\nu}\hspace{0.02cm}D_{\mu}D_{\nu}.
\]
In these expressions the matrices $\beta_{\mu}$ satisfy the trilinear relation (\ref{eq:1r}). A similar situation will take place and for the spin-3/2 case, where the operator $L(D)$  should be meant as the Bhabha operator (\ref{eq:1f}) with the replacement $\partial_{\mu} \rightarrow D_{\mu}$ and the divisor should be taken in the form (\ref{eq:1d}) with a similar replacement of derivatives. To circumvent the difficulty connected with noncommutativity of these two operators in constructing the path integral representation for the spin-3/2 massive particle propagator in the presence of an external gauge field, we can proceed in complete analogy to the spin-1 case.\\
\indent
In the case of the Bhabha-Madhavarao theory as a basic element of the construction, we take the fourth root of the fourth order wave operator
\[
\hat{\cal L}(z,D) =
A\biggl[\frac{\!1}{\,\epsilon^{1/2}(z)}\,\eta_{\mu}^{(\pm\hspace{0.02cm}3/2)}(z)
\hspace{0.02cm}D_{\mu} +
\Bigl(\hspace{0.02cm}{\cal P}_{3/2}^{(\pm)}(q) +
{\cal P}_{3/2}^{(\mp)}(q)\!\hspace{0.02cm}\Bigr)\hspace{0.02cm}m\biggr].
\]
Let us assume that the operator is a para-Fermi operator (parastatistics of order three). In this case it is not difficult to write an analog of the Fock-Schwinger proper-time representation for the inverse operator $\hat{\cal L}^{-1}$:
\begin{equation}
\frac{1}{\hat{\cal L}} \equiv \frac{\hat{\cal L}^{3}}{\hat{\cal L}^{4}} =
\label{eq:9q}
\end{equation}
\[
=\hspace{0.02cm}
i\!\int\limits_{0}^{\infty}\!d\hspace{0.02cm}\tau\!
\int\!\frac{d^{\,3}\chi}{\tau^{3}}\;\hspace{0.02cm}
{\rm e}^{\displaystyle{-\hspace{0.02cm}i\hspace{0.01cm}\tau\bigl (\hat{H}(z) - i\hspace{0.01cm}\epsilon\hspace{0.02cm}\bigr)
+
\frac{1}{2}\,\bigl(\hspace{0.02cm}\tau\hspace{0.02cm}[\hspace{0.03cm}\chi,\hat{\cal L}\hspace{0.02cm}]
+
\frac{1}{4}\,\tau^{2\,}[\hspace{0.03cm}\chi,\hat{\cal L}\hspace{0.02cm}]^{\hspace{0.02cm}2}\hspace{0.03cm}
-
\frac{5}{12}\,\tau^{3\,}[\hspace{0.03cm}\chi,\hat{\cal L}\hspace{0.02cm}]^{\hspace{0.02cm}3}\hspace{0.03cm}
\bigr)}},\quad
\epsilon\rightarrow +\hspace{0.01cm}0,\hspace{10cm}
\]
where
\[
\hat{H}(z) \equiv \hat{\cal L}^{\hspace{0.01cm}4}(z,D)
\]
and $\chi$ is a para-Grassmann variable of order $p\!\hspace{0.02cm}=\hspace{0.01cm}\!3$ (i.e. $\chi^{4}\!=\!0$) with the rules of an integration \cite{omote_1979, ohnuki_1980}
\[
\int\!d^{\,3}\chi  = \int\!d^{\,3}\chi\,[\hspace{0.03cm}\chi,\hat{\cal L}\hspace{0.03cm}]
=\int\!d^{\,3}\chi\,[\hspace{0.03cm}\chi,\hat{\cal L}\hspace{0.03cm}]^{\hspace{0.03cm}2} = 0,
\quad
\int\!d^{\,3}\chi\,[\hspace{0.03cm}\chi,\hat{\cal L}\hspace{0.03cm}]^{\hspace{0.03cm}3} = -8\hspace{0.02cm}\hat{\cal L}^{3}.
\]
We consider that the para-Grassmann variable $\chi$ and the operator $\hat{\cal L}$ conform to the
following rules of commutation:
\[
[\hspace{0.03cm}[\hspace{0.03cm}[\hspace{0.03cm}\chi,\hat{\cal L}\hspace{0.04cm}],
\hat{\cal L}\hspace{0.03cm}], \hat{\cal L}\hspace{0.03cm}]= 0,
\quad
[\hspace{0.03cm}[\hspace{0.03cm}[\hspace{0.02cm}\chi,\hat{\cal L}\hspace{0.04cm}], \chi\hspace{0.03cm}],\chi\hspace{0.03cm}] = 0
\]
and so on. As a proper para-supertime here it is necessary to take a tetrad $(\tau, \chi, \chi^2,\chi^3)$. The expression (\ref{eq:9q}) can be taken as the starting one for the construction of the desired path integral representation with the use of an appropriate system of coherent states. On constructing it is necessary to passage to the limit $z \rightarrow q$.

\section{Conclusion}
\setcounter{equation}{0}
\label{section_10}

In the present work we have set up the formalism needed to construct a fourth root of the fourth order wave operator within the framework of Bhabha-Madhavarao spin\hspace{0.01cm}-\hspace{0.01cm}3/2 theory. The fundamental point here is the introduction of the so-called deformed commutator, Eq.\,(\ref{eq:5ee}). By means of (\ref{eq:5ee}) a new set of the matrices $\eta_{\mu}$ was defined, instead of the original matrices $\beta_{\mu}$. One of our aims was to show that the fourth-order wave operator (and the relevant wave equation for the wave function $\varphi(x; z)$) can be obtained as a formal limit of some first-order differential operator to the fourth power, which is singular with respect to the deformation parameter $z$ when the latter approaches the primitive fourth root of unity $q$.\\
\indent
Unfortunately, we did not completely succeeded in reaching the purpose outlined above. Instead of an expected fourth order wave equation for the function $\varphi_0(x)$ in the expansion (\ref{eq:7y}) we obtain in the limit $z \rightarrow q$ a self-consistent system of two fourth order wave equations, Eqs.\,(\ref{eq:7j}) and (\ref{eq:7k}), for the functions $\varphi_{0}(x)$ and $\varphi_{\frac{1}{2}}(x)$ (more exactly, for their projections $\varphi_{0}^{(\pm\hspace{0.02cm}3/2)}(x;q)$ and $\varphi_{\frac{1}{2}}^{(\mp\hspace{0.02cm}3/2)}(x;q)$). The immediate reason of this circumstance is that in the expansion of the operator $\bigl[\hat{\cal L}^{(3/2)}(z; \partial)\bigr]^{4}$ in the term
\begin{equation}
\Bigr[A\hspace{0.03cm}\eta_{\mu}^{(\pm\hspace{0.02cm}3/2)}(z) A\hspace{0.03cm}\eta_{\nu}^{(\pm\hspace{0.02cm}3/2)}(z)
A\hspace{0.03cm}\eta_{\lambda}^{(\pm\hspace{0.02cm}3/2)}(z)
A\hspace{0.03cm}\eta_{\sigma}^{(\pm\hspace{0.02cm}3/2)}(z)
\Bigr]\partial_{\mu}\hspace{0.03cm}\partial_{\nu}
\hspace{0.03cm}\partial_{\lambda}\hspace{0.03cm}\partial_{\sigma},
\label{eq:10q}
\end{equation}
by virtue of the property (\ref{eq:6q}), all the contributions linear in $\delta$ vanish in the limit $z \rightarrow q$. It was for this reason that we have been forced to take the singular factor $1/\epsilon^{1/2}(z)$ in the definition of the first order operator $\hat{\cal L}^{(3/2)}(z; \partial)$ instead of the singular factor $1/\epsilon^{1/4}(z)$ (that it would seem more natural\footnote{\,Let us recall that in the DKP theory in constructing the cubic root of the third order wave equation we have used the singular factor $1/\epsilon^{1/3}(z)$ (the expression (6.1) in \cite{markov_2015}).}) since in (\ref{eq:10q}) the first nonzero terms in the expansion in $\delta$ are proportional to $\delta^2$. A negative consequence of such a choice is the survival of the singular contribution in the expansion of the operator $\bigl[\hat{\cal L}^{(3/2)}(z; \partial)\bigr]^{4}$, the expression (\ref{eq:7f}). The requirement of vanishing the singular contribution leads in turn to the necessity of introducing by hand the additional condition (\ref{eq:7s}):
$$
\varphi_{0}^{(\mp\hspace{0.02cm}3/2)}(x;q) \equiv 0.
$$
All the preceding finally results in a chain of equations for the functions $\varphi_0(x)$, $\varphi_{\frac{1}{2}}(x),\, \ldots$ in the expansion (\ref{eq:7y}).\\
\indent
The property (\ref{eq:6q}) for the matrices $\eta_{\mu}^{(\pm\hspace{0.02cm}3/2)}(q)$ is obviously too severe. The reason of this is perhaps a rather simplified choice of the representation (\ref{eq:3t}) for the matrix $\Omega$, the immediate consequence of which is the simple commutation rules (\ref{eq:3o}). Now it is not clear how we can improve the formalism suggested in the present paper so as to obtain a wave equation instead of a system of the wave equations of the fourth order in $\partial_{\mu}$. Any of attempts of an extension of this approach involves a drastic increase in the complexity of the theory and as a consequence leads to its ineffectiveness. Presumably, here it is necessary to invoke some new additional considerations of algebraic character.


\section*{\bf Acknowledgments}

This  work  was  supported  in  part  by the Council for Grants of the President  of  Russian  Foundation  for  state  support of the leading scientific schools, project NSh-8081.2016.9.


\begin{appendices}
\numberwithin{equation}{section}

\section{The $A_{\xi}$ matrix algebra}
\label{appendix_A}

In this Appendix we write out the basic relations for matrices $\xi_{\mu}$ of the $A_{\xi}$-algebra \cite{madhavarao_1946}:
\begin{align}
&\,\xi_{\mu}^{2} = \xi_{\mu} + \frac{3}{4},   \label{ap:A1}\\
&(\xi_{\mu}\hspace{0.02cm} \xi_{\nu} + \xi_{\nu}\hspace{0.02cm} \xi_{\mu}) + 2\hspace{0.02cm}\xi_{\mu}\hspace{0.02cm}\xi_{\nu}\hspace{0.02cm}\xi_{\mu} = -\frac{1}{2}\,\xi_{\nu},
\quad (\mu\neq\nu)  \label{ap:A2}  \\[0.6ex]
&(\xi_{\mu}\hspace{0.02cm}\xi_{\nu}\hspace{0.02cm}\xi_{\lambda}
-
\xi_{\lambda}\hspace{0.02cm}\xi_{\nu}\hspace{0.02cm}\xi_{\mu})
=
(\xi_{\nu}\hspace{0.02cm}\xi_{\lambda}\hspace{0.02cm}\xi_{\mu}
-
\xi_{\mu}\hspace{0.02cm}\xi_{\lambda}\hspace{0.02cm}\xi_{\nu})
=
(\xi_{\lambda}\hspace{0.02cm}\xi_{\mu}\hspace{0.02cm}\xi_{\nu}
-
\xi_{\nu}\hspace{0.02cm}\xi_{\mu}\hspace{0.02cm}\xi_{\lambda}),
\quad (\mu\neq\nu\neq\lambda)   \label{ap:A3} \\[1.6ex]
&\,\xi_{\mu}\hspace{0.02cm}(\xi_{\nu}\hspace{0.02cm}\xi_{\lambda}\hspace{0.02cm}\xi_{\sigma} - \xi_{\sigma}\hspace{0.02cm}\xi_{\lambda}\hspace{0.02cm}\xi_{\nu})
=
(\xi_{\nu}\hspace{0.02cm}\xi_{\lambda}\hspace{0.02cm}\xi_{\sigma} - \xi_{\sigma}\hspace{0.02cm}\xi_{\lambda}\hspace{0.02cm}\xi_{\nu})\hspace{0.02cm}\xi_{\mu},
\quad (\mu\neq\nu\neq\lambda\neq\sigma).   \label{ap:A4}
\end{align}
The conditions (\ref{ap:A2}) and (\ref{ap:A3}) can be presented in a somewhat different more compact form:
\begin{align}
&\hspace{3.75cm}\{\xi_{\mu}\hspace{0.02cm},\{\xi_{\mu}\hspace{0.02cm},\xi_{\nu}\}\} = \xi_{\nu},
\qquad (\mu\neq\nu)\hspace{6.6cm} ({\rm A}.2^{\hspace{0.02cm}\prime}) \notag \\[0.8ex]
&\hspace{3.8cm}[\hspace{0.03cm}\xi_{\mu}\hspace{0.02cm},\{\xi_{\nu}\hspace{0.02cm},\xi_{\lambda}\}] = 0,
\;\;\qquad (\mu\neq\nu\neq\lambda). \hspace{5.8cm} ({\rm A}.3^{\hspace{0.02cm}\prime}) \notag
\end{align}
In the $D = 4$ dimension Euclidean space, the total number of independent elements of the algebra is equal to 42 and the center of the algebra consists of three elements
\begin{equation}
I_{\xi}, \qquad P_{2} - P_{1}, \qquad P_{4} - 2\hspace{0.02cm}P_{3},
 \label{ap:A5}
\end{equation}
where $I_{\xi}$ is the unity matrix of the $A_{\xi}$-algebra and
\begin{equation}
P_{1} = \sum \xi_{\mu}, \quad
P_{2} = \sum \xi_{\mu}\hspace{0.02cm}\xi_{\nu}, \quad
P_{3} = \sum \xi_{\mu}\hspace{0.02cm}\xi_{\nu}\hspace{0.02cm}\xi_{\lambda}, \quad
P_{4} = \sum \xi_{\mu}\hspace{0.02cm}\xi_{\nu}\hspace{0.02cm}\xi_{\lambda}\hspace{0.02cm}\xi_{\sigma}
 \label{ap:A6}
\end{equation}
the indices is being unequal in each of the summations. We mention that in paper \cite{venkatachaliengar_1954} this algebra was investigated for an arbitrary $D$ and, in particular, it was shown that the center of the algebra is generated by a single element $\theta$, as it was defined by the expression (\ref{eq:3q}). Further, the $A_{\xi}$-\hspace{0.01cm}algebra has three irreducible representations of degree 1, 4 and 5, respectively. For completeness below we give an explicit form of the matrix representation of degree 4 of $A_{\xi}$ in which the matrix $\xi_{4}$ is diagonal \cite{madhavarao_1946}
\[
\xi_{1} =
\left(
\begin{array}{lrrl}
c & \frac{\!1}{2} & 0 & c \\
\frac{\!1}{2} & -s & 0 & s \\
0 & 0 & -\frac{\!1}{2} & 0 \\
c & s & 0 & 0
\end{array}
\right),
\qquad
\xi_{2} =
\left(
\begin{array}{rrll}
-s & 0 & \frac{\!1}{2} & s \\
0 & -\frac{\!1}{2} & 0 & 0 \\
\frac{\!1}{2} & 0 & c & c \\
s & 0 & c & 0
\end{array}
\right),
\]
\vspace{0.3cm}
\[
\hspace{0.4cm}
\xi_{3} =
\left(
\begin{array}{rlrl}
-\frac{\!1}{2} & 0 & 0 & 0 \\
0 & c & \frac{\!1}{2} & c \\
0 & \frac{\!1}{2} & -s & s \\
0 & c & s & 0
\end{array}
\right),
\qquad
\xi_{4} =
\left(
\begin{array}{rrrl}
-\frac{\!1}{2} & 0 & 0 & 0 \\
0 & -\frac{\!1}{2} & 0 & 0 \\
0 & 0 & -\frac{\!1}{2} & 0 \\
0 & 0 & 0 & \frac{3}{2}
\end{array}
\right),
\]
where
$$
s\equiv \sin\frac{\pi}{10} = \frac{\sqrt{\hspace{0.02cm}5} - 1}{4}, \qquad
c\equiv \cos\frac{2\hspace{0.015cm}\pi}{10} = \frac{\sqrt{\hspace{0.02cm}5} + 1}{4}.
$$
In paper \cite{madhavarao_1946} an explicit form of the representation of degree 5 of $A_{\xi}$, in which the matrix $\xi_4$ is diagonal, was also derived. Besides, the scheme of obtaining nondiagonal representations was presented, and the spurs of the elements of the basis of the $A_{\xi}$-\hspace{0.01cm}algebra in the three irreducible representations was calculated.


\section{\bf The solution of algebraic system (\ref{eq:3d})}
\label{appendix_B}
\numberwithin{equation}{section}

Let us write out once again the system (\ref{eq:3d}) for the unknown parameters $m$ and $n$:
\begin{equation}
\left\{
\begin{split}
&19\hspace{0.03cm}m^{2} + n^{2} + 4\hspace{0.02cm}m\hspace{0.02cm}n + 2\hspace{0.02cm}m = 0,\\
&\;m^{2} + m\hspace{0.02cm}n + \frac{1}{15}\,n = 0.
\end{split}
\label{ap:B1}
\right.
\end{equation}
It is easy to find one nontrivial solution of this system. For this purpose we express the product $m\hspace{0.02cm}n$ from the second equation and substitute it into the first equation. After simple algebraic transformations we obtain
$$
m\bigl(15\hspace{0.02cm}m + 2\bigr) + n\biggl(n - \frac{4}{15}\biggr) = 0.
$$
Here, the left-hand side is vanishing if one sets
\begin{equation}
m = -\frac{2}{15}, \qquad n = \frac{4}{15}\,.
\label{ap:B2}
\end{equation}
By direct substitution of this solution into (\ref{ap:B1}) we check that it is valid.\\
\indent
Let us derive the other solutions of the system (\ref{ap:B1}). From the second equation we obtain the parameter $m$ as a function of $n$. Here, we have two possibilities:
\begin{equation}
m^{\pm} = \frac{1}{2}\biggl( - n \pm \sqrt{n^{2} - \frac{4}{15}\,n}\;\biggr).
\label{ap:B3}
\end{equation}
The substitution of this relation into the first equation in (\ref{ap:B1}) leads in turn to
$$
\frac{17}{2}\,n^{2} - \frac{34}{15}\,n =
\biggl(\pm\hspace{0.02cm}\frac{15}{2}\,n \hspace{0.03cm}\mp\hspace{0.03cm} 1\biggr)\sqrt{n^{2} - \frac{4}{15}\,n}.
$$
Here, we have rearranged the last term to the right-hand side. After squaring, simple algebraic transformations lead us to the following expression
\begin{equation}
\biggl(\frac{17}{4}\biggr)^{\!\!2}n^{2}\biggl(n - \frac{4}{15}\biggr)^{\!\!2} =
\biggl(\frac{15}{4}\biggr)^{\!\!2}n\hspace{0.02cm}
\biggl(n - \frac{2}{15}\biggr)^{\!\!2}\biggl(n - \frac{4}{15}\biggr).
\label{ap:B4}
\end{equation}
This equation is of the fourth degree of nonlinearity in $n$. However, considering $n \neq 0$, one can reduce the equation of fourth order to that of third order. Further we see that $n = 4/15$ is really a root of the system (\ref{ap:B1}), and the value $m = -2/15$ is the only value corresponding to the $n = 4/15$, by virtue of vanishing the subradical expression in (\ref{ap:B3}).\\
\indent
Thus, considering $n\neq 0$ and $n\neq 4/15$, we can reduce equation (\ref{ap:B4}) to the quadratic one
$$
n^{2} - \frac{4}{15}\,n - \frac{1}{16} = 0
$$
which has the solutions
$$
n_{1} = \frac{5}{12}, \qquad n_{2} = -\frac{3}{20}.
$$
Thereby, by virtue of the relation (\ref{ap:B3}), we result in the other four possible solutions of the system (\ref{ap:B1}) (in addition to the solution (\ref{ap:B2})):
\begin{align}
&n_{1} = \frac{5}{12}, \qquad\;\; m_{1}^{+} = -\frac{1}{12}\,, \notag\\[1ex]
&n_{1} = \frac{5}{12}, \qquad\;\; m_{1}^{-} = -\frac{1}{3}\,, \notag\\[1ex]
&n_{2} = -\frac{3}{20}, \qquad m_{2}^{+} = \frac{1}{5}\,, \notag\\[1ex]
&n_{2} = -\frac{3}{20}, \qquad m_{2}^{-} = -\frac{1}{20}\,. \notag
\end{align}
By direct substitution of these solutions in (\ref{ap:B1}), we find that only the first and the last of them obey this system. Thus, we remain only with three nontrivial solutions as they was written out in (\ref{eq:3f}). A similar analysis of the solutions can be also performed for the choice $l\equiv l_{\rm I\!\hspace{0.02cm}I}= 1/4$ in the initial system (\ref{eq:3s}).


\section{\bf An explicit form of the matrix $A^3$}
\label{appendix_C}
\numberwithin{equation}{section}

In this Appendix we analyze the general structure of the matrix $A^3$ and derive finally a nontrivial relation between the matrices $A^{3}$ and $A$. The two following expressions
\begin{equation}
\begin{split}
&A\hspace{0.03cm} = \alpha\hspace{0.015cm}I  + \beta\hspace{0.04cm}\Omega
+ \gamma\hspace{0.04cm}\Omega^{\hspace{0.02cm}2} + \delta\hspace{0.03cm}\Omega^{\hspace{0.02cm}3},\\[1ex]
&A^{2} = a\hspace{0.015cm}I  + b\hspace{0.04cm}\Omega
+ c\hspace{0.06cm}\Omega^{\hspace{0.02cm}2} + d\hspace{0.04cm}\Omega^{\hspace{0.02cm}3}
\end{split}
\label{ap:C1}
\end{equation}
are the starting ones in determining an explicit form of the required matrix $A^3$. Here, as the coefficients ($\alpha,\hspace{0.02cm}\beta,\hspace{0.02cm}\gamma,\hspace{0.02cm}\delta$) we can take, for example, the coefficients which are given by formulas (\ref{eq:4t}), (\ref{eq:4p}), and (\ref{eq:2j}), correspondingly, and as the coefficients $(a,\,b,\,c,\,d)$ we can use those in formulae (\ref{eq:2f}) and (\ref{eq:2d}). A somewhat cumbersome multiplication of the matrices $A$ and $A^{2}$ with the use of the characteristic equation (\ref{eq:2r}) and its consequences (\ref{eq:2t}) leads to the following expression:
\begin{align}
A^{3} &= \Bigl(\alpha\hspace{0.02cm}a - \frac{9}{16}\,\beta\hspace{0.015cm}d - \frac{9}{16}\,\gamma\hspace{0.015cm}c - \frac{9}{16}\,\delta\hspace{0.015cm}b
-\frac{45}{32}\,\delta d\Bigr) I  \notag\\
&+\! \Bigl(\alpha\hspace{0.015cm}b + \beta\hspace{0.015cm}a - \frac{9}{16}\,\gamma\hspace{0.015cm}d -
\frac{9}{16}\,\delta\hspace{0.015cm}c\Bigr)\hspace{0.015cm}\Omega
 \label{ap:C2}\\
&+\! \Bigl(\alpha\hspace{0.015cm}c + \beta\hspace{0.015cm}b + \frac{5}{2}\,\beta\hspace{0.015cm}d +
\gamma\hspace{0.015cm}a + \frac{5}{2}\,\gamma\hspace{0.015cm}c + \frac{5}{2}\,\delta\hspace{0.015cm}b
+ \frac{91}{16}\,\delta\hspace{0.015cm}d \Bigr)\hspace{0.015cm}\Omega^{\hspace{0.03cm} 2}
\notag\\
&+\! \Bigl(\alpha\hspace{0.015cm}d + \beta\hspace{0.01cm}c + \gamma\hspace{0.015cm}b +
\frac{5}{2}\,\gamma\hspace{0.015cm}d + \delta\hspace{0.01cm}a + \frac{5}{2}\,\delta\hspace{0.01cm}c
\Bigr)\hspace{0.015cm}\Omega^{\hspace{0.03cm} 3}.
 \notag
\end{align}
To be specific, in this Appendix we will be concerned only with the solution (I) from a general set of possible values for the parameters $c,\,b$ and $d$ in (\ref{eq:2d}), namely,
\begin{equation}
(\textrm{I}): \qquad \quad    c_{1} = -\hspace{0.02cm}4\hspace{0.03cm}a,\qquad\quad
b_{1}^{+} = 18\hspace{0.03cm}a, \qquad\quad
d_{1}^{+} =  -\hspace{0.02cm}8\hspace{0.03cm}a.
\label{ap:C3}
\end{equation}
Let us consider the coefficient of the unity matrix $I$ in (\ref{ap:C2}). Substituting the values (\ref{ap:C3}) into this coefficient, we find
\begin{equation}
I: \qquad \qquad    \biggl(\hspace{0.02cm}\frac{9}{2}\,\beta +\hspace{0.02cm} \frac{9}{8}\,\delta\hspace{0.01cm}\biggr)\hspace{0.02cm}a.
\hspace{2cm}
\label{ap:C4}
\end{equation}
Further, as the parameters $\beta$ and $\delta$ in the preceding expression we take the first pair of the values from (\ref{eq:4p}):
\begin{equation}
\left\{
\begin{array}{l}
\beta_{1}^{(\pm)} = \biggl[\hspace{0.03cm}\displaystyle\frac{\!\!2}{3^3} \hspace{0.03cm}+\hspace{0.03cm} (\pm\hspace{0.02cm} i)
\biggl(2 - \displaystyle\frac{\!\!2}{3^3}\biggr) \biggr]\hspace{0.02cm}\alpha, \\[3ex]
\delta_{1}^{(\pm)} = \biggl[-\displaystyle\frac{\!8}{3^3} \hspace{0.03cm}+\hspace{0.03cm} (\pm\hspace{0.02cm} i)
\biggl(-\displaystyle\frac{\!8}{3^2} + \displaystyle\frac{\!8}{3^3}\biggr) \biggr]\hspace{0.02cm}
\alpha.
\end{array}
\right.
\label{ap:C5}
\end{equation}
Substitution of (\ref{ap:C5}) into (\ref{ap:C4}) gives
\begin{equation}
I: \qquad \quad    \biggl(\hspace{0.02cm}\frac{9}{2}\,\beta +\hspace{0.02cm} \frac{9}{8}\,\delta\hspace{0.01cm}\biggr)\hspace{0.02cm}a
=
(\pm i)\hspace{0.02cm}8\hspace{0.02cm}\alpha\hspace{0.02cm}a =
\frac{1}{m}\,(-\alpha).
\label{ap:C6}
\end{equation}
At the last step we have taken into account that
$$
a \equiv a_{\rm I} = \pm\, i\hspace{0.02cm}\frac{1}{8\hspace{0.02cm}m}\,,
$$
by virtue of (\ref{eq:2f}). The reasonings completely similar to the previous ones result in the following expressions for the coefficients of the matrices $\Omega$, $\Omega^{\hspace{0.03cm}2}$, and $\Omega^{\hspace{0.03cm}3}$:
\begin{align}
&\Omega\,: \qquad \frac{1}{m}\hspace{0.02cm}\biggl[\hspace{0.03cm}\displaystyle\frac{\!\!2}{3^3} \hspace{0.03cm}
-
\hspace{0.03cm} (\pm\hspace{0.02cm} i)
\biggl(2 - \displaystyle\frac{\!\!2}{3^3}\biggr) \biggr]\hspace{0.03cm}(-\alpha),
\label{ap:C7}\\[1ex]
&\Omega^{\hspace{0.03cm}2}\!: \qquad \frac{1}{m}\,(-\gamma),
\label{ap:C8}\\[1ex]
&\Omega^{\hspace{0.03cm}3}\!: \qquad \frac{1}{m}\hspace{0.02cm}\biggl[-\displaystyle\frac{\!8}{3^3} \hspace{0.03cm} - \hspace{0.03cm}
(\pm\hspace{0.02cm} i)
\biggl(-\displaystyle\frac{\!8}{3^2} + \displaystyle\frac{\!8}{3^3}\biggr) \biggr]\hspace{0.03cm}
(-\alpha).
\label{ap:C9}
\end{align}
Further we consider the permissible values of the parameter $\alpha$. The parameter is subject to the condition:
$$
\alpha^{4} = \frac{1}{4}\biggl(\frac{9}{8}\biggr)^{\!\!4}\frac{\!1}{m^2}\,.
$$
This condition defines four permissible values for the parameter $\alpha$, which we denote as
\begin{align}
&\alpha^{(\pm)}_{\Re} = \pm\,\frac{1}{\sqrt{2}}\,\biggl(\frac{9}{8}\biggr)\frac{\!1}{m^{1/2}}\,,\\
\notag
&\alpha^{(\pm)}_{\Im} = (\pm i)\,\frac{1}{\sqrt{2}}\,\biggl(\frac{9}{8}\biggr)\frac{\!1}{m^{1/2}}\,.
\notag
\end{align}
Here, our concern is only with the last two values $\alpha^{(\pm)}_{\Im}$, which are pure imaginary and for which the conjugation rule
$$
\alpha^{*} = -\alpha
$$
is true. In the choice of the pure imaginary values for the parameter $\alpha$, in formulae (\ref{ap:C6})\,--\,(\ref{ap:C9}) we can replace $(- \alpha)$ by $\alpha^{\ast}$ (we recall that the parameter $\gamma$ in (\ref{ap:C8}) is connected with the $\alpha$ by the relation $\gamma= - 4/9\hspace{0.03cm} \alpha$ and, therefore, in this expression it is necessary to replace $(- \gamma)$ by $\gamma^{\ast}$). Further, by virtue of the definitions of the parameters $\beta_1$ and $\delta_1$,
Eq.\,(\ref{ap:C5}), the expressions (\ref{ap:C7}) and (\ref{ap:C9}) in fact represent those for $\beta_1^{\ast}/m$ and $\delta_{1}^{\ast}/m$, correspondingly. With allowance made for all of the preceding, if one chooses $\alpha\equiv\alpha_{\Im}^{(\pm)}$, we finally obtain
$$
A^{3}\hspace{0.03cm} = \frac{1}{m}\,\bigl(\alpha^{\ast}\hspace{0.015cm}I  + \beta^{\ast}\hspace{0.04cm}\Omega +
\gamma^{\ast}\hspace{0.04cm}\Omega^{\hspace{0.02cm}2} + \delta^{\ast}\hspace{0.04cm}\Omega^{\hspace{0.02cm}3}\bigr).
$$
Comparing this expression with the expression for $A$ in (\ref{ap:C1}) and taking into account hermitian character of $\Omega$, we see that the matrix $A^3$ satisfies the following relation:
$$
A^{3} = \frac{1}{m}\,A^{\dagger}.
$$
The relation represents in effect a matrix analog of the relation between the primitive fourth roots $q$ and $q^{3}$: $q^{\ast}=q^3$. We recall for comparison that in the spin-1 case \cite{markov_2015} for a similar matrix $A$ we have obtained the relation in the form
$$
A^{2} = \frac{1}{m^{1/3}}\,A^{\dagger},
$$
which in turn is a matrix analog of the relation between the primitive cubic roots $q$ and $q^2$:
$q^{\ast} = q^{2}$.


\section{Proof of the identity (\ref{eq:8e})}
\label{appendix_D}
\setcounter{equation}{0}

Let us present a product of four covariant derivations $D_{\mu}$ in an identical form
\begin{equation}
D_{\mu_{1}}D_{\mu_{2}}D_{\mu_{3}}D_{\mu_{4}} = \bigl\{\!\hspace{0.02cm}D_{\mu_{1}},D_{\mu_{2}},D_{\mu_{3}},D_{\mu_{4}}\!\hspace{0.02cm}\bigr\}
\label{ap:D1}
\end{equation}
\[
-\,D_{\mu_{1}}\sideset{}{'}\sum_{({\cal P})}\! D_{\mu_{2}}D_{\mu_{3}}D_{\mu_{4}}
-
\sum_{({\cal P})}D_{\mu_{2}}D_{\hat{\mu}_{1}}D_{\mu_{3}}D_{\mu_{4}}
-
\sum_{({\cal P})}D_{\mu_{2}}D_{\mu_{3}}D_{\hat{\mu}_{1}}D_{\mu_{4}}
-
\sum_{({\cal P})}D_{\mu_{2}}D_{\mu_{3}}D_{\mu_{4}}D_{\hat{\mu}_{1}}.
\]
Here, the symbol $\sum_{({\cal P})}$ denotes summation over all permutations of free indices
$\mu_{2},\,\mu_{3}$ and $\mu_{4}$. There is no permutation over the index $\hat{\mu}_{1}$ (with hat above). Besides, in the second term on the right-hand side the prime on the summation symbol indicates that the term with the ``right'' order of the vector indices, i.e. the term $D_{\mu_2} D_{\mu_3} D_{\mu_4}$, is discarded.\\
\indent
Let us consider the third term on the right-hand side of (\ref{ap:D1}). We rearrange the covariant derivative $D_{\hat{\mu}_1}$ to take it outside the permutation sign $\sum_{({\cal P})}$ :
\begin{align}
\sum_{({\cal P})}D_{\mu_{2}}D_{\hat{\mu}_{1}}D_{\mu_{3}}D_{\mu_{4}}
&=
D_{\mu_{1}}\!\sum_{({\cal P})}D_{\mu_{2}}D_{\mu_{3}}D_{\mu_{4}}
+
\sum_{({\cal P})}\,[\hspace{0.015cm}D_{\mu_{2}},D_{\hat{\mu}_{1}}\hspace{0.015cm}]
D_{\mu_{3}}D_{\mu_{4}} \notag\\[0.6ex]
&\equiv
D_{\mu_{1}}\!\sum_{({\cal P})}D_{\mu_{2}}D_{\mu_{3}}D_{\mu_{4}}
- i\hspace{0.015cm}e \sum_{({\cal P})}F_{\hat{\mu}_{1}\mu_{2}}D_{\mu_{3}}D_{\mu_{4}}.
\notag
\end{align}
A similar manipulation with the fourth and fifth terms gives
\begin{align}
&\sum_{({\cal P})}D_{\mu_{2}}D_{\mu_{3}}D_{\hat{\mu}_{1}}D_{\mu_{4}}
=
D_{\mu_{1}}\!\sum_{({\cal P})}D_{\mu_{2}}D_{\mu_{3}}D_{\mu_{4}}
-
i\hspace{0.015cm}e \sum_{({\cal P})}F_{\hat{\mu}_{1}\mu_{2}}D_{\mu_{3}}D_{\mu_{4}}
-
i\hspace{0.015cm}e \sum_{({\cal P})}D_{\mu_{2}}F_{\hat{\mu}_{1}\mu_{3}}D_{\mu_{4}},
\notag\\[0.6ex]
&\sum_{({\cal P})}D_{\mu_{2}}D_{\mu_{3}}D_{\mu_{4}}D_{\hat{\mu}_{1}}
=
D_{\mu_{1}}\!\sum_{({\cal P})}D_{\mu_{2}}D_{\mu_{3}}D_{\mu_{4}}
-
i\hspace{0.015cm}e \sum_{({\cal P})}F_{\hat{\mu}_{1}\mu_{2}}D_{\mu_{3}}D_{\mu_{4}}
-
i\hspace{0.015cm}e \sum_{({\cal P})}D_{\mu_{2}}F_{\hat{\mu}_{1}\mu_{3}}D_{\mu_{4}} -
\notag\\[0.6ex]
&-
i\hspace{0.015cm}e \sum_{({\cal P})}D_{\mu_{2}}D_{\mu_{3}}F_{\hat{\mu}_{1}\mu_{4}}.
\notag
\end{align}
Taking into account the expressions above, the identity (\ref{ap:D1}) can be rewritten in a somewhat different form:
\begin{equation}
D_{\mu_{1}}D_{\mu_{2}}D_{\mu_{3}}D_{\mu_{4}} = \bigl\{\!\hspace{0.02cm}D_{\mu_{1}},D_{\mu_{2}},D_{\mu_{3}},D_{\mu_{4}}
\!\hspace{0.02cm}\bigr\}
-\,\biggl(\!D_{\mu_{1}}\!\sideset{}{'}\sum_{({\cal P})}\!D_{\mu_{2}}D_{\mu_{3}}D_{\mu_{4}}
+
3\hspace{0.01cm}D_{\mu_{1}}\!\sum_{({\cal P})}D_{\mu_{2}}D_{\mu_{3}}D_{\mu_{4}}
\!\biggr)
\label{ap:D2}
\end{equation}
\vspace{-0.2cm}
\[
+\;
3\hspace{0.01cm}i\hspace{0.01cm}e
\sum_{({\cal P})}F_{\hat{\mu}_{1}\mu_{2}}D_{\mu_{3}}D_{\mu_{4}}
+
2\hspace{0.015cm}i\hspace{0.015cm}e
\sum_{({\cal P})}D_{\mu_{2}}F_{\hat{\mu}_{1}\mu_{3}}D_{\mu_{4}}
+
i\hspace{0.015cm}e \sum_{({\cal P})}D_{\mu_{2}}D_{\mu_{3}}F_{\hat{\mu}_{1}\mu_{4}}.
\]
\indent
Now we turn to an analysis of the completely symmetrized product of three covariant derivatives, namely $\sum_{({\cal P})} D_{\mu_2} D_{\mu_3} D_{\mu_4}$. By analogy with (\ref{ap:D1}) we present it in an identical form:
\[
\sum_{({\cal P})}D_{\mu_{2}}D_{\mu_{3}}D_{\mu_{4}}
=
D_{\mu_{2}}\!\sum_{({\cal P})}D_{\mu_{3}}D_{\mu_{4}}
+
\sum_{({\cal P})}D_{\mu_{3}}D_{\hat{\mu}_{2}}D_{\mu_{4}}
+
\sum_{({\cal P})}D_{\mu_{3}}D_{\mu_{4}}D_{\hat{\mu}_{2}},
\]
where on the right-hand side there is no permutation over the index $\hat{\mu}_{2}$. The reasoning similar to the  above-mentioned one gives for each term
\begin{align}
&D_{\mu_{2}}\!\hspace{0.02cm}\sum_{({\cal P})}D_{\mu_{3}}D_{\mu_{4}} =
2\hspace{0.01cm}D_{\mu_{2}}D_{\mu_{3}}D_{\mu_{4}} - i\hspace{0.01cm}eD_{\mu_{2}}F_{\mu_{3}\mu_{4}},
\notag \\
&\sum_{({\cal P})}D_{\mu_{3}}D_{\hat{\mu}_{2}}D_{\mu_{4}} =
2\hspace{0.01cm}D_{\mu_{2}}D_{\mu_{3}}D_{\mu_{4}} - i\hspace{0.01cm}eD_{\mu_{2}}F_{\mu_{3}\mu_{4}} -
i\hspace{0.01cm}eF_{\mu_{2}\mu_{3}}D_{\mu_{4}},
\notag \\
&\sum_{({\cal P})}D_{\mu_{3}}D_{\mu_{4}}D_{\hat{\mu}_{2}} =
2\hspace{0.01cm}D_{\mu_{2}}D_{\mu_{3}}D_{\mu_{4}} - i\hspace{0.01cm}eD_{\mu_{2}}F_{\mu_{3}\mu_{4}} -
i\hspace{0.01cm}eF_{\mu_{2}\mu_{3}}D_{\mu_{4}}
-i\hspace{0.01cm}eD_{\mu_{3}}F_{\mu_{2}\mu_{4}}
\notag
\end{align}
and thus,
\[
\sum_{({\cal P})}D_{\mu_{2}}D_{\mu_{3}}D_{\mu_{4}}
=
6\hspace{0.01cm}D_{\mu_{2}}D_{\mu_{3}}D_{\mu_{4}} - 3\hspace{0.01cm}i\hspace{0.015cm}eD_{\mu_{2}}F_{\mu_{3}\mu_{4}} -
2\hspace{0.01cm}i\hspace{0.015cm}eF_{\mu_{2}\mu_{3}}D_{\mu_{4}}
-i\hspace{0.01cm}eD_{\mu_{3}}F_{\mu_{2}\mu_{4}}.
\]
By considering the identity above, the expression in parentheses in (\ref{ap:D2}) takes the final form:
\begin{equation}
D_{\mu_{1}}\!\sideset{}{'}\sum_{({\cal P})}\!D_{\mu_{2}}D_{\mu_{3}}D_{\mu_{4}}
+
3\hspace{0.01cm}D_{\mu_{1}}\!\sum_{({\cal P})}D_{\mu_{2}}D_{\mu_{3}}D_{\mu_{4}}
\label{ap:D3}
\end{equation}
\[
=
(5 + 3\cdot 6)\hspace{0.02cm}D_{\mu_{1}}D_{\mu_{2}}D_{\mu_{3}}D_{\mu_{4}}
-
12\hspace{0.015cm}i\hspace{0.01cm}eD_{\mu_{2}}F_{\mu_{3}\mu_{4}} -
8\hspace{0.015cm}i\hspace{0.01cm}eF_{\mu_{2}\mu_{3}}D_{\mu_{4}}
-4\hspace{0.015cm}i\hspace{0.01cm}eD_{\mu_{3}}F_{\mu_{2}\mu_{4}}.
\]
\indent
It remains to analyze the last three terms in (\ref{ap:D2}). It is already impossible to transform the next to last term without an explicit differentiation of the strength tensor $F_{\mu_1\hspace{0.02cm}\mu_3}$, and therefore this term remains unchanged (recall that all the expressions above are considered as the operator ones). For two remaining terms we can write
\begin{align}
\sum_{({\cal P})}F_{\hat{\mu}_{1}\mu_{2}}D_{\mu_{3}}D_{\mu_{4}}
&=
F_{\mu_{1}\mu_{2}}\{D_{\mu_{3}},D_{\mu_{4}}\}
+
F_{\mu_{1}\mu_{3}}\{D_{\mu_{2}},D_{\mu_{4}}\}
+
F_{\mu_{1}\mu_{4}}\{D_{\mu_{2}},D_{\mu_{3}}\} \label{ap:D4} \\
&= 2\hspace{0.015cm}\bigl(F_{\mu_{1}\mu_{2}}D_{\mu_{3}}D_{\mu_{4}}
+
F_{\mu_{1}\mu_{3}}D_{\mu_{2}}D_{\mu_{4}}
+
F_{\mu_{1}\mu_{4}}D_{\mu_{2}}D_{\mu_{3}}
\bigr) \notag \\[1ex]
&-i\hspace{0.015cm}e\hspace{0.02cm}\bigl(F_{\mu_{1}\mu_{2}}F_{\mu_{3}\mu_{4}}
+
F_{\mu_{1}\mu_{3}}F_{\mu_{2}\mu_{4}}
+
F_{\mu_{1}\mu_{4}}F_{\mu_{2}\mu_{3}}
\bigr) \notag
\end{align}
and similarly
\begin{align}
\sum_{({\cal P})}D_{\mu_{2}}D_{\mu_{3}}F_{\hat{\mu}_{1}\mu_{4}}
&=
2\hspace{0.02cm}\bigl(D_{\mu_{2}}D_{\mu_{3}} F_{\mu_{1}\mu_{4}} +
D_{\mu_{2}}D_{\mu_{4}}F_{\mu_{1}\mu_{3}}
+
D_{\mu_{3}}D_{\mu_{4}}F_{\mu_{1}\mu_{2}}
\bigr)\hspace{0.7cm} \label{ap:D5} \\[1ex]
&-i\hspace{0.02cm}e\hspace{0.02cm}\bigl(F_{\mu_{1}\mu_{2}}F_{\mu_{3}\mu_{4}}
+
F_{\mu_{1}\mu_{3}}F_{\mu_{2}\mu_{4}}
+
F_{\mu_{1}\mu_{4}}F_{\mu_{2}\mu_{3}}
\bigr). \notag
\end{align}
We tried to achieve the maximum ordering of the indices in writing these expressions. Substituting (\ref{ap:D3})\,--\,(\ref{ap:D5}) into (\ref{ap:D2}) and collecting the terms similar in structure, we lead to (\ref{eq:8e}).


\section{Interaction terms with the spin structures}
\label{appendix_E}
\setcounter{equation}{0}

In this Appendix we present an explicit form of the contraction of the matrix structure
\begin{equation}
\eta_{\mu_{1}}^{(\pm\hspace{0.02cm}3/2)}(q)\hspace{0.02cm} \eta_{\mu_{2}}^{(\mp\hspace{0.02cm}3/2)}(q)
\eta_{\mu_{3}}^{(\pm\hspace{0.02cm}3/2)}(q)\hspace{0.02cm} \eta_{\mu_{4}}^{(\mp\hspace{0.02cm}3/2)}(q)
\label{ap:E1}
\end{equation}
with the remaining four contributions on the right-hand side of the identity (\ref{eq:8e}). The contraction of the structure (\ref{ap:E1}) with the first contribution containing the terms of the $D_{\mu_1} D_{\mu_2} F_{\mu_3 \mu_4\!}$-\hspace{0.02cm}type, namely,
$$
12\hspace{0.015cm}i\hspace{0.01cm}eD_{\mu_{1}}D_{\mu_{2}}F_{\mu_{3}\mu_{4}}
+
4\hspace{0.015cm}i\hspace{0.005cm}eD_{\mu_{1}}D_{\mu_{3}}F_{\mu_{2}\mu_{4}}
+
2\hspace{0.015cm}i\hspace{0.005cm}e\bigl(D_{\mu_{2}}D_{\mu_{3}}F_{\mu_{1}\mu_{4}}
+
D_{\mu_{2}}D_{\mu_{4}}F_{\mu_{1}\mu_{3}}
+
D_{\mu_{3}}D_{\mu_{4}}F_{\mu_{1}\mu_{2}}\bigr)
$$
can be resulted in the following form:
\begin{align}
i\hspace{0.01cm}e D_{\mu_{1}} D_{\mu_{2}} F_{\mu_{3} \mu_{4}\!}
\Big\{\!&-2\hspace{0.02cm}\eta_{\mu_{1}}^{(\pm\hspace{0.02cm} 3/2)}(q)S_{\mu_2\mu_3}^{\rm (I)}(q) \hspace{0.02cm}\eta_{\mu_{4}}^{(\mp\hspace{0.02cm} 3/2)}(q)
+
2\hspace{0.02cm}  \eta_{\mu_{1}}^{(\pm\hspace{0.02cm} 3/2)}(q) S_{\mu_2\mu_4}^{\rm (I)}(q) \hspace{0.02cm}\eta_{\mu_{3}}^{(\mp\hspace{0.02cm} 3/2)}(q)
\label{ap:E2} \\[1ex]
&- S_{\mu_1\mu_3}^{\rm (I\!\hspace{0.025cm}I)}(q)
\bigl(
\eta_{\mu_{2}}^{(\pm\hspace{0.02cm} 3/2)}(q)\hspace{0.02cm}\eta_{\mu_{4}}^{(\mp\hspace{0.02cm} 3/2)}(q)
+
\eta_{\mu_{4}}^{(\pm\hspace{0.02cm}3/2)}(q)\hspace{0.02cm}\eta_{\mu_{2}}^{(\mp\hspace{0.02cm} 3/2)}(q) \bigr)
\notag\\[1ex]
&+ S_{\mu_1\mu_4}^{\rm (I\!\hspace{0.025cm}I)}(q)
\bigl(
\eta_{\mu_{2}}^{(\pm\hspace{0.02cm} 3/2)}(q)\hspace{0.02cm}\eta_{\mu_{3}}^{(\mp\hspace{0.02cm} 3/2)}(q)
+
\eta_{\mu_{3}}^{(\pm\hspace{0.02cm}3/2)}(q)\hspace{0.02cm}\eta_{\mu_{2}}^{(\mp\hspace{0.02cm} 3/2)}(q)\bigr)
\notag\\[1ex]
&+ 6\hspace{0.02cm}  \eta_{\mu_{1}}^{(\pm\hspace{0.02cm} 3/2)}(q)\hspace{0.02cm} \eta_{\mu_{2}}^{(\mp\hspace{0.02cm} 3/2)}(q) \bigl( {\cal P}_{3/2}^{(\pm\hspace{0.02cm})}(q) S_{\mu_{3} \mu_{4}}(q) \bigr)
+
\bigl( {\cal P}_{3/2}^{(\pm\hspace{0.02cm})}(q) S_{\mu_{3}\mu_{4}}(q) \bigr)\eta_{\mu_{1}}^{(\pm\hspace{0.02cm} 3/2)}(q)\hspace{0.02cm} \eta_{\mu_{2}}^{(\mp\hspace{0.02cm} 3/2)}(q) \!\Big\}.\notag
\end{align}
Here, the expressions for the spin structures $S_{\mu\nu}^{\rm (I)}(q),
S_{\mu\nu}^{\rm (I\!\hspace{0.025cm}I)}(q)$ and $\bigl({\cal P}^{(\pm)}(q)S_{\mu \nu}(q)\bigr)$ are defined by formulas (\ref{eq:6u}) and (\ref{eq:6eee}), correspondingly.\\
\indent
Further, a similar contraction (E.1) with the second contribution containing the terms of the  $F_{\mu_1 \mu_2} D_{\mu_3} D_{\mu_4\!\hspace{0.02cm}}$-\hspace{0.01cm}type:
$$
6\hspace{0.015cm}i\hspace{0.005cm}e\bigl(F_{\mu_{1}\mu_{2}}D_{\mu_{3}}D_{\mu_{4}}
+
F_{\mu_{1}\mu_{3}}D_{\mu_{2}}D_{\mu_{4}}
+
F_{\mu_{1}\mu_{4}}D_{\mu_{2}}D_{\mu_{3}}\bigr)
$$
gives us
\begin{align}
3\hspace{0.015cm}i\hspace{0.005cm}e F_{\mu_{3}  \mu_{4}} D_{\mu_{1}} D_{\mu_{2}}
\Big\{\!
&- \!S_{\mu_1\mu_3}^{\rm (I\!\hspace{0.025cm}I)}(q)
\bigl(
\eta_{\mu_{2}}^{(\pm\hspace{0.02cm} 3/2)}(q)\hspace{0.02cm}\eta_{\mu_{4}}^{(\mp\hspace{0.02cm} 3/2)}(q)
+
\eta_{\mu_{4}}^{(\pm\hspace{0.02cm} 3/2)}(q)\hspace{0.02cm}\eta_{\mu_{2}}^{(\mp\hspace{0.02cm} 3/2)}(q)
\bigr)
\label{ap:E3} \\[1ex]
&+ S_{\mu_1\mu_4}^{\rm (I\!\hspace{0.025cm}I)}(q)
\bigl(
\eta_{\mu_{2}}^{(\pm\hspace{0.02cm} 3/2)}(q)\hspace{0.02cm}\eta_{\mu_{3}}^{(\mp\hspace{0.02cm} 3/2)}(q)
+
\eta_{\mu_{3}}^{(\pm\hspace{0.02cm} 3/2)}(q)\hspace{0.02cm}\eta_{\mu_{2}}^{(\mp\hspace{0.02cm} 3/2)}(q)
\bigr)  \notag\\[1ex]
&+ \bigl( {\cal P}_{3/2}^{(\pm\hspace{0.02cm})}(q) S_{\mu_{3}\mu_{4}}(q) \bigr)\eta_{\mu_{1}}^{(\pm\hspace{0.02cm} 3/2)}(q)\hspace{0.02cm} \eta_{\mu_{2}}^{(\mp\hspace{0.02cm} 3/2)}(q) \hspace{0.02cm}\!\Big\}.  \notag
\end{align}
The third contribution containing the terms of the $
D_{\mu_1}F_{\mu_3 \mu_4}D_{\mu_2\!\hspace{0.02cm}}$-\hspace{0.01cm}type
$$
8\hspace{0.015cm}i\hspace{0.005cm}eD_{\mu_{1}}F_{\mu_{2}\mu_{3}}D_{\mu_{4}}
+
2\hspace{0.015cm}i\hspace{0.005cm}e\sum_{({\cal P})} D_{\mu_{2}}F_{\hat{\mu}_{1}\mu_{3}}D_{\mu_{4}}
$$
for the contraction with (\ref{ap:E1}) can be recast in the following form:
\begin{align}
i\hspace{0.005cm}e D_{\mu_{1}} F_{\mu_{3} \mu_{4}} D_{\mu_{2}}\Bigl\{\!
&- \!S_{\mu_1\mu_3}^{\rm (I\!\hspace{0.025cm}I)}(q)
\bigl(\eta_{\mu_{2}}^{(\pm\hspace{0.02cm} 3/2)}(q)\hspace{0.02cm}
\eta_{\mu_{4}}^{(\mp\hspace{0.02cm} 3/2)}(q)
+
\eta_{\mu_{4}}^{(\pm\hspace{0.02cm} 3/2)}(q)\hspace{0.02cm}
\eta_{\mu_{2}}^{(\mp\hspace{0.02cm} 3/2)}(q)\bigr)
\label{ap:E4} \\[1ex]
&+ S_{\mu_1\mu_4}^{\rm (I\!\hspace{0.025cm}I)}(q)
\bigl(\eta_{\mu_{2}}^{(\pm\hspace{0.02cm} 3/2)}(q)\hspace{0.02cm}
\eta_{\mu_{3}}^{(\mp\hspace{0.02cm} 3/2)}(q)
+
\eta_{\mu_{3}}^{(\pm\hspace{0.02cm} 3/2)}(q)\hspace{0.02cm}
\eta_{\mu_{2}}^{(\mp\hspace{0.02cm} 3/2)}(q)\bigr)
\notag \\[1ex]
&- S_{\mu_2\mu_3}^{\rm (I\!\hspace{0.025cm}I)}(q)
\bigl(\eta_{\mu_{1}}^{(\pm\hspace{0.02cm} 3/2)}(q)\hspace{0.02cm}
\eta_{\mu_{4}}^{(\mp\hspace{0.02cm} 3/2)}(q)
+
\eta_{\mu_{4}}^{(\pm\hspace{0.02cm} 3/2)}(q)\hspace{0.02cm}
\eta_{\mu_{1}}^{(\mp\hspace{0.02cm} 3/2)}(q)\bigr)
\notag \\[1ex]
&+S_{\mu_2\mu_4}^{\rm (I\!\hspace{0.025cm}I)}(q)
\bigl(\eta_{\mu_{1}}^{(\pm\hspace{0.02cm} 3/2)}(q)\hspace{0.02cm}
\eta_{\mu_{3}}^{(\mp\hspace{0.02cm} 3/2)}(q)
+
\eta_{\mu_{3}}^{(\pm\hspace{0.02cm} 3/2)}(q)\hspace{0.02cm}
\eta_{\mu_{1}}^{(\mp\hspace{0.02cm} 3/2)}(q)\bigr)
\notag \\[1ex]
&+ \bigl( {\cal P}_{3/2}^{(\pm\hspace{0.02cm})}(q) S_{\mu_{3}\mu_{4}}(q) \bigr) \bigl( \eta_{\mu_{1}}^{(\pm\hspace{0.02cm} 3/2)}(q)  \, \eta_{\mu_{2}}^{(\mp\hspace{0.02cm} 3/2)}(q) + \eta_{\mu_{2}}^{(\pm\hspace{0.02cm} 3/2)}(q)  \, \eta_{\mu_{1}}^{(\mp\hspace{0.02cm} 3/2)}(q) \bigr)
 \notag\\[1ex]
&+ 4\hspace{0.02cm}\eta_{\mu_{1}}^{(\pm\hspace{0.02cm} 3/2)}(q)
\bigl( {\cal P}_{3/2}^{(\mp\hspace{0.02cm})}(q) S_{\mu_{3}\mu_{4}}(q) \bigr) \eta_{\mu_{2}}^{(\mp\hspace{0.02cm} 3/2)}(q) \!\Bigr\}.
\notag
\end{align}
Finally, the last contribution in the identity (\ref{eq:8e}) that does not contain the covariant derivatives, namely,
$$
4\hspace{0.015cm}e^{2}\Bigl\{\!\,F_{\mu_{1}\mu_{2}}F_{\mu_{3}\mu_{4}}
+ F_{\mu_{1}\mu_{3}}F_{\mu_{2}\mu_{4}} + F_{\mu_{1}\mu_{4}}F_{\mu_{2}\mu_{3}}\!\Bigr\},
$$
gives for the contraction with (\ref{ap:E1}) the following perfectly symmetric expression:
\begin{align}
&\frac{e^2 }{2}\,F_{\mu_{1} \mu_{2}} F_{\mu_{3} \mu_{4}}\! \Bigl\{\!\hspace{0.02cm}
\bigl({\cal P}_{3/2}^{(\pm\hspace{0.02cm})}(q) S_{\mu_{1}\mu_{2}}(q)\bigr)
\bigl({\cal P}_{3/2}^{(\pm\hspace{0.02cm})}(q) S_{\mu_{3}\mu_{4}}(q)\bigr)
\!+\!
\bigl({\cal P}_{3/2}^{(\pm\hspace{0.02cm})}(q) S_{\mu_{3}\mu_{4}}(q)\bigr)
\bigl({\cal P}_{3/2}^{(\pm\hspace{0.02cm})}(q) S_{\mu_{1}\mu_{2}}(q)\bigr)
\label{ap:E5} \\[1ex]
&+ \bigl( \eta_{\mu_{1}}^{(\pm\hspace{0.02cm} 3/2)}(q)\hspace{0.02cm}
\eta_{\mu_{3}}^{(\mp\hspace{0.02cm} 3/2)}(q)
+
\eta_{\mu_{3}}^{(\pm\hspace{0.02cm} 3/2)}(q)\hspace{0.02cm}
\eta_{\mu_{1}}^{(\mp\hspace{0.02cm} 3/2)}(q)\bigr)
\bigl(\eta_{\mu_{2}}^{(\pm\hspace{0.02cm} 3/2)}(q)\hspace{0.02cm}
\eta_{\mu_{4}}^{(\mp\hspace{0.02cm} 3/2)}(q)
+
\eta_{\mu_{4}}^{(\pm\hspace{0.02cm} 3/2)}(q)\hspace{0.02cm}
\eta_{\mu_{2}}^{(\mp\hspace{0.02cm} 3/2)}(q)\bigr)
\notag \\[1.5ex]
&+ \bigl( \eta_{\mu_{2}}^{(\pm\hspace{0.02cm} 3/2)}(q)\hspace{0.02cm}
\eta_{\mu_{4}}^{(\mp\hspace{0.02cm} 3/2)}(q)
+
\eta_{\mu_{4}}^{(\pm\hspace{0.02cm} 3/2)}(q)\hspace{0.02cm}
\eta_{\mu_{2}}^{(\mp\hspace{0.02cm} 3/2)}(q)\bigr)
\bigl(\eta_{\mu_{1}}^{(\pm\hspace{0.02cm} 3/2)}(q)\hspace{0.02cm}
\eta_{\mu_{3}}^{(\mp\hspace{0.02cm} 3/2)}(q)
+
\eta_{\mu_{3}}^{(\pm\hspace{0.02cm} 3/2)}(q)\hspace{0.02cm}
\eta_{\mu_{1}}^{(\mp\hspace{0.02cm} 3/2)}(q)\bigr)
\notag \\[1ex]
&- \bigl(\eta_{\mu_{2}}^{(\pm\hspace{0.02cm} 3/2)}(q)\hspace{0.02cm}
\eta_{\mu_{3}}^{(\mp\hspace{0.02cm} 3/2)}(q)
+
\eta_{\mu_{3}}^{(\pm\hspace{0.02cm} 3/2)}(q)\hspace{0.02cm}
\eta_{\mu_{2}}^{(\mp\hspace{0.02cm} 3/2)}(q)\bigr)
\bigl(\eta_{\mu_{1}}^{(\pm\hspace{0.02cm} 3/2)}(q)\hspace{0.02cm}
\eta_{\mu_{4}}^{(\mp\hspace{0.02cm} 3/2)}(q)
+
\eta_{\mu_{4}}^{(\pm\hspace{0.02cm} 3/2)}(q)\hspace{0.02cm}
\eta_{\mu_{1}}^{(\mp\hspace{0.02cm} 3/2)}(q) \bigr)
\notag \\[1.5ex]
&- \bigl(\eta_{\mu_{1}}^{(\pm\hspace{0.02cm} 3/2)}(q)\hspace{0.02cm}
\eta_{\mu_{4}}^{(\mp\hspace{0.02cm} 3/2)}(q)
+
\eta_{\mu_{4}}^{(\pm\hspace{0.02cm} 3/2)}(q)\hspace{0.02cm}
\eta_{\mu_{1}}^{(\mp\hspace{0.02cm} 3/2)}(q)\bigr)
\bigl(\eta_{\mu_{2}}^{(\pm\hspace{0.02cm} 3/2)}(q)\hspace{0.02cm}
\eta_{\mu_{3}}^{(\mp\hspace{0.02cm} 3/2)}(q)
+
\eta_{\mu_{3}}^{(\pm\hspace{0.02cm} 3/2)}(q)\hspace{0.02cm}
\eta_{\mu_{2}}^{(\mp\hspace{0.02cm} 3/2)}(q) \bigr)\!\Bigr\}.
\notag
\end{align}
In deriving the expressions (\ref{ap:E2})\,--\,(\ref{ap:E5}) we have used the identity:
$$
\eta_{\nu} \eta_{\mu} \eta_{\lambda} - \eta_{\lambda} \eta_{\mu} \eta_{\nu} = \eta_{\mu} [\eta_{\nu}, \eta_{\lambda}] - \bigl(\hspace{0.02cm}[\eta_{\mu}, \eta_{\nu}]\eta_{\lambda} -
[\eta_{\mu}, \eta_{\lambda}] \eta_{\nu} \bigr)
$$
and the nilpotency property for the $\eta$\hspace{0.01cm}-matrices, Eq.\,(\ref{eq:6q}).
\end{appendices}

\end{document}